\PassOptionsToPackage{dvipsnames}{xcolor}
\documentclass[sigconf, nonacm, pdfa]{acmart}
\newcommand\vldbvolume{17}
\newcommand\vldbissue{10}

\newcommand\vldbauthors{\authors}
\newcommand\vldbtitle{\shorttitle} 
 
\newcommand\vldbpagestyle{empty} 

\usepackage{enumitem}
\usepackage{cleveref}
\usepackage{pifont}
\usepackage{graphicx}
\usepackage{balance}  
\usepackage[linesnumbered,vlined,ruled]{algorithm2e}
\usepackage{lipsum}
\usepackage{amsmath}
\usepackage{physics}
\usepackage{subfigure}
\usepackage{tabularx}
\usepackage{color}
\usepackage{colortbl}
\usepackage{float}
\usepackage{listings}
\usepackage{multirow}
\usepackage[normalem]{ulem}
 \usepackage{longtable}
\usepackage{enumitem}
\usepackage{multirow}
\usepackage{booktabs}
\usepackage{capt-of}
\usepackage{pifont}
\usepackage{ulem}
\usepackage{soul}
\usepackage{cancel}
\usepackage{bm}
\usepackage{url}
\usepackage{caption}
\usepackage{tikz}

\def\BibTeX{{\rm B\kern-.05em{\sc i\kern-.025em b}\kern-.08em
    T\kern-.1667em\lower.7ex\hbox{E}\kern-.125emX}}

 \graphicspath{{./Graph/}, {./Fig/}, {./Legend/}}

\long\def\comment#1{}

\newcounter{example}[section]
\renewcommand{\theexample}{\nthesection.\arabic{example}}

\newcounter{definition}[section]
\renewcommand{\thedefinition}{\nthesection.\arabic{definition}}

\newcounter{theorem}[section]
\renewcommand{\thetheorem}{\nthesection.\arabic{theorem}}

\newcounter{lemma}[section]
\renewcommand{\thelemma}{\nthesection.\arabic{lemma}}
        
\newcounter{proposition}[section]
\renewcommand{\theproposition}{\nthesection.\arabic{proposition}}

\newcounter{remark}[section]
\renewcommand{\theremark}{\nthesection.\arabic{remark}}

\newcommand{\nthesection}{\arabic{section}}

\newcommand{\stitle}[1]{\vspace{1ex} \noindent{\bf #1}}

\newcommand{\green}[1]{\textcolor{green}{}}

\newcommand{\kw}[1]{{\ensuremath {\mathsf{#1}}}\xspace}
\newcommand{\kwnospace}[1]{{\ensuremath {\mathsf{#1}}}}

\newcommand{\IMDB}{\kw{IMDB}}

\usepackage{algorithmic}
\usepackage{graphicx}
\usepackage{textcomp}
\usepackage{enumitem}
\usepackage{cleveref}
\usepackage{pifont}
\usepackage{cleveref}
\usepackage{makecell}
\usepackage{balance}
\usepackage{booktabs}
\usepackage{adjustbox}
\usepackage{multirow}
\usepackage{fancyvrb}
\usepackage[dvipsnames]{xcolor}
\usepackage[breakable]{tcolorbox} 

\crefformat{section}{\S#2#1#3} 
\crefformat{subsection}{\S#2#1#3}
\long\def\comment#1{}

\newcommand{\algocomment}[1]{\footnotesize $\rhd$ \emph{#1}}
\definecolor{LightCyan}{rgb}{0.88,1,1}
\definecolor{LightRed}{rgb}{1,0.88,1}
\definecolor{LightYellow}{rgb}{1,1,0.88}
\definecolor{LightGray}{gray}{0.8}

\usepackage{caption}
\usepackage{subfigure}
\usepackage{listings}

\lstdefinestyle{sqlstyle}{
    language=SQL,
    backgroundcolor=\color{white},
    basicstyle=\ttfamily,
    keywordstyle=\color{blue},
    commentstyle=\color{green},
    stringstyle=\color{red},
    showstringspaces=false,
    numbers=none,
    breaklines=true,
}

\newcommand{\loss}{\mathcal{L}}

\newcommand{\HashJoin}{{{\sl HashJoin}}\xspace}
\newcommand{\MergeJoin}{{{\sl MergeJoin}}\xspace}
\newcommand{\NestLoop}{{{\sl NestLoopJoin}}\xspace}
\newcommand{\SeqScan}{{{\sl SeqScan}}\xspace}

\newcommand{\Oracle}{{{\sl Oracle}}\xspace}
\newcommand{\Postgres}{{{\sl PostgreSQL}}\xspace}
\newcommand{\db}{{{\sl \textit{DB2}}}\xspace}
\newcommand{\bao}{{{\sl Bao}}\xspace}
\newcommand{\hybrid}{{{\sl HybridQO}}\xspace}
\newcommand{\QueryInstruct}{{{\sc QInstruct}}\xspace}
\newcommand{\LLMQO}{{\text{LLM-QO}}\xspace}
\newcommand{\QIT}{{\sc Qit}\xspace}
\newcommand{\QDPO}{{\sc Qdpo}\xspace}
\newcommand{\imdb}{\kw{IMDB}}
\newcommand{\dsb}{\kw{DSB}}

\newcommand{\tpcds}{\kwnospace{TPC}\mbox{-}\kw{DS}}
\newcommand{\job}{\kwnospace{JOB}\mbox{-}\kw{light}}

\begin{document}

\title{Can Large Language Models Be Query Optimizer for Relational Databases?}

\author{Jie Tan}
\affiliation{%
	\institution{The Chinese University of Hong Kong}
	\city{}
	\country{}
}
\email{jtan@se.cuhk.edu.hk}

\author{Kangfei Zhao}
\affiliation{%
	\institution{The Chinese University of Hong Kong}
	\city{}
	\country{}
}
\email{zkf1105@gmail.com}

\author{Rui Li}
\affiliation{%
	\institution{The Chinese University of Hong Kong}
	\city{}
	\country{}
}
\email{lirui@se.cuhk.edu.hk}

\author{Jeffrey Xu Yu}
\affiliation{%
	\institution{The Chinese University of Hong Kong}
	\city{}
	\country{}
}
\email{yu@se.cuhk.edu.hk}

\author{Chengzhi Piao}
\affiliation{%
	\institution{Hong Kong Baptist University}
	\city{}
	\country{}
}
\email{czpiao@comp.hkbu.edu.hk}

\author{Hong Cheng}
\affiliation{%
	\institution{The Chinese University of Hong Kong}
 \city{}
	\country{}
}
\email{hcheng@se.cuhk.edu.hk}

\author{Helen Meng}
\affiliation{%
	\institution{The Chinese University of Hong Kong}
 \city{}
	\country{}
}
\email{hmmeng@se.cuhk.edu.hk}

\author{Deli Zhao}
\affiliation{%
	\institution{DAMO Academy, Alibaba group, Hupan Lab}
 \city{}
	\country{}
}
\email{zhaodeli@gmail.com}

\author{Yu Rong}
\affiliation{%
	\institution{DAMO Academy, Alibaba group, Hupan Lab}
 \city{}
	\country{}
}
\email{yu.rong@hotmail.com}

\begin{abstract}
Query optimization, which finds the optimized execution plan for a given query, is a complex planning and decision-making problem within the exponentially growing plan space in database management systems (DBMS). 
Traditional optimizers heavily rely on a certain cost model constructed by various heuristics and empirical tuning, probably leading to generating suboptimal plans. 
Recent developments of Large Language Models (LLMs) have demonstrated their potential in solving complex planning and decision-making problems, such as arithmetic and programmatic tasks. 
In this paper, we try to explore the potential of LLMs in handling query optimization and propose a tentative LLM-based query optimizer dubbed \LLMQO, established on \Postgres's execution engine. 
In \LLMQO, we formulate query optimization in an autoregressive fashion which directly generates the execution plan without explicit plan enumeration. 
To investigate the essential input of \LLMQO, we design a customized data recipe named \QueryInstruct to collect the training data from various optimizers and serialize the database's meta data, queries and corresponding plans into a textual format. 
Based on \QueryInstruct, we implement a two-stage fine-tuning pipeline, Query Instruction Tuning (\QIT) and Query Direct Preference Optimization (\QDPO), to empower the capability of general-purpose LLMs in handling query optimization. 
In our experiments, \LLMQO can generate valid and high-quality plans and consistently outperforms both traditional and learned optimizers on three query workloads. 
Our findings verify that LLMs can be derived as query optimizers where generalization, efficiency and adaptivity deserve further research efforts. 

\end{abstract}

\maketitle
\pagestyle{\vldbpagestyle}
\begingroup\small\noindent\raggedright\textbf{PVLDB Reference Format:}\\
\vldbauthors. \vldbtitle. PVLDB, 
\endgroup
\begingroup
\renewcommand\thefootnote{}\footnote{\noindent

\noindent This work is licensed under the Creative Commons BY-NC-ND 4.0 International License. Visit \url{https://creativecommons.org/licenses/by-nc-nd/4.0/} to view a copy of this license. For any use beyond those covered by this license, obtain permission by emailing \href{mailto:info@vldb.org}{info@vldb.org}. Copyright is held by the owner/author(s). Publication rights licensed to the VLDB Endowment. \\
\raggedright Proceedings of the VLDB Endowment, Vol. \vldbvolume, No. \vldbissue\ %
}\addtocounter{footnote}{-1}\endgroup



\comment{Large language models (LLMs) have achieved impressive success across several fields, but their proficiency in understanding and resolving query optimization problems is less explored. 
To bridge this gap, we introduce QueryInstruct, a novel and comprehensive instruction-tuning dataset designed to equip language models with the ability to generate logical plans using SQL semantics and data distributions in the schema. 
Utilizing QueryInstruct, we build PlanLLM, an open-source language model capable of generating efficient plans for various query types. 
To enhance the model's capability and reliability, we incorporate the Direct Preference Optimization (DPO) framework into the training procedure. 
The enhanced model, PlanLLM-DPO, achieves near-optimal performance across several benchmarks. It reduces the plan execution time of PostgreSQL's native optimizer by up to 10\%.
Moreover, our research delves into the delicate balance between training data volume and model performance, highlighting the potential for overfitting with increased data. 
We also examine the transferability of the model's capabilities across different database schemas and query templates, highlighting its adaptability and practical application potential. Our investigation provides a new blueprint and valuable insights for developing LLMs specialized in query optimization challenges.
}
\section{Introduction}
Query optimizer is a fundamental component in database management systems (DBMS), which has been continuously studied over decades~\cite{DBLP:books/sp/KimRB85, DBLP:books/mk/FreytagMV91, DBLP:journals/csur/JarkeK84}. 
Generic query optimizers follow a `plan enumeration and search' paradigm, which enumerate candidate plans from an exponentially growing plan space and employ a cost model to find the plan with minimal cost~\cite{DBLP:conf/sigmod/SelingerACLP79, DBLP:journals/csur/Graefe93, DBLP:conf/job/Leis18, DBLP:conf/pods/Chaudhuri98}. 
The cost model in query optimizers relies on various heuristics, assumptions, and hyper-parameters which are empirically calibrated and tuned over years under certain systems and hardware configurations. Recent studies have explored machine learning techniques to construct learning-based search algorithms, such as reinforcement learning algorithms, to replace dynamic programming to generate plans~\cite{DBLP:journals/corr/abs-1808-03196, DBLP:conf/kdd/0008Y000CZ022, DBLP:conf/icde/Yu0C020, DBLP:journals/pvldb/MarcusNMZAKPT19}, and inject learning-based strategies into generic query optimizers to steer the process of candidate plan selection~\cite{DBLP:conf/sigmod/MarcusNMTAK21, DBLP:journals/pvldb/YuC0L22, DBLP:journals/pvldb/ZhuCDCPWZ23}. 
This `plan enumeration and search' paradigm makes a good trade-off between plan space exploration and execution efficiency, as it is computationally infeasible to calculate costs for all possible plans using the cost model. However, the performance upper bound may be constrained by the candidate plans generated in the initial enumeration phase, potentially missing the better plan due to insufficient exploration of the plan space. 

\begin{figure}
    \centering
    \begin{subfigure}[An Overview of \LLMQO]{%
        \includegraphics[width=1.05\linewidth]{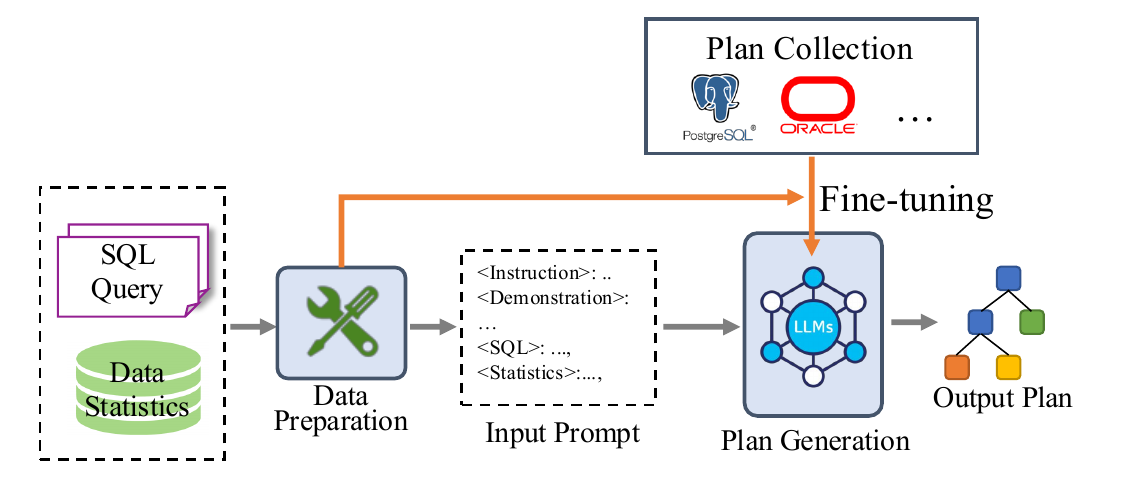}
  \label{fig:intro:overview:plan}
     }\end{subfigure}
     \vspace{-1ex}
    \begin{subfigure}[Execution Time Comparsion with Traditional Optimizers on \dsb]{%
        \includegraphics[width=1\linewidth]{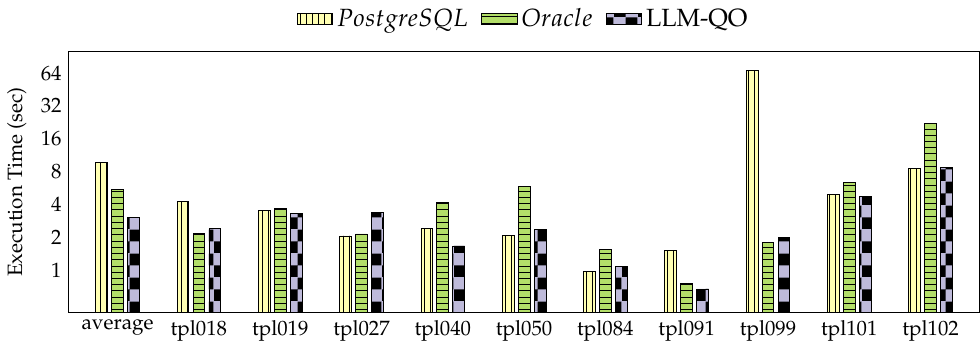}
  \label{fig:intro:overview:performance}
     }\end{subfigure}
    \vspace{1ex}
    \caption{The Framework and Performance of \LLMQO}
    \label{fig:intro_overview}
\end{figure}
Recently, we have witnessed the great success of Large Language Models (LLMs)~\cite{DBLP:journals/pvldb/LaoWLWZCCTW24, zhao2024surveylargelanguagemodels} in solving planning and decision-making problems. Pretrained on massive datasets, LLMs exhibit surprising emergent abilities for solving specific tasks~\cite{wei2022emergent}. Furthermore, after fine-tuned with sophisticated task-specific data, these models can accept the in-context information from inputs and conduct rigorous logical reasoning process to solve complex problems, such as mathematical~\cite{openai2024openaio1card} and programming tasks~\cite{jiang2024surveylargelanguagemodels}. Notably, recent database studies have also begun to explore using LLMs to address  problems in DBMS, including knob tuning~\cite{DBLP:journals/pvldb/LaoWLWZCCTW24} and query rewriting~\cite{DBLP:journals/pvldb/AroraYENHTR23, DBLP:journals/corr/abs-2312-17449}, etc.
Impressed by such abilities, it is worth reconsidering query optimization from an alternative perspective: Is it feasible to directly generate execution plans by LLMs without explicit plan enumeration?
The advanced reasoning capabilities of LLMs, developed through extensive pretraining on massive datasets, may facilitate a comprehensive exploration of the plan space, potentially generating new and efficient plans. 

Despite the fascinating envison, deriving LLMs to serve as a query optimizer faces multiple intertwined challenges.
In the context of generative AI, query optimization task should shift from the `plan enumerate and search' paradigm to an `autoregressive generation' paradigm. This is a completely new task for general-purpose LLMs since unlike existing optimizers, they are unaware of information of involved database instances and expert knowledge for query optimization such as the operator types, algorithms and complexity.
Therefore, the principal problem we try to solve in this paper concentrates on enabling LLMs to comprehend query planning task and generate corresponding execution plan for an input SQL query, where the plan should be valid and efficient at the largest extent.
Specifically, in the data level, LLMs should be trained by high-quality ingredients relevant to query optimization, e.g., the description of the task, available operators, the query to be processed, the meta data in the catalog and the query plans etc.
Necessary ingredients should be delivered contextually by a clear and informative format, i.e., a data recipe, that aligns to the textual input-output specification of LLMs. 
In the algorithmic level, we need effective and efficient training algorithms to fine-tune LLMs with reasonable learning objectives. The algorithms should facilitate general-purpose LLMs to absorb the knowledge from the data recipe and gradually approach an expert optimizer to generate valid and efficient plans while preserving the emergent ability of LLMs. 
\comment{
\ding{172} \textbf{Problem formulation:} In the context of LLMs, query optimization task should shifts from the `plan enumerate and search' paradigm to an `auto-regressive generation' paradigm, which is challenging. The objective for how the LLM produces the correct plans should be figured out. 
\ding{173} \textbf{Data preparation:}  LLMs should be trained by high-quality ingredients relevant to query optimization. The essential information of query optimization task -- including the data schema, target query, data distribution, available operators, DBMS characteristics and hardware configurations -- should be organized into a clear and informative format——a data recipe——that aligns with the textual input-output paradigm of LLMs.
\ding{174} \textbf{Model finetuning:} Effective and efficient training algorithms are needed to inject knowledge from data recipe into LLMs, will enhance the generalization capability of LLMs on query optimization task, especially for out-of-distribution queries.
}

To this end,  we propose a tentative LLM-based framework dubbed \LLMQO, built upon \Postgres's execution engine. Fig.~\ref{fig:intro:overview:plan} depicts the overview of \LLMQO.
\LLMQO is formulated to `distill' the knowledge from the experiences of multiple existing optimizers and aspire to inferring better plans via the generalization and emergency of LLMs. 
In the perspective of data, we design a data preparing pipeline for collecting training data from a query workload, encapsulated in a customized data recipe named \QueryInstruct. 
To be specific, 
\QueryInstruct incorporates the input query, the target query plans collected from multiple optimizers, meta data of the database instance and a planning demonstration, and converts them into a textual format. 
This protocol is delicately designed, maintaining a necessary and minimal context to generate valid and efficient plans and save the usage of tokens for online generation. 
In the perspective of algorithm, we implement a two-stage fine-tuning pipeline in \LLMQO to empower the capabilities of general-purpose LLMs in handling the query optimization task. The first stage, named Query Instruction Tuning (\QIT), aims to initiate an LLM capable of generating valid plans.
The second stage, Query Direct Preference Optimization (\QDPO), further refines the model's expertise on query optimization by training it to differentiate between good and ordinary execution plans. 
In a nutshell, this two-stage training workflow guides LLMs to distill and synthesize the behavior of multiple query
optimizers, in a parameterized fashion, so that earn the potential
of being an overall best optimizer. 

The experimental results demonstrate that \LLMQO achieves an improvement of 5.6\%, 8.6\%, and 68.7\% in average execution time on \imdb, \job, and \dsb, respectively, compared with \Postgres built-in optimizer. Fig.~\ref{fig:intro:overview:performance} presents the per-template comparison of \LLMQO with \Postgres and \Oracle built-in optimizers on \dsb.
The contribution of this paper is summarized as follows.
\begin{itemize}[leftmargin=*]
    \item 
    We take the first step towards formulating query optimization as a sequence generation task, which can be solved by generative LLMs. As an innovative trial, we propose a framework \LLMQO built on \Postgres's execution engine, which deploys and fine-tunes general-purpose LLMs for query optimization in DBMS. 
    \item We develop \QueryInstruct,  a data recipe including data collecting methodology from multiple optimizers and textualization protocol that standardizes the input and output of \LLMQO. Building upon \QueryInstruct, we design a two-stage training workflow to fine-tune general-purpose LLMs, enabling distillation and incorporation behaviors from multiple query optimizers.
    \item We conduct comprehensive experiments for \LLMQO on three query sets, which demonstrates that \LLMQO achieves the capability to generate valid and high-quality plans and outperforms traditional optimizers and learned optimizers, in general cases and out-of-distribution queries. The ablation studies and parameter studies evaluate different possible design choices, empirically justifying the effectiveness of \LLMQO.
    \item Based on the insights and lessons learned from our trials, we verify that LLMs have the potential as optimizers to generate high-quality plans for DBMS. We identify future research directions in generalization, efficiency and adaptivity that will enhance the flexibility and usability of LLMs in query optimization. 
\end{itemize}

\stitle{Roadmap.} The rest of this paper is organized as follows. \cref{sec:related} introduces the related work. In~\cref{sec:background} we give the preliminary and formulate the problem. 
\cref{sec:overview} presents the overview of \LLMQO, which includes the data preparation pipeline and the training pipeline. We then elaborate on the two pipelines in \cref{sec:method:queryinstruct} and \cref{sec:method:training} respectively. We report the experimental studies in \cref{sec:exp}. \cref{sec:conclusion} concludes the paper and points out future research directions.

\section{Related Work}
\label{sec:related}



\stitle{Learning-based Query Optimization.} Database community has witnessed the recent boom of learned query optimization, which can be broadly divided into two categories, de novo learned optimization and steered learned optimization.
De nove learned query optimization leverages learning-based search algorithms, e.g., reinforcement learning (RL) algorithms to replace the dynamic programming in traditional optimizers to generate query plans.
DQ~\cite{DBLP:journals/corr/abs-1808-03196} adopts a policy-based RL algorithm for searching cost-efficient join orders, while RTOS~\cite{DBLP:conf/icde/Yu0C020} and JOGGER~\cite{DBLP:conf/kdd/0008Y000CZ022} adopt a value-based RL algorithm where neural networks such as MLP, TreeLSTM~\cite{DBLP:conf/acl/TaiSM15}, graph representation learning model the intermediate states of joins. 
Neo~\cite{DBLP:journals/pvldb/MarcusNMZAKPT19} and Balsa~\cite{DBLP:conf/sigmod/YangC0MLS22} provide end-to-end solutions for query optimization.
Specifically, 
Neo employs Tree Convolutions~\cite{DBLP:conf/aaai/MouLZWJ16} to estimate the latencies of an execution plan based on a given sub-plan and the optimal plan is generated via a best-first search in the plan space in a bottom-up fashion.
Balsa adopts the same model architecture as Neo but bootstraps the model from a logical-only cost model, bypassing an expert optimizer or expensive query execution.
The second category, steered learned query optimization injects learning-based strategies into generic query optimizers to steer the process of candidate plan selection.
Bao~\cite{DBLP:conf/sigmod/MarcusNMTAK21}, HybridQO~\cite{DBLP:journals/pvldb/YuC0L22} and Lero~\cite{DBLP:journals/pvldb/ZhuCDCPWZ23} tune the native  optimizers with different knobs to generate a set of candidate plans. 
Specifically, Bao exploits a multi-arm bandit algorithm to search for per-query optimization hints, where a Tree Convolution model is trained as the value function. 
HybridQO combines learned and cost-based optimization, interpreting hints as prefix trees with partial join orders to guide the prediction of the final join order. 
Lero scales the estimated cardinality of sub-queries with different factors to produce different candidate
plans and applies learning-to-rank to decide the preference of plans.


\stitle{LLM Prompting.}
LLMs have demonstrated remarkable capabilities through various prompting techniques. Early works established prompt engineering to elicit desired behaviors from LLMs without updating model parameters, while in-context learning~~\cite{DBLP:conf/emnlp/MinLHALHZ22} emerged as a powerful learning framework where models learn from few-shot exemplars within the prompt. 
Chain-of-Thought prompting (CoT)~\cite{DBLP:conf/nips/KojimaGRMI22} further enables LLMs to break down complex reasoning tasks into intermediate steps, significantly improving their performance on mathematical and logical problems. These methodologies have proven complementary, with recent research showing that combining structured prompting with CoT reasoning and in-context examples can enhance model performance across diverse tasks, from arithmetic reasoning to complex inference.

\comment{
Prompting is a crucial mechanism for human-computer interaction, facilitating the explicit communication of clear task descriptions to LLMs, which subsequently generate responses aligned with user expectations through analogical learning. 
The content of a prompt may include instructions, questions, multiple demonstrations with specified output formats, as well as additional requirements such as complex reasoning processes. 
Few-shot prompting, often termed in-context learning~\cite{DBLP:conf/emnlp/MinLHALHZ22}, is a method in which LLMs learn to complete a task with only a few examples, without the need for weight updates/retraining.
In contrast to few-shot prompting, zero-shot prompting uses zero examples. There are several established standalone zero-shot techniques, as well as methods that combine zero-shot prompting with other concepts, such as Chain of Thought~\cite{DBLP:conf/nips/KojimaGRMI22}.
zero-shot-CoT involves appending a thought
inducing phrase like "Let’s think step by step." to the prompt, which enables the model to break down complex queries into logical steps, improving the clarity and coherence of the responses.
The framework Chain of Table~\cite{wang2024chainoftable} dynamically plans a reasoning chain by in-context learning for tabular data reasoning, where the intermediate operations manipulate temporary tables. 
}

\comment{Zero-shot prompting.1 This involves querying LLMs with a prompt that hasn’t been
seen in the training data of the model. Such
prompts typically provide specific task instructions along with the main query. Given the
sensitivity of LLMs to the structure and content of prompts, careful prompt engineering is
crucial to achieve optimal performance.
Few-shot learning. Often referred to as incontext learning, few-shot learning is a technique where LLMs are provided with a handful of examples to guide their responses. Zero-shot prompting can be considered a subset of this, where no examples are given. In few-shot learning, these examples are integrated
into the prompt template, serving as context
to instruct the model on how to resp
}
\comment{
This module interprets the user’s request into instructions that LLMs can easily follow. It first extracts the request intent, generates the prompt by inserting the query intent into the prepared prompt template, and inputs the prompt to LLM to handle the request.} 


\stitle{LLM Fine-tuning}. 
%
%
%
%
\comment{fine-tuning involves extending the training of the LLMs using additional, task-specific data. This is particularly beneficial when such tailored datasets are available.
}
Pre-trained language models (PLMs), trained on extensive corpora, have demonstrated exceptional capabilities across numerous NLP tasks~\cite{DBLP:journals/corr/abs-2310-06825, DBLP:journals/corr/abs-2302-13971,DBLP:journals/corr/abs-2205-01068,DBLP:journals/corr/abs-2311-16867}.
However, deploying these models for specific applications requires additional refinement through instruction tuning and preference alignment.
Instruction-tuning~\cite{DBLP:conf/iclr/WeiBZGYLDDL22,DBLP:conf/nips/DuboisLTZGBGLH23,DBLP:conf/nips/WangIDHKCWMSBH23} enables models to follow task descriptions given in natural language, facilitating generalization to new tasks. 
Nevertheless, the instruction-tuned models may still generate inappropriate or unethical responses.
To address this problem, specialized training strategies are proposed to further align these models with human values, typically implemented by reinforcement learning with human feedback (RLHF)~\cite{DBLP:journals/corr/abs-1909-08593,DBLP:conf/nips/StiennonO0ZLVRA20,DBLP:conf/nips/Ouyang0JAWMZASR22}.
The RLHF approach involves training a reward model using pairwise preference data, which guides the optimization of a policy model.
To streamline this process, direct preference optimization (DPO)~\cite{DBLP:conf/nips/RafailovSMMEF23} has emerged as an innovative solution, integrating reward modeling directly into the preference learning stage, thereby eliminating the need for two separate models and simplifying the training pipeline.   
\comment{


}

While LLMs exhibit powerful capabilities, their large size poses significant challenges for fine-tuning. 
Low-Rank Adaptation (LoRA) \cite{DBLP:conf/iclr/HuSWALWWC22} is a prevailing parameter-efficient technique for fine-tuning customized LLMs. By introducing a low-rank approximation that updates only a small collection of parameters, LoRA substantially reduces computational overhead while mitigating overfitting. 


\comment{Pretraining involves training the model on a large corpus to maximize the log-likelihood of predicting the next token based on the preceding text.

While LLMs gain broad linguistic knowledge through pre-training, they often require additional expertise for specific tasks. Supervised Fine-Tuning (SFT) addresses this by further training the model with data that is more relevant to the downstream task. Often, such data will comprise instructions and an appropriate response (i.e., instruction fine-tuning). 

Fine-tuning is a process in machine learning that involves adapting a pre-trained model to perform specific tasks by training it on a smaller, task-relevant dataset. 

Supervised fine-tuning (SFT) can align models with human preferences. However, as the probability
of preferred outputs increases, so does the likelihood of undesirable ones, leading to hallucinations.}

\comment{
Recently, LLMs with Reinforcement Learning from Human Feedback (RLHF) can further generate a response that is well aligned with human values.
To generate more reliable outputs, Reinforcement Learning from Human Feedback (RLHF) (Christiano
et al., 2017; Ouyang et al., 2022) has been introduced for LLM alignment. This approach involves
training a reward model with comparison data and then using this reward model to optimize the policy
model. The final performance heavily depends on the quality of the reward model, and the training
pipeline is quite complex.

Direct Preference Optimization (DPO) is an emerging technique that optimizes language models directly based on human preferences, bypassing the need for complex reward models and traditional reinforcement learning processes. By leveraging human feedback data, DPO streamlines the model fine-tuning process, enhancing efficiency and stability while demonstrating superior performance across various tasks compared to existing methods. 
To simplify this process, Direct Preference Optimization (DPO) (Rafailov et al., 2024) was proposed,
which directly uses pair-wise preference data for model optimization. This transition significantly
streamlines the training pipeline.
}









\stitle{LLM for Database Management.} The recent success of LLMs incites their utilization in DBMS to support various tasks.  
For database knob tuning, $\lambda$-Tune~\cite{DBLP:journals/pvldb/LaoWLWZCCTW24} leverages LLMs to optimize database knobs in response to varying workloads  adaptively. GPTuner~\cite{DBLP:conf/sigmod/GiannakourisT24} utilizes LLMs to read manuals as domain knowledge for optimizing database parameter configurations.
For tabular data cleaning and analysis, Table-GPT~\cite{DBLP:journals/pacmmod/LiHYCGZF0C24} 
introduces a `table-tuning' paradigm that enhances the performance of LLMs on table-understanding tasks, and Sui et al.~\cite{DBLP:conf/wsdm/SuiZZH024} develop a benchmark to evaluate the table understanding capabilities of LLMs. 
NL2VIS~\cite{DBLP:journals/pacmmod/Wu00SWZ0024} assesses the potential of LLMs in transforming natural language descriptions of tabular data into visual representations.  
Furthermore, there have been substantial efforts to leverage LLMs in query processing tasks, including text-to-SQL, query planning, and query rewriting.  
Numerous studies~\cite{DBLP:conf/nips/LiHQYLLWQGHZ0LC23, DBLP:journals/corr/abs-2306-00739, DBLP:journals/corr/abs-2406-08426} have highlighted the potential of LLMs in the text-to-SQL task. 
GPT-DB~\cite{DBLP:journals/pvldb/Trummer23} builds a pipeline with GPT-4 to automatically generate query-specific code for SQL query processing from natural language instructions. 
CAESURA~\cite{DBLP:conf/cidr/UrbanB24} uses GPT-4 to translate natural language queries into multi-modal query plans containing relational operators and Python UDFs.
LLM-R\textsuperscript{2}~\cite{DBLP:journals/corr/abs-2404-12872} proposes a new query rewriting system that utilizes LLMs to recommend rewriting rules by few-shot learning.
Evaporate~\cite{DBLP:journals/pvldb/AroraYENHTR23} uses in-context learning of LLMs to generate structured views for semi-structured data lakes. 
Despite the progress, there remains a gap in exploring the potential of LLMs in query optimization. 

\comment{
Modern database management systems (DBMS) expose hundreds of confgurable knobs to control system behaviours. Determining the appropriate values for these knobs to improve DBMS performance.
Recent advancements in database tuning systems have explored the integration of large language models to enhance performance optimization. 
$\lambda$-Tune~\cite{DBLP:journals/pvldb/LaoWLWZCCTW24} presents an approach that leverages large language models (LLMs) to optimize database systems adaptively based on varying workloads. 
$\lambda$-Tune~\cite{DBLP:journals/pvldb/LaoWLWZCCTW24}
exploits information from an input set of queries, in order to tune an input database system in a workload-adaptive manner.

GPTuner~\cite{DBLP:conf/sigmod/GiannakourisT24} introduces a manual-reading approach that leverages domain knowledge extensively to optimize the tuning process, employing a novel framework based on GPT-guided Bayesian optimization to suggest effective knob configurations while reducing tuning costs.

Table-GPT~\cite{DBLP:journals/pacmmod/LiHYCGZF0C24} introduces a "table-tuning" paradigm to enhance the performance of language models like GPT-3.5 and ChatGPT on table-related tasks. By synthesizing diverse training data from real tables, the model is fine-tuned to improve its understanding and manipulation of tabular data, achieving notable advancements in tasks such as data transformation, question answering, and error detection.

Inspired by the capability of LLMs, researchers have investigated whether LLMs can be used for data cleaning tasks such as data imputation where LLMs repair dirty or missing values in data entries

Trummer~\cite{DBLP:journals/pvldb/Trummer23a} investigates the ability of large language models (LLMs) to predict correlations between data columns by analyzing their names. 

\cite{DBLP:journals/pacmmod/Wu00SWZ0024} first explores the use of large language models (LLMs) to automatically generate data visualizations from natural language descriptions and evaluates the effectiveness of LLMs in producing relevant visual representations and identifies key factors that enhance their performance in the visualization task.

Text-to-SQL aims at automatically translating natural language questions into SQL queries. BIRD(a large-scale benchmark dataset)~\cite{DBLP:conf/nips/LiHQYLLWQGHZ0LC23} is proposed
to narrow the gap between Text-to-SQL research and its real-world deployment.

GPT-DB~\cite{DBLP:journals/pvldb/Trummer23} presents a system that utilizes GPT-4 to automatically generate code for SQL processing in general-purpose programming languages like Python, based on user-provided natural language instructions. 

DB-GPT~\cite{DBLP:journals/dase/ZhouSL24} introduces an automated prompt strategy utilizing LLMs for query rewriting and index tuning. 

LLM-R\textsuperscript{2}\cite{DBLP:journals/corr/abs-2404-12872} proposes a DB-based SQL query rewrite pipeline enhanced by a LLM. It employs a contrastive model to learn query representations and select effective demonstrations, which optimizes the LLM's rule selection for rewriting.

CAESURA~\cite{DBLP:conf/cidr/UrbanB24} emphasizes multi-modal query planning and introduces a prototype built upon GPT-4 that translates natural language queries into executable query plans. In contrast to conventional relational query planners, the generated query plans can integrate complex operators that are adept at handling diverse modalities.

CAESURA~\cite{DBLP:conf/cidr/UrbanB24} goes outside the scope of code generation to generate reasoning plans
for natural language multi-modal queries.
}





\comment{
GPTuner~\cite{DBLP:conf/sigmod/GiannakourisT24} utilizes domain knowledge and a GPT-guided Bayesian optimization framework to improve knob tuning efficiency and reduce costs. Both studies underscore the potential of LLMs in enhancing database tuning processes.}

\comment{
CAESURA: we propose Language-Model-Driven Query Planning, a new paradigm of query planning that uses Language Models to translate natural language queries into executable query plans.
Different from relational query planners, the resulting query plans
can contain complex operators that are able to process arbitrary
modalities. As part of this paper, we present a first GPT-4 based
prototype called CAESURA and show the general feasibility of this
idea on two datasets}

\comment{





}

\comment{
\begin{table}[t]
\footnotesize
\centering
\caption{Frequently-used Notations}
\label{tab:notation}
\begin{tabular}{c|c}
\toprule
Notation & Description                 \\\midrule
$q$ & a SQL query \\
$p$ & a query plan \\
$\pi$ & The LLM model \\
$x$, $y$ & The input and output of LLM \\
$\mathcal{L}$ & The loss function of LLM fine-tuning \\
 \bottomrule
\end{tabular}
\end{table}
}

\section{Preliminary \& Problem Statement}
\label{sec:background} 
In this section, we formulate the research problem of utilizing LLMs for query optimization and introduce the foundation of LLMs. 
\subsection{Problem Statement}
\label{sec:background:problem_statement} 
Given a database with multiple tables,
for a SQL query $q$, traditional query optimizers transform the SQL query into a tree-shape execution plan $p$ as illustrated in Fig.~\ref{fig:example:plan}.
The plan $p$ specifies a join order and the corresponding algorithm used in each operator. 
The basic paradigm of traditional optimizers is to enumerate candidate plans and then search for the optimal plan based on a cost model by dynamic programming. 
Typically, the cost model leverages database statistical information to estimate the execution cost of candidate plans. 

In this paper, for a database $\mathcal{D}$, we aim to build 
a model based on general-purpose LLMs 
to infer the query plan directly. 
By formulating a SQL query $q$ and pertinent information for query planning (e.g., the database schema and statistics of the database) into a sequence of tokens $\bm{x}$ in a textual space  $\mathcal{X}$,
the LLM $\pi: \mathcal{X} \rightarrow \mathcal{Y}$, directly infers its execution plan as a textual response $\bm{y}$ in the output space $\mathcal{Y}$.
The generated response should be (1) accessible; the sequential output tokens $\bm{y}$ can be easily transformed into a physical plan that can be evaluated by the execution engine, (2) valid; the output query plan should be a valid plan that is consistent with the semantics of the query, and (3) efficient; the performance of the plan should approach or outperform the plan from existing well-designed optimizers. 
In the following, we will use `generated response' or `generated plan' interchangeably when the context is clear. 
The model $\pi$ is trained by fine-tuning an off-the-shelf LLM, e.g., LLaMA~\cite{DBLP:journals/corr/abs-2302-13971}, given a set of SQL queries as a training workload.  
In this paper, we consider selection-project-join (SPJ) queries, with one join condition connecting two tables. 
At the current stage, we support three join operators: \NestLoop, \MergeJoin, and \HashJoin for joining tables, while the table scan operator \SeqScan is utilized for selection.
We aim at building an LLM for a static database instance on a specific execution engine. Generalization on multiple databases and hardware configurations, complex query types, dynamic query workloads and dynamic relational data are our further work.

\subsection{Large Language Models}
LLMs are fundamentally built on Transformer~\cite{DBLP:conf/nips/VaswaniSPUJGKP17} architectures, by scaling self-attention layers with massive parameter spaces from billions to trillions of parameters.
The landscape of LLMs is divided between closed-source models, such as GPT-4~\cite{DBLP:journals/corr/abs-2303-08774} and Claude, developed by commercial institutions, which are only accessible via remote API calls, and open-source alternatives like LLaMA~\cite{DBLP:journals/corr/abs-2302-13971} and Falcon~\cite{DBLP:conf/nips/PenedoMHCACPAL23}, which enable broader research exploitation and customization.  
In this paper, we fine-tune open-source LLMs to build an LLM-based query optimizer. 

The inference of LLMs performs via the process of next token prediction, where the model generates a response by predicting the next token based on the input context and previous output in an autoregressive process. 
Formally, given the input as a sequence of tokens $\bm{x} = [x^1, \cdots, x^n]$, and the output sequence $\bm{y} = [y^1, \cdots, y^T]$, model $\theta$  parameterizes the conditional distribution of $\bm{y}$ given $\bm{x}$ as 
\begin{align}
\label{eq:cond_prob}
    p_{\theta}(\bm{y} \mid \bm{x})= p_{\theta}( y^1, y^2 \cdots, y^T \mid \bm{x}) 
    = \prod_{t=1}^{T} p_{\theta}\left(y^{t} \mid 
    \bm{x} , y^{1:t-1}\right)
\end{align}
Given a set of input-output pairs $\mathbb{D} = \{ (\bm{x}, \bm{y})\}$ as training data, LLMs can be trained/fined-tuned like typical language models by maximum likelihood estimation, which minimizes the negation of the logarithmic likelihood in Eq.~\eqref{eq:mle_loss}.
\begin{align}
\label{eq:mle_loss}
    \loss({\theta}) &=
   \mathop{\mathbb{E}}_{(\bm {x}, \bm{y}) \sim \mathbb{D}} - \log p_{\theta} (\bm{y} \mid \bm{x}) \\
   &= \mathop{\mathbb{E}}_{(\bm {x}, \bm{y}) \sim \mathbb{D}}    -  \Big[ \sum_{t=1}^T \log p_{\theta}(y^t \mid \bm{x},  y^{1:t-1})\Big]\text{,}
\end{align}
To unleash the reasoning potential of open-source LLMs on a specific task, researchers customize their training data in a specific format, named a \emph{data recipe}~\cite{DBLP:conf/sigmod/Chen0MCPGGXLGLD24} which is a mixture of data of heterogeneous types and from different sources. 
As the importance of ingredients to dishes, the quality, quantity and diversity of training data significantly affect the performance of LLMs. 
\comment{
Recent advances in the field of artificial intelligence (AI) have enabled the development of increasingly powerful linguistic models capable of generating text fluently
and coherently. Among these models, “large language models” (LLMs) stand out for
their imposing size and their ability to learn enormous amounts of textual data. Models
like GPT-3.5 [1] and GPT-4 [2] developed by OpenAI [3], or Bard, created by Google [4],
have billions of parameters and have demonstrated impressive comprehension skills and
language generation.
Recent advances in NLP have led to the development of powerful LLMs capable of performing a wide range of tasks. However, despite their impressive capabilities, general-purpose LLMs often struggle with domain-specific tasks that require specialized knowledge

}
\comment{
Autoregressive models are a type of statistical or machine learning model that predicts the next value in a sequence based on the previous values in that sequence. These models assume that the future values in the sequence are dependent on the past values and use this dependency to make predictions. In the context of natural language processing, autoregressive models are often applied to generate text or make predictions based on previous words in a sentence. These models learn the statistical patterns and dependencies in the training data and then use that knowledge to generate coherent and contextually relevant text. Autoregressive language models, such as GPT (Generative Pre-trained Transformer), GPT-2, and GPT-3, have gained significant attention for their ability to generate high-quality text and perform a variety of language-related tasks.
LLMs serve as the core framework for text generation. In essence, these models consist of large pretrained transformer architectures specifically designed to predict the next word (or, more accurately, the next token) based on a given input text. Since they generate one token sequentially, creating new sentences involves a more complex approach than merely invoking the model; it requires implementing autoregressive generation techniques.
}

\comment{
\subsection{Query Optimization Problem} 
\label{sec:preliminary:problem_definition}
The Query Optimization task can be formulated as
a sequence-to-sequence task where the input $x$ includes an instruction that specifies the query optimization task, a SQL query, and the statistical information of its schema, and the output is a planning path $y$ leading to the plan $p$.
A generated planning path $\hat{y}$ is regarded as correct if the extracted final plan $\hat{p}$ is valid execution plan. Formally, the
labeled dataset for a query optimization problem solving
task with instances can be represented as:
\begin{equation}
\mathbb{D} = \left\{ (x, y, p)^{i} \right\}_{i=1}^l
\end{equation}


The goal of plan generation with a sequence-to-sequence model is to generate a planning path $y$ with length T conditioned on the input text sequence $x$. 
Given a training set $\mathbb{D}$, the neural plan generation model parameterizes the conditional distribution $p_{\theta}(y|x)$ as follows:
\begin{equation}
\label{eq:p_function}
    p_{\theta}(y \mid x)=\prod_{t=1}^{T} p_{\theta}\left(y_{t} \mid  x , y_{1:t-1}\right),
\end{equation}

By adopting Maximum Likelihood Estimation (MLE) as the learning approach, the training objective of the neural plan generation model can be defined as follows: 
\begin{equation}
    \label{eq:sft_loss}
    \mathcal{L}_{\text{MLE}}({\theta})=
    \mathop{\mathbb{E}}_{(x, y) \sim \mathbb{D}}      \Big[ \sum_{t=1}^T \log p_{\theta}(y_t | x,        y_{1:t-1})\Big]\text{,}
\end{equation}
where $T$ is the length of the planning path $y$ and we use $y_t$ to represent the $t$-th token in $y$.
}

\section{Overview of LLM-QO}
\label{sec:overview}

\begin{figure*}
  \centering
  \includegraphics[width=1\textwidth]{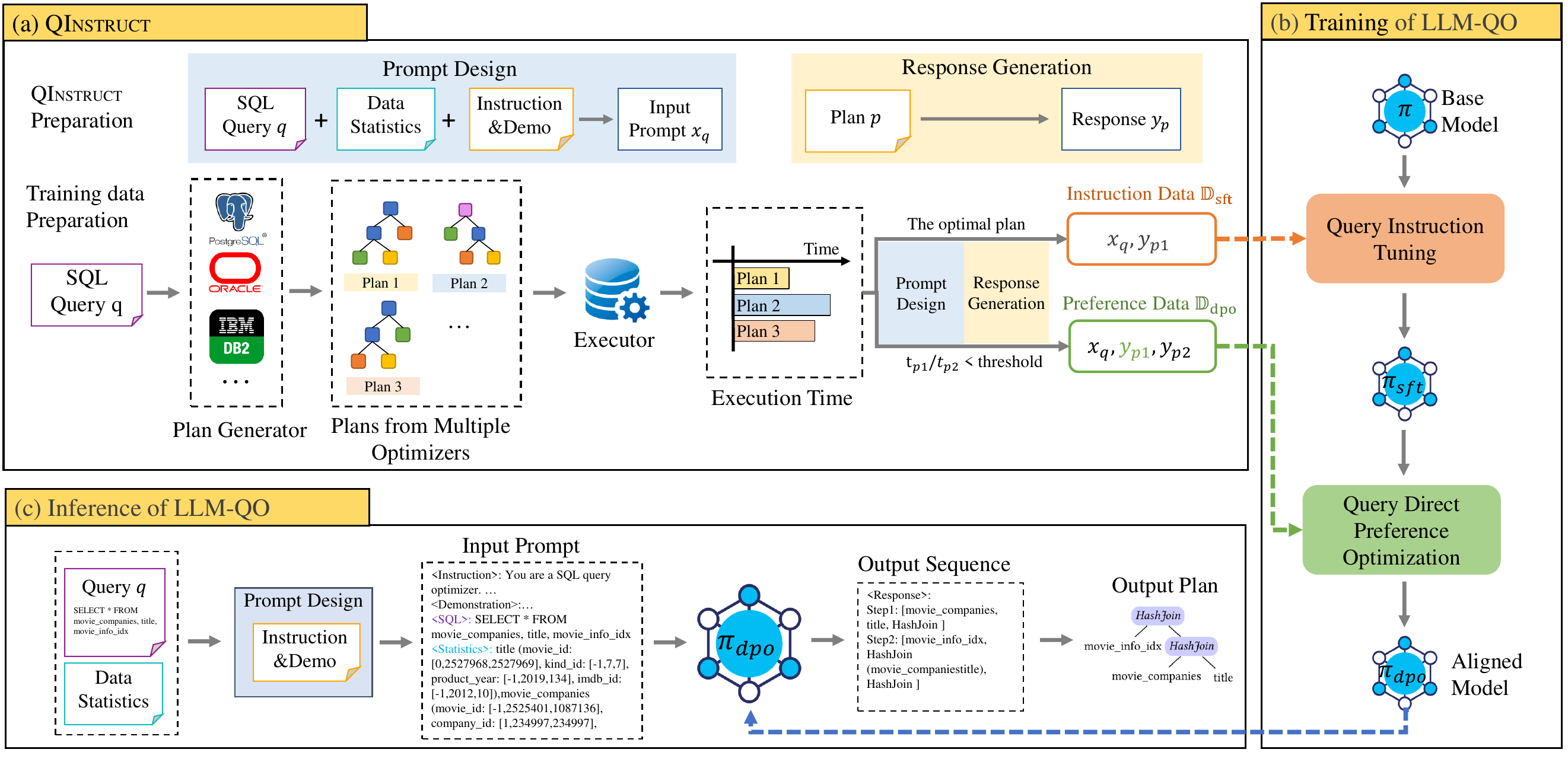}
  \caption{The overview of \LLMQO, which consists of: (a) \QueryInstruct and the data preparing pipeline,  (b)The training pipeline for LLMs, and (c) The inference pipeline.
  }
  \label{fig:pipline}
\end{figure*}

We present the overview of our approach, \LLMQO, whose framework is illustrated in Fig.~\ref{fig:pipline}.
\LLMQO comprises two main pipelines, a data preparing pipeline and a training pipeline for LLMs. 

The data preparation pipeline is a crucial initial step in \LLMQO. It generates the training data regarding a data recipe, named \QueryInstruct, which will be used to fine-tune the LLM in the training pipeline.
For data preparation, the gist of designing \QueryInstruct is to keep the data informative and concise, so that feeding an LLM with sufficient instructions and knowledge while saving token usage.
Serving as the foundation of \LLMQO training, \QueryInstruct includes five elements as follows: \ding{172} Query: the SQL query that needs optimization. 
\ding{173} Instruction of query optimization: guidelines and instructions on how to optimize the given queries and explanations of the input.
\ding{174} Auxiliary information: including the database schema and relevant statistics that provide context and support for query optimization.
\ding{175} High-quality plans: efficient execution plans of the query that are serialized and converted into a textural format. 
\ding{176} Planning demonstration: one concrete planning example to alleviate the hallucination of the LLM.
It is worth noting that the query, instruction, auxiliary information, and demonstration compose the prompt of the LLM, and the textualized execution plan is the target response used for model fine-tuning. 
Given a query workload, the execution plans can be collected from traditional query optimizers, especially the optimizers of commercial DBMS, which have been delicately designed over decades. 
We will elaborate on the construction of \QueryInstruct in ~\cref{sec:method:queryinstruct}. 

The LLM training pipeline is designed to fine-tune an LLM with the training data from a query workload, which is generated regarding \QueryInstruct in the first pipeline. Considering the intrinsic difficulty of the problem, we adopt a two-stage training workflow to gradually enhance the reasoning capability of an LLM for query planning.
The first stage initiates an LLM the ability to generate an execution plan for a given query by Instruction Tuning. This stage mainly aims at bootstrapping a general-purpose LLM to generate accessible and valid plans. 
Subsequently, the second stage further improves the planning capability of an LLM by training it to 
distinguish a good plan from ordinary plans.
Through this stage, the LLM is expected to infer efficient execution plans. 
In a nutshell,  this two-stage training workflow guides the LLM to distill and synthesize the behavior of multiple query optimizers, in a parameterized fashion, so that earn the potential of being an overall best optimizer. We will delve into the training pipeline of \LLMQO in ~\cref{sec:method:training}.

In inference stage, as shown in Fig.~\ref{fig:pipline}(c), a test query and corresponding auxiliary information are inserted into an input prompt, following the formulation of \QueryInstruct. Afterwards, an  inference calling is issued on the fine-tuned LLM, by feeding the  prompt into the model and obtaining the response. Finally, the textual response is transformed into an execution plan for query  evaluation.


\section{Formulation of \QueryInstruct}
\label{sec:method:queryinstruct}
We first introduce the formulation of our data recipe, \QueryInstruct, followed by the procedure of  training data collection. 

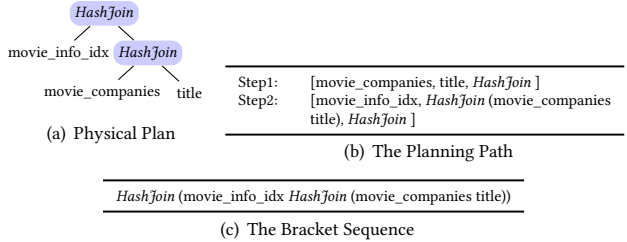
\begin{figure}[t]
\centering
\begin{subfigure}[Physical Plan]{%
\scriptsize  
\begin{tikzpicture}[sibling distance=5.5em, level distance = 2.5em,
  every node/.style = { shape= rectangle, rounded corners,
    draw =none , align=center,
   }]]
  \node[fill=blue!20] { \HashJoin } 
      child {
        node { movie\_info\_idx }
      }
      child {
        node[fill=blue!20] { \HashJoin }
          child { node{ movie\_companies } }
          child {
              node { title }
                }
            }
  ;
\end{tikzpicture}
  \label{fig:example:plan}
}\end{subfigure}
\begin{subfigure}[The Planning Path]{%
\scriptsize  
\begin{tabular}{m{0.6cm} p{4.0cm}}
\toprule
Step1: & [movie\_companies, title, \HashJoin] \\
Step2: & [movie\_info\_idx, \HashJoin(movie\_companies title), \HashJoin] \\
 \bottomrule
\end{tabular}
  \label{fig:example:planpath}
}\end{subfigure}
\begin{subfigure}[The Bracket Sequence]{%
\scriptsize  
\begin{tabular}{c l}
\toprule
\HashJoin(movie\_info\_idx \HashJoin(movie\_companies title)) \\ 
\bottomrule

\end{tabular}
  \label{fig:example:bracket_sequence}
}\end{subfigure}
\caption{A plan with two textual representations}
\label{fig:example}
\end{figure}

\begin{figure}
    \centering
    \footnotesize
    \begin{subfigure}{}
        \centering
        \begin{tcolorbox}[colback=yellow!10!white, colframe=blue!75!black]
            \textbf{INSTRUCTION:} You are a SQL query optimizer. You will be given a multi-table SQL query {\color{Orchid} \texttt{<SQL>}} and the statistics of the tables involved in the query {\color{BlueGreen} \texttt{<Statistics>}}. The statistics include the minimum value, maximum value, and the count of distinct values for each column of each table in the query, in the format of [min, max, distinct count]. Your task is to generate the optimal execution plan for the given SQL query. You should represent the execution plan using a bracket sequence, where \HashJoin, \NestLoop, or \MergeJoin are used to join the tables in the SQL query. Let's think step by step and show your reasoning before showing the final result. {\color{Peach} \texttt{<Planning Demonstration>}: \texttt{<SQL>}: ..., \texttt{<Statistics>}:..., \texttt{<Response>}:...}
            
            \textbf{INPUT:}
            \begin{Verbatim}[commandchars=\\\{\}]
\textcolor{Orchid}{<SQL>}: \textcolor{blue}{SELECT} * \textcolor{blue}{FROM} movie_companies, title, movie_info_idx  
\textcolor{blue}{WHERE} title.movie_id = movie_companies.movie_id \textcolor{blue}{AND} 
title.movie_id = movie_info_idx.movie_id  \textcolor{blue}{AND} 
movie_companies.company_type_id = 1 \textcolor{blue}{AND} 
title.product_year < 1904 \textcolor{blue}{AND} title.product_year > 58;
\textcolor{BlueGreen}{<Statistics>}: 
title (movie_id: [0,2527968,2527969], kind_id: [-1,7,7], 
product_year: [-1,2019,134], imdb_id: [-1,2012,10]), 
movie_companies (movie_id: [-1,2525401,1087136], 
company_id: [1,234997,234997], company_type_id: [1,2,2]), 
movie_info_idx (movie_info_idx_id: [0,1380033,1380034], 
movie_id: [-1,2525449,459876]).
            \end{Verbatim}          
        \end{tcolorbox} 
        \label{fig:query_instruct:prompt}
    \end{subfigure}%
    
    \vspace{-3ex}
    \begin{subfigure}{}
        \centering
        \label{fig:query_instruct:response}
        \begin{tcolorbox}[colback=yellow!10!white, colframe=blue!75!black]
            \textbf{RESPONSE:}\\
            Step1: [movie\_companies, title, \HashJoin], \\
            Step2: [movie\_info\_idx, \HashJoin(movie\_companies title), \HashJoin], \\
            Therefore, the final answer is: \\ \HashJoin (movie\_info\_idx \HashJoin(movie\_companies itle)).
        \end{tcolorbox}
        \label{fig:query_instruct:response}
    \end{subfigure}
    \caption{The format of \QueryInstruct}
    \label{fig:query_instruct} 
\end{figure}

\subsection{\QueryInstruct Preparation}
\label{sec:method:query_instruct:prepare}
For a SQL query in the training workload, \LLMQO generates its data recipe for the two-stage training, which is conducted offline.
To start with, we present the formulation of the prompt of \LLMQO, which is composed of the input prompt and a response, as shown in Fig.~\ref{fig:query_instruct}. 

\stitle{Prompt Design.}
The prompt, as a user-specific input of LLM, includes a query, the instruction, the auxiliary information and a planning demonstration.
Fig.~\ref{fig:query_instruct} illustrates the input prompt of \LLMQO, for both training and inference. The first part of the prompt is the instruction that requires the LLM to mimic as a SQL query optimizer. 
There are two placeholders, {\color{Orchid}\texttt{<SQL>}} and {\color{BlueGreen} \texttt{<Statistics>}}, indicating the raw text of the input query and corresponding statistics involved by the query.  
The contents of these two placeholders are attached below the prompt template. 
Specifically, to help LLM understand the semantics of the statistics, the prompt concisely describes the per-column statistics in a bracket format. In addition, we specify the plan output in a sequence of bracket representation, which we will introduce later, as well as the the join operators supported in the system. 
To initiate the reasoning capability of LLM, we encourage the LLM to generate a plan by  explicit instruction `think step by step'~\cite{DBLP:conf/nips/KojimaGRMI22}. 
Considering different LLMs support different maximum length of input tokens, users can further enrich the database statistics, e.g., the histograms, distinct values, estimated 
cardinalities, etc, which are available in the database catalog. 

At the end of the instruction, we add a one-shot planning demonstration identified by a placeholder {\color{Peach}\texttt{<Planning Demonstration>}} as shown in Fig.~\ref{fig:query_instruct}. 
It is challenging for general LLMs to consistently generate valid plans as users can issue any syntactically correct queries. In our extensive trials, LLMs tend to infer invalid plans when the involved tables and columns are unseen in the training queries. Therefore, we further introduce in-context learning into the prompt design to alleviate invalidation, which is also used in table understanding~\cite{DBLP:journals/pacmmod/LiHYCGZF0C24} and query rewriting~\cite{DBLP:journals/corr/abs-2404-12872} tasks.
Here, a planning demonstration serves as an illustrative example to align LLM to our query optimization task.
The demonstration is assembled by an exemplar {\color{Peach}\texttt{<SQL>}}, the pertinent statistical information {\color{Peach}\texttt{<Statistics>}}, and the output plan of the exemplar query {\color{Peach}\texttt{<Response>}}.
To avoid invalid generation to the largest extent, we plug an exemplar query into the demonstration, which possesses the same query template, i.e., the same set of tables and join conditions, as the input query. 
At the current stage, that is the most effective strategy in our extensive attempts on various test queries from multiple workloads. 
It is worth noting that the plans in the response are unnecessary to be well performed. We find that LLMs only refer to the occurrence of tables and columns, and will generate a new plan in most cases.

\stitle{Response Generation.}
The curial design of \QueryInstruct is to represent the tree-shape execution plan as a textual sequence, which preserves the intact execution strategy, i.e., the join order and the physical operators for LLMs to comprehend. 
A straightforward way is to represent the plan in a bracket format. For instance, Fig.~\ref{fig:example:bracket_sequence} presents the bracket representation of the query plan in Fig.~\ref{fig:example:plan}, where the bracketed tokens $opt(opd_1, opd_2)$ denote a join operation $opt$ on a left sub-plan and a right sub-plan. 
The sub-plans $opd_1, opd_2$ are either base table names or nested bracket representations. 
Although the bracket representation consumes minimal tokens, it fails to stimulate the multi-step reasoning capability of LLMs in the context of query optimization.
%
To this end, we devise a structured hierarchical representation, planning path, for an execution plan $p$. For a query involving $n$ tables, its planning path comprises $n - 1$ sequential steps $[s_1, s_2, \cdots, s_{n-1}]$ where each step $s_i$ denotes one non-leaf (join) operator. We use a sequence of tokens $s = [ opd_1, opd_2, opt ]$ to represent one step, where $opd_1$, $opd_2$ are either base table name or intermediate step and $opt$ is the name of the physical operator.  
The order of the steps in the planning path is determined by the post-order traversal of the tree-shape execution plan, which also aligns with the bottom-up query execution semantic.
As an example, Fig.~\ref{fig:example:planpath} demonstrates the corresponding planning path for the query in Fig.~\ref{fig:example:plan}. 
The complete response includes the planning path as reasoning steps and the bracket representation as the final answer as shown in Fig.~\ref{fig:query_instruct}. 
\LLMQO will transform the response into an execution plan for query execution.

\subsection{Training Data Preparation}
\label{sec:queryinstruct:PDG}
Formulated by the data recipe \QueryInstruct, \LLMQO prepares the training data from a query workload of a database instance. 
In the two-stage training pipeline, 
we use instruction data $(\bm{x}, \bm{y})$ and preference data $(\bm{x}, \bm{y}_{w}, \bm{y}_{l})$ to train the model by Query Instruction Tuning in the first stage and Query Direct Preference Optimization in the second stage, respectively, where $\bm{x}$ denotes the input prompt of a query, 
$\bm{y}, \bm{y}_{w}, \bm{y}_{l}$ denote expected responses of LLM. 
In particular, for the preference data, $\bm{y}_{w}$ contains a well-performed execution plan and $\bm{y}_{l}$ contains an ordinary plan by contrast, for a same query in $\bm{x}$, reflecting a planning preference we wish the LLM to absorb and deliver. 
We concentrate on the preparation of the training data and defer the technical details of our training algorithms to~\cref{sec:method:training}.

\stitle{Execution Plan Collection.} As training data, original execution plans can be collected from existing query optimizers, which are available in query logs. As Fig.~\ref{fig:pipline} depicts, \LLMQO collects execution plans from multiple DBMSs, e.g., \Postgres, \Oracle, based on the real performance of the execution plans. 
It is worth mentioning that we also tried using the cost of the plan, derived from traditional cost models, as the evaluation metric for selecting training plans. However, as is well-known, the real performance of execution plans is inconsistent with their cost, which compromises the efficiency of plans generated by LLMs.

\begin{algorithm}[t]
	\small
    \caption{Preference Data Generation}
	\label{alg:pdg}
	\DontPrintSemicolon
    \SetKwData{Procedure}{Procedure}
	\SetKwData{Up}{up}  \SetKwInOut{Input}{Input} \SetKwInOut{Output}{Output}
    \SetKwInOut{Initialize}{Initialize}
    
	\Input{SQL query $q$, $k$ optimizers,  threshold $r_0$}

	\Output{$\mathbb{D}_{\mathrm{dpo}}$ of $q$}

    Initialize $\mathbb{D}_{\mathrm{dpo}} \leftarrow \emptyset$, $p^* \leftarrow \varnothing$, $t^* \leftarrow \infty$\;
    \For{$ i \leftarrow 1$ to $k$}{
    \label{line:plan_and_time:start}  
        generate plan $p_i$ by the $i$-th optimizer \;
        obtain the execution time $t_i$ of $p_i$ in the execution engine \;
        \If {$t_i < t^*$}{ \label{line:plan_compare}
            $p^* \leftarrow p_i$, $t^* \leftarrow t_i$\;
        }
    } \label{line:plan_and_time:end}  
    generate input $\bm{x}$ for $q$ and output $\bm{y}_w$ for $p^{*}$ as Fig.~\ref{fig:query_instruct} \; \label{line:prompt_generate}
        
        \For{$ i \leftarrow 1$ to $k$}{
    \label{line:data_collection:start}
            \If {${t^{*}}/{t_i} < r_0$}{ \label{line:gap_compute}
                generate output $\bm{y}_l$ for $p_i$\;
                add ($\bm{x}$, $\bm{y}_w$, $\bm{y}_l$) into $\mathbb{D}_{\mathrm{dpo}}$ \;
                }
        }\label{line:data_collection:end}
\Return{$\mathbb{D}_\mathrm{dpo}$} \;
\end{algorithm}

\stitle{Preference Plan Generation.}
To obtain a pair of preferred and dispreferred plans, \LLMQO collects the execution plans from multiple query optimizers of different DBMSs and selects those plans with a performance gap. 
For one query $q$ in a training workload, the procedure for generating preference data is detailed in Algorithm~\ref{alg:pdg}.
Given $k$ optimizers, first, we collect the $k$ corresponding plans $\{p_1, \cdots, p_k\}$ and obtain the execution time of plan $p_i$ in a certain execution engine, denoted as $t_i$ (line~\ref{line:plan_and_time:start}-\ref{line:plan_and_time:end}). Here, the optimal plan $p^*$ with a minimum execution time $t^*$ is chosen as the preferred plan in $\bm{y}_w$ (line~\ref{line:plan_compare}-\ref{line:plan_and_time:end}).   
Subsequently, we generate the prompt for the query and the preferred response $\bm{y}_w$ for $p^*$ (line~\ref{line:prompt_generate}).
Finally, we compare the performance of others' plans with $p^*$. If $t^*/t_{i}$ is smaller than a threshold $r_{0} \in (0, 1)$, we choose plan $p_i$ as a dispreferred plan and generate its response $\bm{y}_l$ (line~\ref{line:data_collection:start}-\ref{line:data_collection:end}). 
In this procedure, we keep a single preferred plan $p^*$, instead of selecting preferred and dispreferred plans from round-robin tournaments. 
This will result in a simpler decision-comparing task, making the model performance in the second-stage training more stable.
It is worth mentioning that when additional query optimizers are available, the preference data set $\mathbb{D}_{\mathrm{dpo}}$ can be extended incrementally for $q$. We can extend Algorithm~\ref{alg:pdg} to collect the plan preference pairs between the additional query optimizer and existing $k$ optimizers. Specifically, we add pairs of plans where the optimal plan $p^*$ is from the new optimizer (line~\ref{line:prompt_generate}). 
The new optimizer is expected to further compensate for the unsatisfactory performance of existing $k$ optimizers on some queries, thereby enhancing the overall performance of \LLMQO.

\section{Training of \LLMQO}
\label{sec:method:training}
Based on the data recipe \QueryInstruct, we develop a fine-tuning pipeline in \LLMQO to enhance the capability of general-purpose LLMs in dealing with  query optimization tasks. 
Our fine-tuning strategy for \LLMQO is composed of two stages, as illustrated in Fig.~\ref{fig:pipline}.
To be more specific, the first stage, called Query Instruction Tuning (\QIT), focuses on initiating LLMs into interpreting and resolving the query optimization task by imitating generic query optimizers, which aims at empowering LLMs to generate valid plans. 
The second stage, named as Query Direct Preference Optimization (\QDPO), further enhances the model's proficiency as an intelligent query planner by training it to differentiate between good and plain execution plans.
Both of the two stages are implemented by efficient fine-tuning technique LoRA. 
%
We will present the algorithmic details in this section.

\subsection{Query Instruction Tuning}

The first stage of our training, called Query Instruction Tuning
(\QIT), takes the prompt $\bm{x}$ as input and guides LLMs to infer its corresponding plan $\bm{y}$ as output, which includes the planning path and the bracket final answer. 
\QIT conducts supervised fine-tuning (SFT) on a pre-trained LLM $\pi$ using a training query set $\mathbb{D}_{\mathrm{sft}}$. 
Let $\pi(\bm{y} \mid \bm{x})$ denote the predictive conditional probability computed by Eq.~\eqref{eq:cond_prob}. This stage minimizes the negative log-likelihood as shown in Eq.~\eqref{eq:sft_loss} using stochastic gradient descent, which is consistent with Eq.~\eqref{eq:mle_loss}. 
\begin{align}
    \label{eq:sft_loss}
    \mathcal{L}_{\mathrm{sft}}({\theta}) = \mathop{\mathbb{E}}_{( \bm{x}, \bm{y}) \sim \mathbb{D}_{\mathrm{sft}}} -\Big[ \log \pi(\bm{y} \mid \bm{x})\Big]\text{.}
\end{align}
\QIT is intrinsically supervised imitation learning~\cite{DBLP:journals/csur/HusseinGEJ17} that is a widely-used bootstrap strategy for building complicated agents from scratch.  
For example, the intelligent Go player AlphaGo~\cite{silver2016mastering} learns from a large number of expert game records through supervised learning and then improves its self-play ability through reinforcement learning.
For the query optimization task, the expert experiences are collected from well-designed query optimizers. 
For an input query in $\bm{x}$, \QueryInstruct encapsulates its best plan in $\bm{y}$ among $k$ optimizers. 
Thereby, in the first stage of training, \LLMQO not only learns to resolve the query optimization task step-by-step, but also `distills' the overall best optimizer among the $k$ experts. 



\comment{
\begin{algorithm}[t]
	\small
    \caption{Preference Data Generation}
	\label{alg:pdg}
	\DontPrintSemicolon
    \SetKwData{Procedure}{Procedure}
	\SetKwData{Up}{up}  \SetKwInOut{Input}{Input} \SetKwInOut{Output}{Output}
    \SetKwInOut{Initialize}{Initialize}
	\Input{query prompt set $\{x_1, \ldots, x_n\}$, execution Engine $E$, threshold for preference data selection $r_0$}

	\Output{preference data $\mathbb{D}_{\mathrm{dpo}}$}
    
	\SetKwFunction{Emit}{Emit}
	\SetKwFunction{Check}{Check}
       
	\For{$i \leftarrow 1$ to $n$}{ 
        $(y_{i1}, \ldots, y_{ik})$ ;
     \algocomment{generate $k$ plans from DBMSs for prompt $x_i$} \;

    $(r_{i1}, \ldots, r_{ik}) \,|\, r_{ij} = E(x_i, y_{ij})$ ;
    \algocomment{obtain execution time for each plan}
    
        \For { $j \leftarrow 1$ to $k$ } 
        {
        \For { $l \leftarrow 1$ to $k$ } 
        {
        \If {$j == l$}{
          $continue$ ;
          }
        $r \leftarrow r_{ij}/r_{il}$ ;
        \algocomment{compute the speed-up ratio between the pair of plans $y_{ij}$ and $y_{il}$} \;
        \If{$r > r_0$~}{
        $\mathbb{D_{\text{pref}}} \leftarrow \{\mathbb{D_{\text{pref}}};(x_i,y_{il},y_{ij})\}$ 
        \algocomment{append the accepted sample}
                    }
        }   
        }
        }
	\Return{$\mathbb{D}_{\mathrm{dpo}}$} ; 
\end{algorithm}
}

\begin{algorithm}[t]
	\small
    \caption{Query Direct Preference Optimization}
	\label{alg:dpo}
	\DontPrintSemicolon
	\SetKwData{Up}{up}  \SetKwInOut{Input}{Input} \SetKwInOut{Output}{Output}
    \SetKwInOut{Initialize}{Initialize}
	\Input{training query set $\mathbb{D}_{\mathrm{dpo}}=\{(\bm{x},\bm{y}_w,\bm{y}_l)^{i}\}_{i=1}^N$ , learning rate $\eta$, number of training steps $T$,  model $\pi_\mathrm{sft}$, coefficient $\beta$  }
	\Output{policy model $\pi_\mathrm{dpo}$}
	\SetKwFunction{Emit}{Emit}
	\SetKwFunction{Check}{Check}
	   Initialize $\pi_\mathrm{dpo}$ with $\pi_\mathrm{sft}$
       
          Shuffle the training query set $\mathbb{D}_{\mathrm{dpo}}$ \;
        
	\For{$step \leftarrow 1$ to $T$}{ \label{line:epoch:start}
        
        Sample a batch $\mathcal{B} \sim \mathbb{D}_{\mathrm{dpo}}$ \;
        
        \For { $(x, \bm{y}_w, \bm{y}_l) \in \mathcal{B}$ } 
	{ \label{line:train:start}
    
        Compute the probabilities $\pi_{\mathrm{dpo}}(\bm{y}_w|\bm{x})$ and $\pi_{{\mathrm{dpo}}}(\bm{y}_l|\bm{x})$ \;
        \label{line:train:compute probabilities from policy model}
        
        Compute the probabilities $\pi_{\mathrm{sft}}(\bm{y}_w|\bm{x})$ and $\pi_{\mathrm{sft}}(\bm{y}_l|\bm{x})$ \;
        \label{line:train:compute probabilities from reference model}

        Compute $u(\bm{x}, \bm{y}_w, \bm{y}_l)$ by Eq.~\eqref{eq:u_function} \; \label{line:caculation:function_u}

        Compute the loss $\mathcal{L}_{\mathrm{dpo}}$ by Eq.~\eqref{eq:dpo_function} and accumulate $\loss_{\mathrm{dpo}}$ \;
        \label{line:train:query_loss}
        } 
			
        Update the model $\theta \leftarrow \theta-\eta \nabla_{\theta} \frac{1}{|\mathcal{B}|}rm\mathcal{L}_{\mathbb{dpo}}$ \; \label{line:epoch:end}  
	}
    \Return{$\pi_\mathrm{dpo}$} \;
\end{algorithm}


\subsection{Query Direct Preference Optimization}
The second stage further enhances the LLM's capability for generating more efficient execution plans. 
Here, we deploy Query Direct Preference Optimization (\QDPO)~\cite{DBLP:conf/nips/RafailovSMMEF23} to refine  $\pi_{\mathrm{sft}}$, the model obtained from the first stage. 
The intuition behind this fine-tuning strategy is to maximize the likelihood of the well-performed plans which we prefer LLM to generate,  while minimizing the likelihood of the plain plans which we do not prefer.
Recall that in \cref{sec:queryinstruct:PDG},  the preference plan is generated by Algorithm~\ref{alg:pdg} where given a query in $\bm{x}$,  its preferred and dispreferred plans are in $\bm{y}_w$ and $\bm{y}_l$, respectively. 
For a training query $(\bm{x}, \bm{y}_w, \bm{y}_l)$, \QDPO minimizes the preference loss function in Eq.~\eqref{eq:dpo_function}, 
\begin{align}
    \loss_{\mathrm{dpo}}(\bm{x}, \bm{y}_w, \bm{y}_l;\theta) &=-\left[\log\sigma\left(u(\bm{x}, \bm{y}_w, \bm{y}_l)\right)\right], \label{eq:dpo_function}
\end{align}
where $u$ is a difference of reward that evaluates the difference of utility between the preferred plan $\bm{y}_w$ and the dispreferred plan $\bm{y}_l$, parameterized by the LLM. $\sigma$ is the sigmoid function, 
normalizing the difference of reward into a conditional probability that $\bm{y}_{w}$ is better than $\bm{y}_l$ given $\bm{x}$, i.e., $p(\bm{y}_w \succ \bm{y}_l \mid \bm{x})$.
The difference of reward $u$ is approximated by the LLM to be learned, $\pi_{\mathrm{dpo}}$, using the \QIT model $\pi_{\mathrm{sft}}$ as a reference model as below:
\begin{align}
     u(\bm{x}, \bm{y}_w, \bm{y}_l) &=\beta\log\frac{\pi_{\mathrm{dpo}}(\bm{y}_w\mid \bm{x})}{\pi_{\mathrm{sft}}(\bm{y}_w\mid \bm{x})}-\beta\log\frac{\pi_{\mathrm{dpo}}(\bm{y}_l\mid \bm{x})}{\pi_{\mathrm{sft}}(\bm{y}_l\mid \bm{x})}. \label{eq:u_function} 
 \end{align}
The normalization by $\pi_{\mathrm{sft}}$ in Eq.~\eqref{eq:u_function} is to prevent the model degrading when only optimizing the likelihood $p(\bm{y}_w \succ \bm{y}_l \mid \bm{x})$. $\beta > 0$ is a hyper-parameter to control the divergence of $\pi_{\mathrm{dpo}}$ and $\pi_{\mathrm{sft}}$, where the larger $\beta$, the larger tolerant deviation.

Algorithm~\ref{alg:dpo} presents the training algorithm of \QDPO in \LLMQO{} by stochastic gradient descent. The algorithm iterates over randomly shuffled training queries and processes a batch for a gradient update in   line~\ref{line:train:start}-\ref{line:epoch:end}.  
Specifically, for each sample $(\bm{x},\bm{y}_w,\bm{y}_l)$, we first compute the probabilities $\pi_{\mathrm{dpo}}(\bm{y}_w \mid\bm{x})$ and $\pi_{\mathrm{dpo}}(\bm{y}_l \mid \bm{x})$ by the LLM  $\pi_{\mathrm{dpo}}$ (line~\ref{line:train:compute probabilities from policy model})
, and compute the probabilities $\pi_{\mathrm{sft}}(\bm{y}_w \mid \bm{x})$ and $\pi_{\mathrm{sft}}(\bm{y}_l\mid \bm{x})$ by the reference model $\pi_{\mathrm{sft}}$ (line~\ref{line:train:compute probabilities from reference model}). 
Afterwards, we compute the difference of rewards $u(\bm{x}, \bm{y}_w, \bm{y}_l)$ defined by Eq.~\eqref{eq:u_function}(line~\ref{line:caculation:function_u}), and the loss $\mathcal{L}_{\mathrm{dpo}}$ (line~\ref{line:train:query_loss}). 
Finally, the model parameters are updated with one gradient step based on the gradient of the aggregated DPO loss (line~\ref{line:epoch:end}).

\section{Experimental Study}
\label{sec:exp}

We introduce the experimental setup in \cref{sec:exp:setup} and test our \LLMQO comprehensively in the following facets: 
\ding{172} Compare the effectiveness of \LLMQO on three datasets with various baselines (\cref{sec:exp:compresult}) 
\ding{173} Explore the capability of model adaptation on unseen templates (\cref{sec:exp:generalization})
\ding{174} Investigate the adaptability of \QDPO with respect to new preference data (\cref{sec:exp:preference adaption})
\ding{175} Conduct an ablation study on prompt design of \QueryInstruct and LLM backbones (\cref{sec:exp:ablation}) 
\ding{176} Investigate the sensitivity of \LLMQO regarding parameter configurations (\cref{sec:exp:parameteranalysis}) 
\ding{177} Analyze the invalid plans generated by \LLMQO (\cref{sec:exp:validity_analysis})
\ding{178} Conduct case studies on visualized plans (\cref{sec:exp:case_study}) 
\ding{179} Evaluate the plan generation time of \LLMQO (\cref{sec:exp:time}). 

\subsection{Experimental Setup}
\label{sec:exp:setup}
\stitle{Datasets.}
We use 2 relational datasets, \imdb and \tpcds. 
For \imdb \cite{DBLP:conf/job/Leis18}, we use 6 relations, including \textsf{title}, \textsf{cast\_info}, \textsf{movie\_info}, \textsf{movie\_companies}, \textsf{movie\_info\_idx}, \textsf{movie\_keyword}.
There are 18 attributes and 6 PK/FK join conditions in total. 
\tpcds benchmark \cite{DBLP:conf/tpcds/Poess02} contains 24 relations and includes both PK/FK joins and non-PK/FK many-to-many joins. We use a scale factor of 5.

\stitle{Queries.}
We use three query sets where two query sets are for \IMDB and one is for \tpcds. For the dataset \imdb, we construct a large query workload including query types supported by existing baselines, i.e., PK/FK join queries without selection conditions on the join attributes. We generate query sets with the number of PF/FK joins, $t$, varying from $0$ to $|T| -1$, where $|T|$ is the number of involved relations. 
To generate a query with $t (t > 0)$ joins, firstly a starting relation is uniformly sampled, and then the query is constructed by traversing from the starting relation over the join graph in $t$ steps. 
Here, for each relation in the sampled join query, additional selection predicates are drawn independently. 
For each $t$, $1,000$ queries are generated for \imdb.
Additionally, we generate a large query set of \job. \job, derived from the Join Order Benchmark (JOB)~\cite{DBLP:journals/pvldb/LeisGMBK015}, comprises 70 hand-written join queries that are suitable for serving as query templates, as each query is different from others in the involved set of relations and/or the selection predicates. 
We generate 100 queries for each template by uniformly drawing the literals of the predicates in their corresponding range.  
\dsb~\cite{ding2021dsb}, built upon the \tpcds benchmark, provides 15 SPJ query templates to assess the performance of DBMS, which enlarges the distinction of queries by introducing various join and selection patterns.  
We use templates tpl018, tpl019, tpl027, tpl040, tpl050, tpl084, tpl091, tpl099, tpl101, and tpl102, and generate 100 queries for each template following the same manner as \job. 
Complicated templates that exceed a time limit of 300 seconds in both \Postgres and \Oracle are excluded. 

\stitle{Baselines.}
We compare our \LLMQO with two traditional optimizers: the \Postgres (12.4), \Oracle (23.0) built-in optimizers, and two learned query optimizers: \bao and \hybrid.

\stitle{Evaluation Metrics.}
The generated plans of all the baselines are transformed into the format of \Postgres's physical plan, and then evaluated by the execution engine of \Postgres.
A \Postgres extension, pg\_hint\_plan~\cite{takaya2001nippon},
is used to transform the generated plans by \LLMQO into the execution plan of \Postgres. 
We use the execution time of the generated plans as our evaluation metric. 

\stitle{Implementation \& Parameter Setting.} 
The learning framework of \LLMQO is built on PyTorch~\cite{pytorch} with Unsloth~\cite{unsloth}
, a LoRA-based fine-tuning framework, which is designed for fast and resource-efficient training of LLMs. We use LLaMa-3-8B~\cite{DBLP:journals/corr/abs-2407-21783} as the default LLM backbone. 
%
The models are trained with the AdamW~\cite{DBLP:conf/iclr/LoshchilovH19} optimizer, utilizing a learning rate of $2 \times 10^{-4}$ and $5 \times 10^{-6}$, for \QIT and \QDPO, respectively. For all datasets, the training steps $T$ are set to 600 and 200 for \QIT and \QDPO, respectively. The training batch size is set as 8.
By default, $r_0$ in Algorithm~\ref{alg:pdg}, $\beta$ in Eq.~\eqref{eq:u_function} are set to 0.95 and 0.1, respectively, and data is split by 80\% for training and 20\% for testing. 
Model training and inference are performed on a Tesla A100 GPU with 80GB memory. 
The experiments for query evaluation are conducted on \Postgres 12.4, 
whose execution engine is deployed on a Linux server with 96 CPUs and 512GB RAM. We set up \Postgres with 128MB shared buffers and 2GB working memory. 
For each query, we clean up the shared buffer of \Postgres and the Linux kernel buffer to fulfill a cold start. 

\subsection{Overall Effectiveness}
\label{sec:exp:compresult}
In this section, we compare \LLMQO with two traditional optimizers \Postgres and \Oracle and two state-of-the-art learned optimizers \bao and \hybrid, across three query workloads. 
For learned optimizers, training and test queries are randomly split by $8:2$. 
Table~\ref{tab:main_results} reports the mean, median, 75th, 95th, and 99th quantiles of the plan execution time. 
The mean and median provide an overview of the optimizers' performance, while other quartiles deliver the performance in long-tail cases. 
It is worth mentioning that all the plans generated by \LLMQO are valid. 
Table~\ref{tab:main_results} shows that \LLMQO 
outperforms both traditional and learned optimizers regarding most metrics across three workloads. 
We can observe that \LLMQO (\QIT) brings an remarkable performance gain over \Postgres across three query sets. 
In addition, \LLMQO (\QDPO) further reduces the average query execution time to 91.4\%, 94.4\%, and 31.3\% of \Postgres on \imdb, \job, and \dsb, respectively.
In general, the 95th and 99th quantiles are reduced by a large margin compared to \Postgres, showcasing that \LLMQO (\QDPO) effectively optimizes query execution time in long-tail scenarios.
In comparison to learned optimizers, \LLMQO outperforms them on average performance across three workloads as shown in Table~\ref{tab:main_results}, as training \LLMQO with \QueryInstruct enhances its ability to discover near optimal plans.
Specifically, \bao and \hybrid cannot surpass \LLMQO in long-tail cases of \imdb and \job, because they rely on coarse-grained hints and the limited plan exploration space makes it difficult for them to approach the optimal solution.
We observe that \bao achieves an impressive performance on \dsb, where \NestLoop is banned in about 76\% of the plans, indicating that certain hints are highly effective in specific workloads. 
It is worth noting that it is feasible to integrate the plan of learned optimizers, collaborating with traditional optimizers to further enhance the effectiveness and robustness of \LLMQO. 
In summary, 
our \LLMQO demonstrates the capability of generating valid and high-quality plans and outperforms both the traditional optimizers and learned optimizers in general cases.

\begin{table*}[!t]
\centering
\small
    \caption{Execution time v.s. different optimizers across query sets. Best results are highlighted.}
\setlength\tabcolsep{2.5pt}
\begin{tabular}{lccccc|ccccc|ccccc}
\toprule
Execution time (sec)   & \multicolumn{5}{c}{{\imdb}}       & \multicolumn{5}{c}{\job} & \multicolumn{5}{c}{\dsb}   \\ \cmidrule(lr){1-1}\cmidrule(lr){2-6}\cmidrule(lr){7-11}\cmidrule(lr){12-16}
Method    & Mean  & Median & 75th  & 95th  & 99th   & Mean & Median & 75th & 95th & 99th  & Mean & Median & 75th & 95th & 99th \\ \midrule
\textbf{\Postgres} & 27.88 & 8.06 & 14.52 & 119.20 & 429.36 & 9.94 & 3.01 & 7.08 & 39.34 & 123.17 & 9.64 & 3.45 & 5.03 & 43.91 & 136.27\\ 
\textbf{\Oracle}     & 26.49 & 8.99 & \cellcolor{LightCyan}{13.60} & 108.47 & 389.78 & 10.11 & 3.03 & 10.00 & 38.96 & \cellcolor{LightCyan}{111.21} &5.45 &2.34 & 6.62 & 23.44 & 28.24 \\
\midrule
\bao & 28.10 & 8.05 & 15.36 & 121.71 & 435.53 & 10.07 & 2.71 & 7.14 & 39.46 & 123.60  & 3.09 & 2.48 & 3.99 & 11.74 & 12.21  \\ 
\hybrid & 27.55 & 7.73 & 13.79 & 120.04 & 427.01  & 9.97 & 3.02 & 7.06 & 39.15 & 124.97  & 9.14 & 2.89 & \cellcolor{LightCyan}{3.69} & 43.48 & 136.32  \\ \midrule
\LLMQO (\QIT)   
& 26.68 & 7.71 & 13.79 & 119.10 & 385.48
& 9.73 & 2.27 & 7.10 & 38.94 & 123.10 & 4.65 & 2.37 & 4.43 & 11.66 & 50.60 \\
\% of \textbf{\Postgres}   & 95.7\% & 95.7\% & 95.0\% & 99.9\% & 89.8\%  & 97.9\% & 75.4\% & 100.2\% & 99.0\% & 99.9\%  & 48.2\% & 68.6\% & 88.0\% & 26.6\% & 37.1\%\\

\LLMQO (\QDPO)   & \cellcolor{LightCyan}{25.48} & \cellcolor{LightCyan}{7.65} & 13.70 & \cellcolor{LightCyan}{107.86} & \cellcolor{LightCyan}{373.24}
& \cellcolor{LightCyan}{9.38} & \cellcolor{LightCyan}{2.27} & \cellcolor{LightCyan}{6.96} & \cellcolor{LightCyan}{38.86} & 115.94 & \cellcolor{LightCyan}{3.02} & \cellcolor{LightCyan}{2.33} & 3.74 & \cellcolor{LightCyan}{11.46} & \cellcolor{LightCyan}{11.72} \\
\% of \textbf{\Postgres}  & 91.4\% & 95.0\% & 94.4\% & 90.5\% & 86.9\% & 94.4\% & 75.4\% & 98.3\% & 98.8\% & 94.1\%  & 31.3\% & 67.6\% & 74.4\% & 26.1\% & 8.6\%\\

\bottomrule
\end{tabular}
\label{tab:main_results}
\end{table*}

\begin{table*}[!t]
\setlength\tabcolsep{2.5pt}
\centering
\small
\caption{Performance of out-of-distribution (OOD) templates. Best results are highlighted.}
\begin{tabular}{lccccc|ccccc|ccccc}
\toprule
Execution time (sec)   & \multicolumn{5}{c}{\imdb}       & \multicolumn{5}{c}{\job} & \multicolumn{5}{c}{\dsb}   \\ \cmidrule(lr){1-1}\cmidrule(lr){2-6}\cmidrule(lr){7-11}\cmidrule(lr){12-16}
Method    & Mean  & Median & 75th  & 95th & 99th   & Mean & Median & 75th & 95th & 99th  & Mean & Median & 75th & 95th & 99th \\ \midrule
\Postgres  & 86.59 & 36.73 & 82.64 & 416.37 & 483.57  & 31.09 & 9.11 & \cellcolor{LightCyan}{20.10} & 229.41 & 272.90 & 5.32 & 4.03 & 7.50 & 12.39 & 12.72\\ 
\Oracle        & 79.40 & 34.06 & 76.48 & \cellcolor{LightCyan}{374.90} & \cellcolor{LightCyan}{436.43}  & 30.59 & \cellcolor{LightCyan}{8.72} & 22.87 & {206.98} & {264.63} & 8.84 & \cellcolor{LightCyan}{2.37} & 19.31 & 24.98 & 29.76\\ \midrule
\bao & 86.68 & 36.65 & 82.85 & 417.81 & 484.65 & 31.25 & 9.21 & 20.60 & 229.36 & 273.29  
& 4.41 & 2.50 & 5.49 & 12.08 & 13.02
\\ 
\hybrid & 86.27 & 36.86 & 83.05 & 411.35 & 479.15 & 31.28 & 9.20 & 20.05 & 230.78 & 273.83
& 4.88 & 3.75 & 6.74 & \cellcolor{LightCyan}{10.84} & \cellcolor{LightCyan}{11.17} \\
\midrule
\LLMQO (\QIT)   & \cellcolor{LightCyan}{79.24} & \cellcolor{LightCyan}{33.12} & \cellcolor{LightCyan}{75.28} & 375.73 & 438.32 
& \cellcolor{LightCyan}{29.93} & 8.74 & 22.40 & \cellcolor{LightCyan}{204.55} & \cellcolor{LightCyan}{263.36}
& \cellcolor{LightCyan}{4.40} & 2.39 & \cellcolor{LightCyan}{5.33} & 11.89 & 12.41\\ 
\% of \textbf{\Postgres}   & 91.5\% & 90.2\% & 91.1\% & 90.2\% & 90.6\%  
& 96.3\% & 95.9\% & 111.5\% & 89.2\% & 96.5\%
& 82.8\% & 59.2\% & 71.1\% & 95.9\% & 97.5\%  \\
\% of \textbf{\Oracle}  & 99.8\% & 97.2\% & 98.4\% & 100.2\% & 100.4\% 
& 97.8\% & 100.2\% & 98.0\% & 98.8\% & 99.5\%
& 49.8\% & 100.9\% & 27.6\% & 47.6\% & 41.7\%\\ 
\bottomrule
\end{tabular}
\vspace{-1ex}
\label{tab:ood_performance}
\end{table*}

\subsection{Generalization to OOD Queries}
\label{sec:exp:generalization}
We study the performance of \LLMQO in OOD scenarios, where test queries are in completely different query templates from those of the training queries, to verify the generalizability of \LLMQO.
For \imdb and \job, we use the most challenging queries, queries with 5 tables as the test queries, and train \LLMQO using queries with $2\sim 4$ tables.
For \dsb, we use three query templates with 5, 6 and 12 joins as test queries, while the remaining templates are used as training queries. 
Table~\ref{tab:ood_performance} shows the plan execution time of \LLMQO (\QIT) compared to baseline optimizers on the three query sets. 
In general, \LLMQO consistently outperforms \Postgres on OOD queries by a statistically large margin, 
indicating that \LLMQO can efficiently explore the plan space to obtain better plans.
In addition, \LLMQO achieves competitive performance compared with the state-of-the-art learned optimizers, \bao and \hybrid.
The coarse-grained hint-based optimization of \bao achieves a low query execution time on \dsb 
by average, however, all the listed quantiles are inferior to those of \LLMQO. Moreover, we observe that the 95th and 99th quantiles of \hybrid are lower than those of \LLMQO.
In addition, we try to further fine-tune LLM in \QDPO stage by using queries with OOD templates and test new queries with these templates, however, the corresponding \LLMQO (\QDPO) suffers from generating a large fraction of invalid plans. 
This phenomenon indicates that a
negative effect occurs when using fully unseen query templates to fine-tune LLMs in \QDPO stage, which suggests that the robustness of LLMs should be further improved by smoothing the transition from \QIT to \QDPO and leveraging more abundant training data.

\subsection{Adapting to New Preferences}
\label{sec:exp:preference adaption}
\begin{figure}
  \centering
 \includegraphics[width=0.45\textwidth]{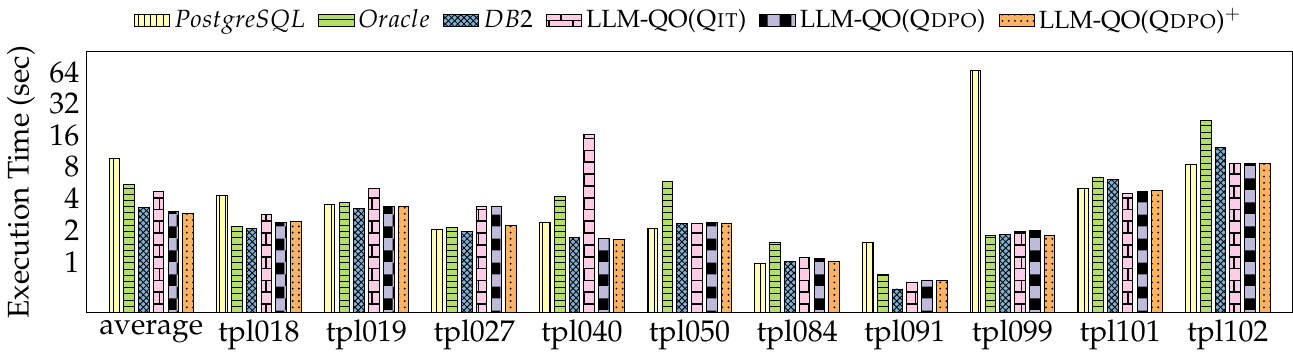}
  \caption{Preference Adaption on \dsb}
  \vspace{-2ex}
\label{fig:exp:preference_adaption}
\end{figure}

To study the adaptability of \LLMQO to possibly new plan preferences through \QDPO stage, we add \db (v12.1) built-in optimizer into Algorithm~\ref{alg:pdg} as an additional source of training plans and construct a new preference training dataset $\mathbb{D}_\mathrm{dpo}^+$ by the optimizers of \Postgres, \Oracle, and  \db  for \dsb. 
Afterwards, we conduct \QDPO training with $\mathbb{D}_\mathrm{dpo}^+$ on \LLMQO (\QIT) and obtain \LLMQO (\QDPO)$^{+}$, in contrast to \LLMQO (\QDPO). 
Fig.~\ref{fig:exp:preference_adaption} compares the average query execution time per template for  the baselines, and the three \LLMQO variants on 200 test queries from \dsb.
In general, we observe that \LLMQO (\QDPO)$^{+}$ achieves the overall best performance across all {\LLMQO}s.
In comparison to \LLMQO (\QIT), \LLMQO (\QDPO) and \LLMQO (\QDPO)$^{+}$ significantly reduce
the execution time on templates tpl018, tpl019, and tpl040, and achieve comparable performance on the remaining templates. 
Fig~\ref{fig:exp:preference_adaption} illustrates that \db outperforms both \Postgres and \Oracle on tpl027. Therefore, the plans generated by \db will be regarded as the preferred plans in $\mathbb{D}_\mathrm{dpo}^+$. \LLMQO (\QDPO)$^{+}$ exhibits performance improvement on tpl027, exceeding that of \LLMQO (\QIT) and \LLMQO (\QDPO), which supports our envision that LLMs can adapt to new preferences and generate more efficient plans in \QDPO stage.
However, we find that \LLMQO (\QDPO)$^{+}$ cannot improve  performance on tpl091, although \db performs better than \Postgres and \Oracle.
This is because the training plans in \QIT stage are from \Oracle for all the queries with tpl091. The behavior of \LLMQO (\QIT) is dominated by \Oracle's preference, and is relatively hard to adapt to more preferences in \QDPO stage. 

\subsection{Ablation Studies}
\label{sec:exp:ablation}
In this section, we conduct ablation studies on different components of \LLMQO, from the perspectives of input features, output format, and LLM backbone to justify the design rationale behind \LLMQO. Table~\ref{tab:ablation} lists the performance of these \LLMQO variants on \dsb.

\subsubsection{Input Features}
As introduced in \cref{sec:method:query_instruct:prepare}, 
the auxiliary statistical information of the database schema and the planning demonstration are important features of \QueryInstruct.  
We remove the statistical information and the one-shot demonstration from the prompt to test their impact on \QueryInstruct respectively. 
As shown in the first part of~Table~\ref{tab:ablation}, compared with the intact version \LLMQO, removing the database statistics feature results in substantial performance degradation.
This result indicates the importance of incorporating data distributions of tables into the plan generation task, which is quite similar to the importance of catalogs to traditional query optimizers. 
In addition, we notice that the withdrawal of the demonstration significantly hurts the performance
on all evaluation metrics. 
This is because the demonstration provides a positive reference for the input query and helps regulate the plan space, and this mechanism is beneficial for plan generation using \LLMQO.
These results further implicitly verify that \LLMQO can understand both the statistics and the demonstration. 

\subsubsection{Output Format}
In \QueryInstruct, we adopt the planning path together with a bracket sequence  as the output format. 
To assess the effectiveness of the planning path, we remove it by only using the bracket answer (Fig.~\ref{fig:example:bracket_sequence})  as an ablation. In the second part of Table~\ref{tab:ablation}, we observe that
the \QIT model yields competitive results, but the performance of the \QDPO model is worse than that of its full version counterpart. 
This finding indicates that the planning path enables the model to differentiate between plain and good plans at a fine-grained level during \QDPO stage, therefore improving the model performance.

\comment{

}
\subsubsection{LLM Backbone}
To explore the capabilities of different LLM backbones for the query optimization task, we adopt Mistral-7B~\cite{jiang2023mistral}, CodeLLaMA-7B~\cite{roziere2023code}, and LLaMA2-7B~\cite{touvron2023llama},  respectively to replace the default LLM, LLaMA3-8B, in \LLMQO. 
As shown in Table~\ref{tab:ablation}, our proposed framework brings improvements across various LLM backbones regarding plan execution time during \QIT and \QDPO stages, affirming its general applicability and superiority.
In addition, the performance of CodeLLaMA-7B and LLaMA2-7B are comparable to that of \LLMQO after \QDPO stage, better than that of Mistral-7B. This phenomenon suggests that LLaMA models may well-suited for the query optimization task. 
It is noteworthy that Mistral-7B after \QDPO stage suffers from the risk of invalid generation.
We observe that the model outputs a large amount of redundant content before performing the plan generation, which is frequently observed in generative models~\cite{DBLP:conf/iclr/HoltzmanBDFC20}.

\begin{table}
\centering
\caption{Ablation studies of \LLMQO on \dsb}
\centering
\begin{adjustbox}{width=0.48\textwidth}
\renewcommand{\arraystretch}{0.8} 
\label{tab:ablation}
\begin{tabular}{@{}l|cccccc@{}}
\toprule
Model Variants     & Mean  & Median & 75th  & 95th & 99th  &invalid\\ 
\midrule
\textbf{\Postgres}     & 9.64 & 3.45 & 5.03 & 43.91 & 136.27 &-
\\ \midrule
Input Features      &  &  &  &  &\\
\quad- \textit{w/o} statistics (\QIT) & 6.28 & 2.41 & 4.25 & 11.74 & 61.91 &-\\
\quad- \textit{w/o} statistics (\QDPO) & 4.58 & 2.39 & 4.24 & 11.63 & 50.03 &-\\
\quad- \textit{w/o} demo (\QIT)  & 6.16 & 2.61 & 4.47 & 32.96 & 62.15 &- \\
\quad- \textit{w/o} demo (\QDPO)  & 5.09 & 2.39 & 4.17 & 11.59 & 37.37 &1  \\
\midrule
Plan Output Format      &  &  &  & &  \\
\quad- \textit{w/o} planning path (\QIT)    & 5.03 & 2.49 & 4.35 & 11.90 & 53.19 &-\\ 
\quad- \textit{w/o} planning path (\QDPO) & 4.24 & 2.54 & 4.24 & 11.60 & 47.59 &-
\\ \midrule
Backbones      &  &  &  & & \\
\quad- CodeLLaMA-7B (\QIT)     & 6.40 & 2.76 & 4.48 & 34.50 & 61.86 &- \\
\quad- CodeLLaMA-7B (\QDPO)      & 3.12 & 2.39 & 3.81 & 11.61 & 11.87 &- \\
\quad- LLaMA2-7B (\QIT)    & 5.15 & 2.50 & 4.24 & 11.61 & 51.78 &-\\
\quad- LLaMA2-7B (\QDPO)  & 3.07 & 2.37 & 3.70 & 11.53 & 11.69 &- \\
\quad- Mistral-7B (\QIT)  & 6.38 & 2.81 & 4.46 & 34.42 & 62.17  &- \\
\quad- Mistral-7B (\QDPO)    & 5.71 & 2.61 & 4.48 & 18.92 & 59.69 &34\\  \midrule
\textbf{Full version}     &  &  &  &  &\\
\quad- \LLMQO (\QIT)     & \textbf{4.65} & \textbf{2.37} & \textbf{4.43} & \textbf{11.60}  &\textbf{50.60} &-\\ 
\quad-  \LLMQO (\QDPO)     & \textbf{3.02} & \textbf{2.33} & \textbf{3.74} & \textbf{11.46} & \textbf{11.72} &-
\\
\bottomrule

\end{tabular}
\end{adjustbox}
\renewcommand{\arraystretch}{1.0}
\vspace{-2.5ex}
\end{table}

\begin{figure*}[t]
\begin{center}
\begin{tabular}[t]{c}
   \subfigure[Varying the scale of \QIT training data]{
   \label{fig:params:scale_sft}
     \includegraphics[width=0.50\columnwidth]{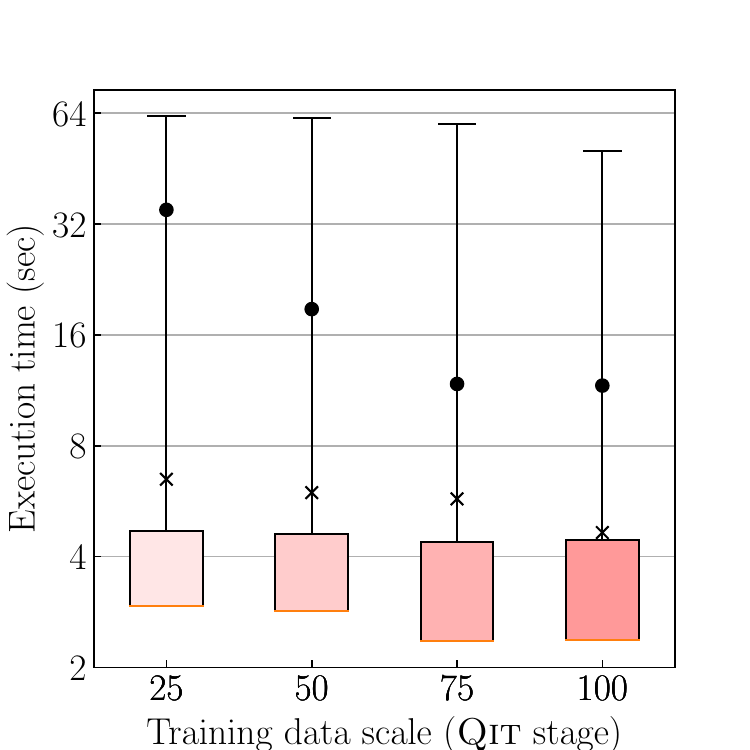} 
    }
    \subfigure[Varying the scale of \QDPO training data]{
      \label{fig:params:scale_dpo}
      \includegraphics[width=0.50\columnwidth]{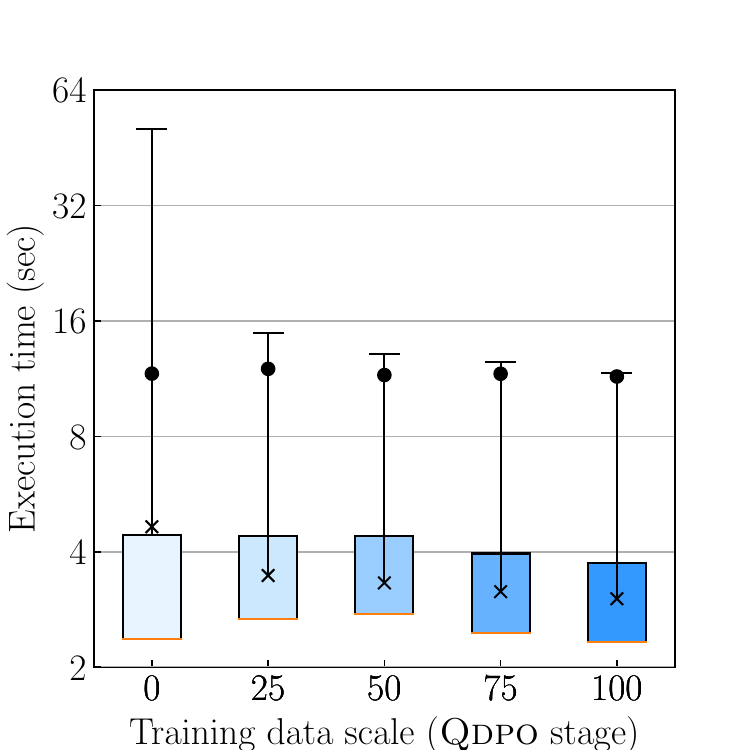} 
    }
    \subfigure[Varying the threshold $r_{0}$ ]{
      \label{fig:params:r_0}
      \includegraphics[width=0.50\columnwidth]{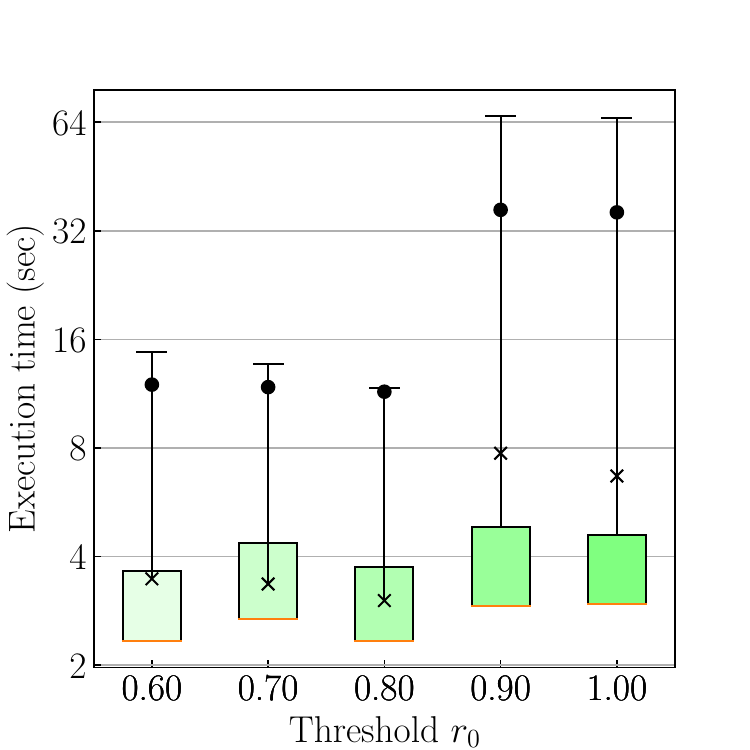}
      }
    \subfigure[Varying the coefficient $\beta$ ]{
      \label{fig:params:beta}
      \includegraphics[width=0.50\columnwidth]{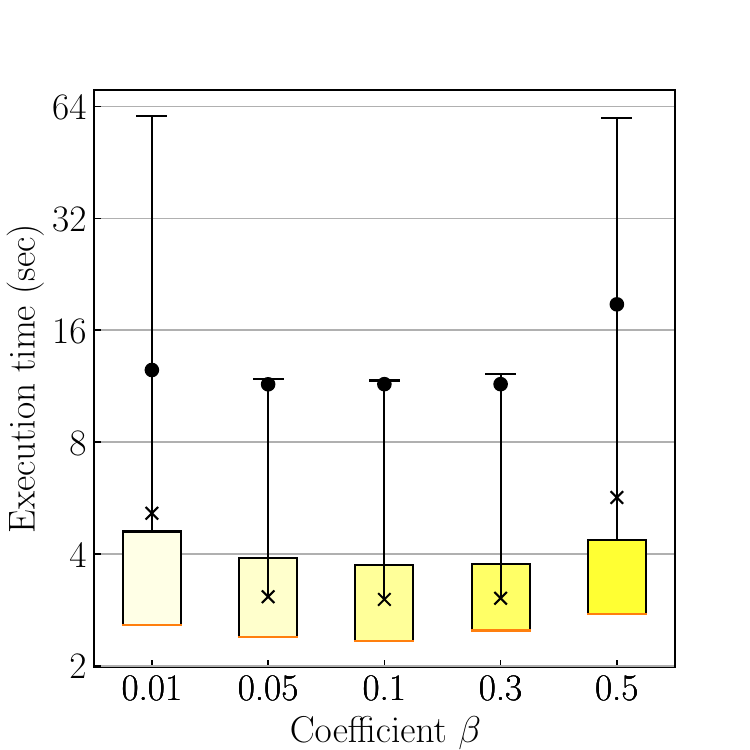} 
    }
\end{tabular}
\end{center}
\caption{The impact of training data scale,  $r_{0}$, and $\beta$ on \dsb}
\label{fig:parameter}
\end{figure*}

\subsection{Parameter Analysis}
\label{sec:exp:parameteranalysis}
We study the parameter sensitivity of \LLMQO w.r.t. three key hyper-parameters: (1) the scale of the training data, (2) the threshold $r_0$ in Algorithm~\ref{alg:pdg} and (3) the coefficient $\beta$ for controlling the divergence of $\pi_\mathrm{dpo}$ from $\pi_\mathrm{sft}$. 
In general, Fig.~\ref{fig:parameter} shows the plan execution time on \dsb for different model variants, where the points `$\times$' and `$\bullet$' denote the mean and 95\% quantiles, respectively. 
%
\subsubsection{Impact of training data scale} 
Fig.~\ref{fig:params:scale_sft}-\ref{fig:params:scale_dpo} show how the scale of training data affects the performance of \LLMQO models on \dsb, in the first and second fine-tuning stages respectively.
LLMs have been observed to possess high data efficiency on text-to-text NLP tasks, achieving competitive performance with a small proportion (e.g., 0.5\%) of full training data~\cite{DBLP:journals/corr/abs-2305-09246}. We study this in the query planning task. 
Specifically, we evaluate the performance of LLaMA3-8B when training on \dsb with different sizes. In the two-stage training, we randomly sample 25\%, 50\%, 75\%, and 100\% of the data from $\mathbb{D}_\mathrm{sft}$ and $\mathbb{D}_\mathrm{dpo}$, respectively. 
In Fig.~\ref{fig:params:scale_sft}, we find a declining trend in execution time as the volume of training data increases for the \QIT models.
Unlike previous observations\cite{DBLP:journals/corr/abs-2305-09246}, the model does not reach its peak performance until using full training data. The reason may be that the structured output of the plan generation task is more challenging for generative LLMs compared to other text-based tasks.
In Fig.~\ref{fig:params:scale_dpo}, we also observe a moderate improvement as the scale of training data increases. 
The data scale $=0$ indicates that the model bypasses \QDPO stage.
We find that the model without DPO fine-tuning cannot achieve results comparable to those achieved by
\LLMQO (\QDPO) models. It is worth noting that the \LLMQO (\QDPO) model can achieve high performance even though the data ratio is as small as 25\%. These findings confirm that the effectiveness of \QDPO stage in \LLMQO when training data is limited.

\comment{
\begin{table}[!t]
\centering
\caption{Performance comparison on the \dsb
dataset with different training data scale in the SFT stage.}
\begin{tabular}{lccccc}
\toprule
Execution time (sec)   & \multicolumn{4}{c}{\textcolor{blue}{DSB}}  \\ \cmidrule(lr){1-1}\cmidrule(lr){2-6}
Data size   & Mean  & Median & 75th  & 95th & 99th \\ \hline \hline
25\% & 6.49 & 2.94 & 4.69 & 35.00 & 62.84 \\ \hline
50\% & 5.97 & 2.85 & 4.61 & 18.81 & 62.15 \\ \hline
75\% 
& 5.74 & 2.36 & 4.38 & 11.78 & 59.97
\\ \hline
100\% & 4.65 & 2.37 &4.43 & 11.66 & 50.60  \\
\bottomrule
\end{tabular}

\label{exp:tab:data scale sft}
\end{table}
}

\comment{
\begin{figure}
  \centering
 \includegraphics[width=0.45\textwidth]{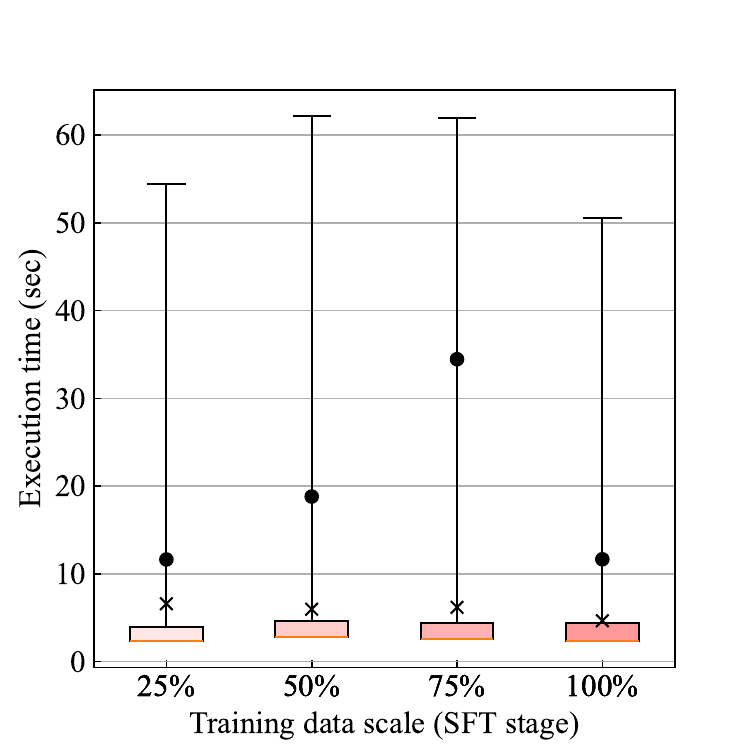}
  \caption{Performance comparison on the \dsb dataset with different training data scale in the SFT stage.}
\label{fig:experiment:scale_up_sft}
\end{figure}
}

\subsubsection{Impact of the threshold $r_0$} 
Recall that \LLMQO introduces the threshold $r_0 \in (0, 1)$ that influences the preference plan generation in \cref{sec:queryinstruct:PDG}.
In Fig.~\ref{fig:params:r_0}, we present the plan execution time for different values of $r_0$, specifically in the range of $\{0.6, 0.7, \cdots, 1.0\}$. 
Algorithm~\ref{alg:pdg} shows that larger values of $r_0$ will increase the volume of preference data. However, setting $r_0$ too large may lead to $\bm{y}_w$ and $\bm{y}_l$ sharing similar quality, potentially impeding the optimization process and model convergence.
When $r_0$ is set between 0.60 and 0.80, the  performance of the model is significantly better than that of 0.90 and 1.00. 
The remarkable performance degradation can be attributed to the fact that in the settings of 0.90 and 1.00, the training data faces a high risk of uncertainty, where the statistical performance dominance is insignificant. 
The best performance is achieved at 0.80 as it increases the size of preference data without compromising its quality.
Therefore, establishing a balanced threshold is essential for optimizing model training and enhancing overall performance.

\begin{table}[t]
	\centering
\caption{The number of invalid cases across three query sets under different \QueryInstruct settings in OOD scenarios}	
        \small 
			\begin{tabular}{l|l|c c c c}
			\toprule
                \multirow{1}{*} {Datasets}  &{\QueryInstruct Variants} 
                
			& \textbf{E1} & \textbf{E2} & \textbf{E3} & {Total} \\ \midrule     
      		\multirow{2}{*}
            {\centering \imdb}	&(a) \textit{w/o} demo          
            &3  &0   &0  &3
            \\
       	& (b) \textit{w/o} planning path            &0 &0  &0  &0  \\

            	& (c) \textit{w/o} both           
                &16 &0  &6  &22
                \\
                     	& (d) \textbf{Full version}            &0 &0  &0  &0  

       \\ \midrule

		\multirow{2}{*}{\centering  \job} &(e) \textit{w/o} demo         
        &10  &0 &0 &10  
         \\
  
		&(f) \textit{w/o} planning path     &0  &0   &1  &1
\\
  
            	& (g) \textit{w/o} both          
               &16 & 0 & 5 & 21 
                
                \\
                     	& (h) \textbf{Full version}            &0 &0  &0  &0  
                      
  \\ \midrule
  		\multirow{2}{*}{\centering \dsb} &(i) \textit{w/o} demo      
        &16 &4    &3   &23
        
        \\
  
		&(j) \textit{w/o} planning path   
                &1 & 0 & 14 & 15
        
        \\
  
            	& (k) \textit{w/o} both          
                &14 & 3 & 12 & 29 
                \\
                     	& (l) \textbf{Full version}            &0 &0  &1  &1  
  
  \\ 
     \bottomrule
		
		\end{tabular}	

	\label{exp:tab:invalid analysis}
        \vspace{-1.0ex}
\end{table}

\subsubsection{Impact of the hyper-parameter $\beta$} 
As described in Algorithm \ref{alg:dpo}, \QDPO stage involves a hyper-parameter $\beta$, which controls the divergence of $\pi_\mathrm{dpo}$ from $\pi_\mathrm{sft}$.
In Fig.~\ref{fig:params:beta}, we vary $\beta$ in its empirical range, \{0.01, 0.05, 0.1, 0.3, 0.5\}, and show the performance of \LLMQO (\QDPO).
Notably, we observe that a large $\beta$, which is correlated with greater knowledge retention, leads to a significant performance drop, indicating its pivotal role in controlling the extent of knowledge  preservation in \QDPO models. 
Conversely, setting $\beta$ to 0.01 substantially weakens the effort in the first-stage training, resulting in a decline in model performance.
The superior performance is achieved with moderate $\beta$ values in 
$\{0.05, 0.1, 0.3\}$.

\subsection{Invalid Plan Analysis}
\label{sec:exp:validity_analysis}
In our experiments, we observe that in OOD scenarios, the plans generated by \LLMQO suffer from different degrees of invalidation across the three query sets.
Therefore, we conduct an in-depth analysis on the effectiveness of the two features in  \QueryInstruct, the planning demonstration and the planning path, in OOD scenarios. 
Table~\ref{exp:tab:invalid analysis} lists the number of invalid cases across the three query sets. 
Specifically, we categorize invalid plans into three common mistakes:
{\textbf{E1) Table Number Mismatch}} denotes that the number of tables in the generated plan is not equal to that of the input query.
{\textbf{E2) Table Mismatch}} indicates that the tables involved in the generated plans do not align with those in the input query. 
{\textbf{E3) Operator Mismatch}} describes general mistakes related to join operators, including bracket mismatchs and missing or redundant operands.
The three categories are not exclusive: a plan may involve multiple types of errors simultaneously.
As shown in Table~\ref{exp:tab:invalid analysis}, both the planning demonstration and the planning path are crucial for reducing invalid generations. 
For \imdb  and \job, using the demonstration can effectively overcome the invalidity issue.
Here, \textbf{E1} is the most frequent error in the `\textit{w/o} both' setting for \imdb and \job, accounting for 74.4\% of the total number. 
As \textbf{E1} comprises redundant and missing table errors, we further observe that missing table errors occur as the majority of \textbf{E1}.
This is because \LLMQO is trained on the queries involving fewer tables while being tested on queries involving more tables, which makes it prone to missing tables.  
Moreover, for complex queries in \dsb, it is necessary to use both the demonstration and the planning path to tackle \textbf{E1}. 
Most errors are \textbf{E1} and \textbf{E3} for \dsb in the `\textit{w/o} both' setting , accounting for 48.2\% and 41.3\% of the total errors, respectively.
Our human analysis shows that \textbf{E3} is typically caused by the complicated inherent structure of plans, which can be significantly alleviated by incorporating the planning path, as shown in Table~\ref{exp:tab:invalid analysis}~(i).

\subsection{Case Study}
\label{sec:exp:case_study}

\begin{figure*}[t]
\centering
\begin{subfigure}[\Postgres (time=190.12s, cost=1137.43)]{%
\scriptsize 
\begin{tikzpicture}[sibling distance=6em, level distance = 2.5em,
  every node/.style = { shape= rectangle, rounded corners,
    draw =none , align=center,
   }]]
  \node[fill=blue!20] { \HashJoin } 
      child {
        node { cast\_info }
      }
      child {
        node[fill=blue!20] {\HashJoin }
          child { node{ movie\_companies} }
          child {
              node[fill=blue!20] { \HashJoin }
                child { node {movie\_keyword} }
                child { node[fill=blue!20] {\HashJoin} 
                             child {node {title}}
                             child {node {movie\_info\_idx}}
                      }
        }
    }
  ;
\end{tikzpicture}
  \label{fig:exp:imdb:postgres}
}\end{subfigure}
%
\begin{subfigure}[\Oracle(time=169.72s, cost=1298.20)]{%
\scriptsize 
\begin{tikzpicture}[sibling distance=6em, level distance = 2.5em,
  every node/.style = { shape= rectangle, rounded corners,
    draw =none , align=center,
   }]]
  \node[fill=blue!20] { \HashJoin } 
      child {
        node[fill=blue!20] {\HashJoin }
          child {
              node[fill=blue!20] { \HashJoin }
                child { node {movie\_keyword} }
                child { node[fill=blue!20] {\HashJoin} 
                             child {node {movie\_info\_idx}}
                             child {node {title}}
                      }
                }
          child { node{ movie\_companies} }
            }
      child {
        node { cast\_info }
      }
      
  ;
\end{tikzpicture}
  \label{fig:exp:imdb:oracle}
}\end{subfigure}
\begin{subfigure}[\QIT (time=176.92s, cost=1946.60)]{%
\scriptsize 
\begin{tikzpicture}[sibling distance=7em, level distance = 2.5em,
  every node/.style = { shape= rectangle, rounded corners,
    draw =none , align=center,
   }]]
  \node[fill=blue!20] { \HashJoin } 
      child {
        node[fill=blue!20] {\HashJoin }
          child {
              node[fill=blue!20] { \HashJoin }
                child { node {movie\_keyword} }
                child { node[fill=blue!20] {\HashJoin} 
                             child {node {movie\_info\_idx}}
                             child {node {movie\_companies}}
                      }
                }
          child { node{ title } }
            }
      child {
        node { cast\_info }
      }
  ;
\end{tikzpicture}
  \label{fig:exp:imdb:qit}
}\end{subfigure}
\begin{subfigure}[\QDPO (time=169.43s, cost=2687.73)]{%
\scriptsize 
\begin{tikzpicture}[sibling distance=6em, level distance = 2.5em,
  every node/.style = { shape= rectangle, rounded corners,
    draw =none , align=center,
   }]]
  \node[fill=blue!20] { \HashJoin } 
      child {
        node[fill=blue!20] {\HashJoin }
          child {
              node[fill=blue!20] { \HashJoin }
                child { node[fill=blue!20] {\HashJoin} 
                             child {node {movie\_info\_idx}}
                             child {node {title}}
                      }
                child { node {movie\_keyword} }
                }
          child { node{ movie\_companies} }
            }
      child {
        node { cast\_info }
      }
      
  ;
\end{tikzpicture}
  \label{fig:exp:imdb:qdpo}
}\end{subfigure}
\caption{Visualization of plans from \Postgres, \Oracle, \LLMQO (\QIT), \LLMQO (\QDPO) on a query of \imdb}
\label{fig:caststudy:example_imdb}

\end{figure*}

\comment{
\begin{figure*}[t]
\centering
\begin{subfigure}[\Oracle (time=6.9s, cost=744)]{%
\scriptsize 
\begin{tikzpicture}[sibling distance=6em, level distance = 2.5em,
  every node/.style = { shape= rectangle, rounded corners,
    draw =none , align=center,
   }]]
  \node[fill=blue!20] { \HashJoin } 
      child {
        node {customer\_ \\
        demographics}
      }
      child {
              node[fill=blue!20] { \HashJoin }
                child { node {store} }
                child { node[fill=blue!20] {\HashJoin} 
                             child {node {data\_dim}}
                             child {
              node[fill=blue!20] { \HashJoin }
              {
              child{node {item}}
              child{node {store\_sales}}
              }
                             }
                      }
        }
  ;
\end{tikzpicture}
  \label{fig:exp:imdb:qdpo}
}\end{subfigure}
\begin{subfigure}[\Postgres (Time=3.1s, Cost=518)]{%
\scriptsize 
\begin{tikzpicture}[sibling distance=7.5em, level distance = 2.5em,
  every node/.style = { shape= rectangle, rounded corners,
    draw =none , align=center,
   }]]
     \node[fill=blue!20] { \HashJoin } 
      child {
        node[fill=blue!20] {\HashJoin }
          child {
              node[fill=blue!20] { \HashJoin }
                child { node[fill=blue!20] {\HashJoin} 
                             child {node {store\_sales}}
                             child {node {data\_dim}}
                      }
                child { node {customer\_demographics} }
                }
          child { node{ store} }
            }
      child {
        node { item }
      }
      ;

\end{tikzpicture}
  \label{fig:example3:subfigure2}
}\end{subfigure}
\begin{subfigure}[\LLMQO(\QIT) \& \LLMQO(\QDPO) (Time=3.3s, Cost=525)]{%
\scriptsize 
\begin{tikzpicture}[sibling distance=6em, level distance = 2.5em,
  every node/.style = { shape= rectangle, rounded corners,
    draw =none , align=center,
   }]]
  \node[fill=green!20] { \NestLoop } 
      child {
        node {item}
            }
      child {
        node[fill=blue!20] {\HashJoin }
          child {
              node[fill=blue!20] { \HashJoin }
                child { node[fill=blue!20] {\HashJoin} 
                             child {node {store\_sales}}
                             child {node {data\_dim}}
                      }
                child { node {customer\_\\
                demographics} }
                }
          child { node{ store} }
      }
  ;
\end{tikzpicture}
  \label{fig:example3:subfigure3}
}\end{subfigure}
\caption{Visualization of plans and planning paths from \Oracle,\Postgres, \LLMQO(\QIT\&\QDPO) on tpl027 of \dsb}
\label{fig:caststudy:example3}
\end{figure*}
}

\comment{
tpl027 45
Oracle 
cost 5616818.46 
Execution Time: 6519.665 ms
HashJoin(customer_demographics HashJoin(store HashJoin(date_dim HashJoin(item store_sales))))

PG 
HashJoin(HashJoin(HashJoin(HashJoin(store_sales date_dim) customer_demographics) store) item)
time 3357.248
time 518168.94 
SFT/DPO
Execution Time: 8799.251 ms
cost 523602.32 
NestLoop(item HashJoin(HashJoin(HashJoin(store_sales date_dim) customer_demographics) store))
LLMQO(DPO)+

HashJoin(item HashJoin(HashJoin(HashJoin(store_sales date_dim) customer_demographics) store))
time 3390.442
cost 518464.11 
}
\comment{
\begin{figure}[t]
\centering
\begin{subfigure}[\Postgres (Time=44s, Cost=298)]{%
\scriptsize 
\begin{tikzpicture}[sibling distance=6em, level distance = 2.5em,
  every node/.style = { shape= rectangle, rounded corners,
    draw =none , align=center,
   }]]
  \node[fill=green!20] { \NestLoop } 
      child {
        node[fill=green!20] { \NestLoop }
          child { node{ warehouse } }
          child {
              node[fill=green!20] { \NestLoop }
                child { node {ship\_mode} }
                child { node[fill=blue!20] {\HashJoin} 
                             child {node {catalog\_sales}}
                             child {node {call\_center}}
                      }
                }
            }
      child {
        node { date\_dim }
      }
  ;
\end{tikzpicture}
  \label{fig:example:subfigure1}
}\end{subfigure}
\begin{subfigure}[\LLMQO (\QIT\&\QDPO)(Time=2s, Cost=304)]{%
\scriptsize 
\begin{tikzpicture}[sibling distance=6em, level distance = 2.5em,
  every node/.style = { shape= rectangle, rounded corners,
    draw =none , align=center,
   }]]
  \node[fill=blue!20] { \HashJoin } 
      child {
        node { call\_center }
      }
      child {
        node[fill=blue!20] {\HashJoin }
          child { node{ warehouse} }
          child {
              node[fill=blue!20] { \HashJoin }
                child { node {date\_dim} }
                child { node[fill=blue!20] {\HashJoin} 
                             child {node {ship\_mode}}
                             child {node {catalog\_sales}}
                      }
        }
    }
  ;
\end{tikzpicture}
  \label{fig:example2:subfigure2}
}\end{subfigure}
\caption{Visualization of plans and planning paths from \Postgres, \LLMQO (\QIT),\LLMQO (\QDPO) on tpl099 of \dsb.}
\label{fig:caststudy:example2}
\end{figure}
}




\begin{figure*}[t]
\centering
\begin{subfigure}[\Postgres (time=4.49s, cost=342.56)]{%
\scriptsize 
\begin{tikzpicture}[sibling distance=6.5em, level distance = 2.5em,
  every node/.style = { shape= rectangle, rounded corners,
    draw =none , align=center,
   }]]
     \node[fill=green!20] { \NestLoop } 
      child {
        node[fill=green!20] {\NestLoop }
          child {
              node[fill=blue!20] { \HashJoin }
                child { node[fill=blue!20] {\HashJoin} 
                             child {node {catalog\_sales}}
                             child {node {data\_dim}}
                      }
                child { node {catalog\_returns} }
                }
          child { node{ warehouse} }
            }
      child {
        node { item }
      }
      ;
\end{tikzpicture}
  \label{fig:exp:dsb:postgres}
}\end{subfigure}
%
%
\begin{subfigure}[\Oracle (time=2.26s, cost=368.00)]{%
\scriptsize 
\begin{tikzpicture}[sibling distance=7.5em, level distance = 2.5em,
  every node/.style = { shape= rectangle, rounded corners,
    draw =none , align=center,
   }]]
     \node[fill=blue!20] { \HashJoin } 
      child {
        node[fill=blue!20] {\HashJoin }
          child {
              node[fill=blue!20] { \HashJoin }
                child { node {date\_dim} }
                child { node[fill=blue!20] {\HashJoin} 
                             child {node {catalog\_returns}}
                             child {node {catalog\_sales}}
                      }
                }
          child { node{ item} }
            }
      child {
        node { warehouse }
      }
      ;

\end{tikzpicture}
  \label{fig:exp:dsb:oracle}
}\end{subfigure}
\begin{subfigure}[\QIT (time=106.08s, cost=1218.60)]{%
\scriptsize 
\begin{tikzpicture}[sibling distance=6em, level distance = 2.5em,
  every node/.style = { shape= rectangle, rounded corners,
    draw =none , align=center,
   }]]
  \node[fill=green!20] { \NestLoop } 
      child {
        node[fill=green!20] {\NestLoop }
          child {
              node[fill=green!20] { \NestLoop }
                child { node {catalog\_returns} }
                child { node[fill=blue!20] {\HashJoin} 
                             child {node {catalog\_sales}}
                             child {node {data\_dim}}
                      }
                }
          child { node{ warehouse} }
      }
    child {
        node {item}
            }
  ;
\end{tikzpicture}
  \label{fig:exp:dsb:qit}
}\end{subfigure}
\begin{subfigure}[\QDPO (time=2.18s, cost=342.31)]{%
\scriptsize 
\begin{tikzpicture}[sibling distance=6em, level distance = 2.5em,
  every node/.style = { shape= rectangle, rounded corners,
    draw =none , align=center,
   }]]
  \node[fill=blue!20] { \HashJoin } 
    child {
        node[fill=blue!20] {\HashJoin }
          child {
              node[fill=blue!20] { \HashJoin }
                child { node {catalog\_returns} }
                child { node[fill=blue!20] {\HashJoin} 
                             child {node {catalog\_sales}}
                             child {node {data\_dim}}
                      }
                }
          child { node{ item} }
      }
    child {
        node {warehouse}
            }
  ;
\end{tikzpicture}
  \label{fig:exp:dsb:qdpo}
}\end{subfigure}

\caption{Visualization of plans from \Postgres, \Oracle, \LLMQO(\QIT), \LLMQO(\QDPO) on a query of \dsb}
\label{fig:caststudy:example_dsb}
\end{figure*}

We conduct a case study to investigate specific plans generated by \Postgres, \Oracle, and \LLMQO.
In Fig.~\ref{fig:caststudy:example_imdb}, we visualize the plans from a query with 5 tables on \imdb. 
As shown in Fig.~\ref{fig:exp:imdb:postgres}, \Postgres selects the query plan with the lowest estimated cost, even though it spends the longest execution time, indicating a misestimation of the plan cost. In further investigation, we identify that the inaccurate cost estimation arises from an underestimation of cardinality. Specifically, for the topmost hash join in Fig.~\ref{fig:exp:imdb:postgres}, the cardinality of the right sub-plan is underestimated by a factor of 30. Typically, a \HashJoin  builds an in-memory hash table on the smaller input to avoid hash table overflow~\cite{graefe1994sort}. However, the underestimation of cardinality causes \Postgres to build the hash table on the right sub-plan that ultimately turns out to incur a larger join size, resulting in a worse real performance. 
In contrast, as shown in Fig.~\ref{fig:exp:imdb:qit}-\ref{fig:exp:imdb:qdpo}, 
\LLMQO generates efficient plans with different join orders, where the hash join builds the hash table on the smaller input table. 

Fig.~\ref{fig:caststudy:example_dsb} illustrates another query with template tpl040 on \dsb. 
In this case, we observe that the\Oracle's plan is superior to that of \Postgres. 
In addition, we find that \LLMQO(\QIT) generates a plan similar to that of \Postgres, but it takes significantly longer to execute. 
In Fig.~\ref{fig:exp:dsb:qit}, the '\NestLoop' operator over catalog\_returns and the intermediate result of the join between catalog\_sales and date\_dim consumes 101.8 seconds, dominating the overall execution time of the plan.
Among the four baselines, \LLMQO (\QDPO) achieves the best performance.
The training of \QDPO enables \LLMQO to learn on the collected preference data from \Oracle and generate a performant plan for the given query, as shown in Fig.~\ref{fig:exp:dsb:qdpo}.
The comparison between \LLMQO (\QIT) and \LLMQO (\QDPO) confirms that \QIT stage empowers the LLM to generate a valid plan through imitating generic optimizers, and \QDPO stage further guides the the generation of more efficient plans by learning from different optimizers. Our proposed autoregressive generation process fosters new and efficient plans.

\begin{figure}
  \centering
 \includegraphics[width=0.36\textwidth]{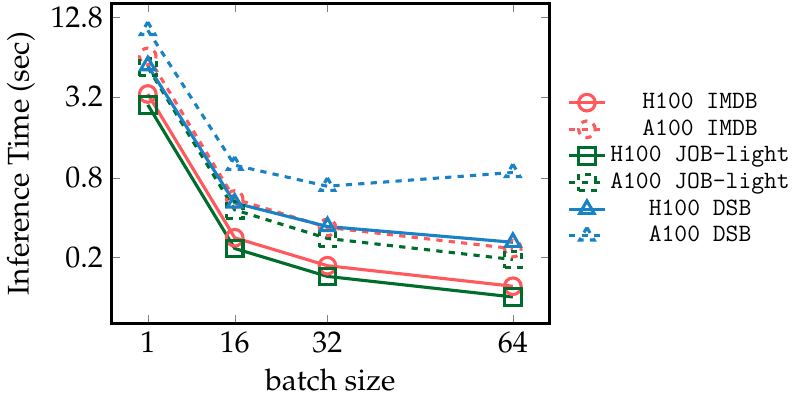}
  \caption{The average inference time on three query sets}
\label{fig:exp:inference_efficiency}
    \vspace{-2.5ex}
\end{figure}

\subsection{Plan Generation Time}
\label{sec:exp:time}
To evaluate the efficiency of plan generation for \LLMQO, we assess the average plan generation latency for input queries across three datasets. 
Fig.~\ref{fig:exp:inference_efficiency} displays the average plan generation latency using LLaMA3-8B on a single A100 or H100 GPU.
The more complex the query, the longer the expected output response and the corresponding generation time.
Among the three datasets, \dsb exhibits the longest inference time because its queries involve more tables and joins compared to \imdb and \job.
In general, we observe that latency decreases as batch size increases due to more efficient utilization of computational and memory resources provided by the hardware.
For \dsb on A100, the inference latency does not decrease when the batch size increases from 32 to 64, as a batch size of 32 is sufficient to exploit GPU resources, given the complicated queries in \dsb.
In summary, complicated queries on large database instances can enjoy more benefits from \LLMQO. Ongoing techniques for speeding up LLM inference, such as KV cache, continuous batching and speculative decoding~\cite{DBLP:conf/icml/LeviathanKM23} will improve the utility of \LLMQO.

\section{Conclusion \& Future Work}
\label{sec:conclusion}
In this paper, we first explore the use of LLMs for SQL query planning by formulating query optimization as a sequence generation task.
To fulfill this desire, we propose an innovative framework, \LLMQO, which absorbs and distills intricate planning strategies from existing optimizers via LLM fine-tuning. 
In \LLMQO, we customize a data recipe \QueryInstruct, based on which a two-stage training workflow is proposed to empower off-the-shelf LLM to generate valid and high-quality execution plans.
We conduct comprehensive experiments on \LLMQO for query evaluation in \Postgres's execution engine, which verify the promise of \LLMQO in both general and OOD settings of query workloads, as well as the effectiveness of our internal designs. 
Our study of \LLMQO paves the way for deploying general LLMs for query planning in database systems. In our investigation, many challenges about generalization, efficiency and adaptivity arise that deserve further exploration. 
First and foremost, we will improve the training of LLMs to enable generalization across multiple database instances, underlying execution engines, query types and logical/physical operators. There is no doubt that this requires abundant computation resources and numerous high-quality training data. 
Second, to better underpin query processing in DBMS, the inference efficiency of \LLMQO should be further improved by systematic optimization. 
Third, \LLMQO may still suffer from the hallucination problem where the input prompt and generation process need to be refined to ensure that the execution plans are coherent with  necessary conditions.
Last but not least, due to the effectiveness of adding \db in \cref{sec:exp:preference adaption}, we plan to add more optimizers to enjoy new preference of plans and adaptivity to dynamic scenarios in \LLMQO.

\clearpage
\balance
\bibliographystyle{ACM-Reference-Format}
\bibliography{bibfile}


\begin{thebibliography}{69}


\ifx \showCODEN    \undefined \def \showCODEN     #1{\unskip}     \fi
\ifx \showDOI      \undefined \def \showDOI       #1{#1}\fi
\ifx \showISBNx    \undefined \def \showISBNx     #1{\unskip}     \fi
\ifx \showISBNxiii \undefined \def \showISBNxiii  #1{\unskip}     \fi
\ifx \showISSN     \undefined \def \showISSN      #1{\unskip}     \fi
\ifx \showLCCN     \undefined \def \showLCCN      #1{\unskip}     \fi
\ifx \shownote     \undefined \def \shownote      #1{#1}          \fi
\ifx \showarticletitle \undefined \def \showarticletitle #1{#1}   \fi
\ifx \showURL      \undefined \def \showURL       {\relax}        \fi
\providecommand\bibfield[2]{#2}
\providecommand\bibinfo[2]{#2}
\providecommand\natexlab[1]{#1}
\providecommand\showeprint[2][]{arXiv:#2}

\bibitem[\protect\citeauthoryear{??}{tak}{[n.d.]}]%
        {takaya2001nippon}
 \bibinfo{year}{[n.d.]}\natexlab{}.
\newblock \bibinfo{title}{Nippon telegraph and telephone corporation}.
\newblock \bibinfo{howpublished}{\url{https://pghintplan.osdn.jp/pg_hint_plan.html}}.
\newblock


\bibitem[\protect\citeauthoryear{??}{pyt}{[n.d.]}]%
        {pytorch}
 \bibinfo{year}{[n.d.]}\natexlab{}.
\newblock \bibinfo{title}{{Pytorch}}.
\newblock \bibinfo{howpublished}{\url{https://github.com/pytorch/pytorch}}.
\newblock


\bibitem[\protect\citeauthoryear{??}{uns}{[n.d.]}]%
        {unsloth}
 \bibinfo{year}{[n.d.]}\natexlab{}.
\newblock \bibinfo{title}{{Unsloth}}.
\newblock \bibinfo{howpublished}{\url{https://github.com/unslothai/unsloth}}.
\newblock


\bibitem[\protect\citeauthoryear{Almazrouei, Alobeidli, Alshamsi, Cappelli, Cojocaru, Debbah, Goffinet, Hesslow, Launay, Malartic, Mazzotta, Noune, Pannier, and Penedo}{Almazrouei et~al\mbox{.}}{2023}]%
        {DBLP:journals/corr/abs-2311-16867}
\bibfield{author}{\bibinfo{person}{Ebtesam Almazrouei}, \bibinfo{person}{Hamza Alobeidli}, \bibinfo{person}{Abdulaziz Alshamsi}, \bibinfo{person}{Alessandro Cappelli}, \bibinfo{person}{Ruxandra Cojocaru}, \bibinfo{person}{M{\'{e}}rouane Debbah}, \bibinfo{person}{{\'{E}}tienne Goffinet}, \bibinfo{person}{Daniel Hesslow}, \bibinfo{person}{Julien Launay}, \bibinfo{person}{Quentin Malartic}, \bibinfo{person}{Daniele Mazzotta}, \bibinfo{person}{Badreddine Noune}, \bibinfo{person}{Baptiste Pannier}, {and} \bibinfo{person}{Guilherme Penedo}.} \bibinfo{year}{2023}\natexlab{}.
\newblock \showarticletitle{The Falcon Series of Open Language Models}.
\newblock \bibinfo{journal}{\emph{CoRR}}  \bibinfo{volume}{abs/2311.16867} (\bibinfo{year}{2023}).
\newblock
\urldef\tempurl%
\url{https://doi.org/10.48550/ARXIV.2311.16867}
\showDOI{\tempurl}
\showeprint[arXiv]{2311.16867}


\bibitem[\protect\citeauthoryear{Arora, Yang, Eyuboglu, Narayan, Hojel, Trummer, and R{\'{e}}}{Arora et~al\mbox{.}}{2023}]%
        {DBLP:journals/pvldb/AroraYENHTR23}
\bibfield{author}{\bibinfo{person}{Simran Arora}, \bibinfo{person}{Brandon Yang}, \bibinfo{person}{Sabri Eyuboglu}, \bibinfo{person}{Avanika Narayan}, \bibinfo{person}{Andrew Hojel}, \bibinfo{person}{Immanuel Trummer}, {and} \bibinfo{person}{Christopher R{\'{e}}}.} \bibinfo{year}{2023}\natexlab{}.
\newblock \showarticletitle{Language Models Enable Simple Systems for Generating Structured Views of Heterogeneous Data Lakes}.
\newblock \bibinfo{journal}{\emph{Proc. {VLDB} Endow.}} \bibinfo{volume}{17}, \bibinfo{number}{2} (\bibinfo{year}{2023}), \bibinfo{pages}{92--105}.
\newblock
\urldef\tempurl%
\url{https://www.vldb.org/pvldb/vol17/p92-arora.pdf}
\showURL{%
\tempurl}


\bibitem[\protect\citeauthoryear{Chaudhuri}{Chaudhuri}{1998}]%
        {DBLP:conf/pods/Chaudhuri98}
\bibfield{author}{\bibinfo{person}{Surajit Chaudhuri}.} \bibinfo{year}{1998}\natexlab{}.
\newblock \showarticletitle{An Overview of Query Optimization in Relational Systems}. In \bibinfo{booktitle}{\emph{Proc. PODS}}, \bibfield{editor}{\bibinfo{person}{Alberto~O. Mendelzon} {and} \bibinfo{person}{Jan Paredaens}} (Eds.). \bibinfo{publisher}{{ACM} Press}, \bibinfo{pages}{34--43}.
\newblock
\urldef\tempurl%
\url{https://doi.org/10.1145/275487.275492}
\showDOI{\tempurl}


\bibitem[\protect\citeauthoryear{Chen, Huang, Ma, Chen, Pan, Ge, Gao, Xie, Liu, Gao, Li, Ding, and Zhou}{Chen et~al\mbox{.}}{2024}]%
        {DBLP:conf/sigmod/Chen0MCPGGXLGLD24}
\bibfield{author}{\bibinfo{person}{Daoyuan Chen}, \bibinfo{person}{Yilun Huang}, \bibinfo{person}{Zhijian Ma}, \bibinfo{person}{Hesen Chen}, \bibinfo{person}{Xuchen Pan}, \bibinfo{person}{Ce Ge}, \bibinfo{person}{Dawei Gao}, \bibinfo{person}{Yuexiang Xie}, \bibinfo{person}{Zhaoyang Liu}, \bibinfo{person}{Jinyang Gao}, \bibinfo{person}{Yaliang Li}, \bibinfo{person}{Bolin Ding}, {and} \bibinfo{person}{Jingren Zhou}.} \bibinfo{year}{2024}\natexlab{}.
\newblock \showarticletitle{Data-Juicer: {A} One-Stop Data Processing System for Large Language Models}. In \bibinfo{booktitle}{\emph{Companion of the 2024 International Conference on Management of Data, {SIGMOD/PODS} 2024, Santiago AA, Chile, June 9-15, 2024}}. \bibinfo{publisher}{{ACM}}, \bibinfo{pages}{120--134}.
\newblock
\urldef\tempurl%
\url{https://doi.org/10.1145/3626246.3653385}
\showDOI{\tempurl}


\bibitem[\protect\citeauthoryear{Chen, Zhang, Zhang, Yang, Hu, Ma, Yanggong, and Zhao}{Chen et~al\mbox{.}}{2023}]%
        {DBLP:journals/corr/abs-2305-09246}
\bibfield{author}{\bibinfo{person}{Hao Chen}, \bibinfo{person}{Yiming Zhang}, \bibinfo{person}{Qi Zhang}, \bibinfo{person}{Hantao Yang}, \bibinfo{person}{Xiaomeng Hu}, \bibinfo{person}{Xuetao Ma}, \bibinfo{person}{Yifan Yanggong}, {and} \bibinfo{person}{Junbo Zhao}.} \bibinfo{year}{2023}\natexlab{}.
\newblock \showarticletitle{Maybe Only 0.5{\%} Data is Needed: {A} Preliminary Exploration of Low Training Data Instruction Tuning}.
\newblock \bibinfo{journal}{\emph{CoRR}}  \bibinfo{volume}{abs/2305.09246} (\bibinfo{year}{2023}).
\newblock
\urldef\tempurl%
\url{https://doi.org/10.48550/ARXIV.2305.09246}
\showDOI{\tempurl}
\showeprint[arXiv]{2305.09246}


\bibitem[\protect\citeauthoryear{Chen, Ye, Zhao, Liu, Deng, Chen, Zhou, and Zheng}{Chen et~al\mbox{.}}{2022}]%
        {DBLP:conf/kdd/0008Y000CZ022}
\bibfield{author}{\bibinfo{person}{Jin Chen}, \bibinfo{person}{Guanyu Ye}, \bibinfo{person}{Yan Zhao}, \bibinfo{person}{Shuncheng Liu}, \bibinfo{person}{Liwei Deng}, \bibinfo{person}{Xu Chen}, \bibinfo{person}{Rui Zhou}, {and} \bibinfo{person}{Kai Zheng}.} \bibinfo{year}{2022}\natexlab{}.
\newblock \showarticletitle{Efficient Join Order Selection Learning with Graph-based Representation}. In \bibinfo{booktitle}{\emph{Proc. {KDD}}}. \bibinfo{publisher}{{ACM}}, \bibinfo{pages}{97--107}.
\newblock
\urldef\tempurl%
\url{https://doi.org/10.1145/3534678.3539303}
\showDOI{\tempurl}


\bibitem[\protect\citeauthoryear{Ding, Chaudhuri, Gehrke, and Narasayya}{Ding et~al\mbox{.}}{2021}]%
        {ding2021dsb}
\bibfield{author}{\bibinfo{person}{Bailu Ding}, \bibinfo{person}{Surajit Chaudhuri}, \bibinfo{person}{Johannes Gehrke}, {and} \bibinfo{person}{Vivek Narasayya}.} \bibinfo{year}{2021}\natexlab{}.
\newblock \showarticletitle{DSB: A decision support benchmark for workload-driven and traditional database systems}.
\newblock \bibinfo{journal}{\emph{Proceedings of the VLDB Endowment}} \bibinfo{volume}{14}, \bibinfo{number}{13} (\bibinfo{year}{2021}), \bibinfo{pages}{3376--3388}.
\newblock


\bibitem[\protect\citeauthoryear{Dubey, Jauhri, Pandey, Kadian, Al{-}Dahle, Letman, Mathur, Schelten, Yang, Fan, Goyal, Hartshorn, Yang, Mitra, Sravankumar, Korenev, Hinsvark, Rao, Zhang, Rodriguez, Gregerson, Spataru, Rozi{\`{e}}re, Biron, Tang, Chern, Caucheteux, Nayak, Bi, Marra, McConnell, Keller, Touret, Wu, Wong, Ferrer, Nikolaidis, Allonsius, Song, Pintz, Livshits, Esiobu, Choudhary, Mahajan, Garcia{-}Olano, Perino, Hupkes, Lakomkin, AlBadawy, Lobanova, Dinan, Smith, Radenovic, Zhang, Synnaeve, Lee, Anderson, Nail, Mialon, Pang, Cucurell, Nguyen, Korevaar, Xu, Touvron, Zarov, Ibarra, Kloumann, Misra, Evtimov, Copet, Lee, Geffert, Vranes, Park, Mahadeokar, Shah, van~der Linde, Billock, Hong, Lee, Fu, Chi, Huang, Liu, Wang, Yu, Bitton, Spisak, Park, Rocca, Johnstun, Saxe, Jia, Alwala, Upasani, Plawiak, Li, Heafield, Stone, and et~al.}{Dubey et~al\mbox{.}}{2024}]%
        {DBLP:journals/corr/abs-2407-21783}
\bibfield{author}{\bibinfo{person}{Abhimanyu Dubey}, \bibinfo{person}{Abhinav Jauhri}, \bibinfo{person}{Abhinav Pandey}, \bibinfo{person}{Abhishek Kadian}, \bibinfo{person}{Ahmad Al{-}Dahle}, \bibinfo{person}{Aiesha Letman}, \bibinfo{person}{Akhil Mathur}, \bibinfo{person}{Alan Schelten}, \bibinfo{person}{Amy Yang}, \bibinfo{person}{Angela Fan}, \bibinfo{person}{Anirudh Goyal}, \bibinfo{person}{Anthony Hartshorn}, \bibinfo{person}{Aobo Yang}, \bibinfo{person}{Archi Mitra}, \bibinfo{person}{Archie Sravankumar}, \bibinfo{person}{Artem Korenev}, \bibinfo{person}{Arthur Hinsvark}, \bibinfo{person}{Arun Rao}, \bibinfo{person}{Aston Zhang}, \bibinfo{person}{Aur{\'{e}}lien Rodriguez}, \bibinfo{person}{Austen Gregerson}, \bibinfo{person}{Ava Spataru}, \bibinfo{person}{Baptiste Rozi{\`{e}}re}, \bibinfo{person}{Bethany Biron}, \bibinfo{person}{Binh Tang}, \bibinfo{person}{Bobbie Chern}, \bibinfo{person}{Charlotte Caucheteux}, \bibinfo{person}{Chaya Nayak}, \bibinfo{person}{Chloe Bi}, \bibinfo{person}{Chris Marra},
  \bibinfo{person}{Chris McConnell}, \bibinfo{person}{Christian Keller}, \bibinfo{person}{Christophe Touret}, \bibinfo{person}{Chunyang Wu}, \bibinfo{person}{Corinne Wong}, \bibinfo{person}{Cristian~Canton Ferrer}, \bibinfo{person}{Cyrus Nikolaidis}, \bibinfo{person}{Damien Allonsius}, \bibinfo{person}{Daniel Song}, \bibinfo{person}{Danielle Pintz}, \bibinfo{person}{Danny Livshits}, \bibinfo{person}{David Esiobu}, \bibinfo{person}{Dhruv Choudhary}, \bibinfo{person}{Dhruv Mahajan}, \bibinfo{person}{Diego Garcia{-}Olano}, \bibinfo{person}{Diego Perino}, \bibinfo{person}{Dieuwke Hupkes}, \bibinfo{person}{Egor Lakomkin}, \bibinfo{person}{Ehab AlBadawy}, \bibinfo{person}{Elina Lobanova}, \bibinfo{person}{Emily Dinan}, \bibinfo{person}{Eric~Michael Smith}, \bibinfo{person}{Filip Radenovic}, \bibinfo{person}{Frank Zhang}, \bibinfo{person}{Gabriel Synnaeve}, \bibinfo{person}{Gabrielle Lee}, \bibinfo{person}{Georgia~Lewis Anderson}, \bibinfo{person}{Graeme Nail}, \bibinfo{person}{Gr{\'{e}}goire Mialon},
  \bibinfo{person}{Guan Pang}, \bibinfo{person}{Guillem Cucurell}, \bibinfo{person}{Hailey Nguyen}, \bibinfo{person}{Hannah Korevaar}, \bibinfo{person}{Hu Xu}, \bibinfo{person}{Hugo Touvron}, \bibinfo{person}{Iliyan Zarov}, \bibinfo{person}{Imanol~Arrieta Ibarra}, \bibinfo{person}{Isabel~M. Kloumann}, \bibinfo{person}{Ishan Misra}, \bibinfo{person}{Ivan Evtimov}, \bibinfo{person}{Jade Copet}, \bibinfo{person}{Jaewon Lee}, \bibinfo{person}{Jan Geffert}, \bibinfo{person}{Jana Vranes}, \bibinfo{person}{Jason Park}, \bibinfo{person}{Jay Mahadeokar}, \bibinfo{person}{Jeet Shah}, \bibinfo{person}{Jelmer van~der Linde}, \bibinfo{person}{Jennifer Billock}, \bibinfo{person}{Jenny Hong}, \bibinfo{person}{Jenya Lee}, \bibinfo{person}{Jeremy Fu}, \bibinfo{person}{Jianfeng Chi}, \bibinfo{person}{Jianyu Huang}, \bibinfo{person}{Jiawen Liu}, \bibinfo{person}{Jie Wang}, \bibinfo{person}{Jiecao Yu}, \bibinfo{person}{Joanna Bitton}, \bibinfo{person}{Joe Spisak}, \bibinfo{person}{Jongsoo Park}, \bibinfo{person}{Joseph Rocca},
  \bibinfo{person}{Joshua Johnstun}, \bibinfo{person}{Joshua Saxe}, \bibinfo{person}{Junteng Jia}, \bibinfo{person}{Kalyan~Vasuden Alwala}, \bibinfo{person}{Kartikeya Upasani}, \bibinfo{person}{Kate Plawiak}, \bibinfo{person}{Ke Li}, \bibinfo{person}{Kenneth Heafield}, \bibinfo{person}{Kevin Stone}, {and} \bibinfo{person}{et al.}} \bibinfo{year}{2024}\natexlab{}.
\newblock \showarticletitle{The Llama 3 Herd of Models}.
\newblock \bibinfo{journal}{\emph{CoRR}}  \bibinfo{volume}{abs/2407.21783} (\bibinfo{year}{2024}).
\newblock
\urldef\tempurl%
\url{https://doi.org/10.48550/ARXIV.2407.21783}
\showDOI{\tempurl}
\showeprint[arXiv]{2407.21783}


\bibitem[\protect\citeauthoryear{Dubois, Li, Taori, Zhang, Gulrajani, Ba, Guestrin, Liang, and Hashimoto}{Dubois et~al\mbox{.}}{2023}]%
        {DBLP:conf/nips/DuboisLTZGBGLH23}
\bibfield{author}{\bibinfo{person}{Yann Dubois}, \bibinfo{person}{Chen~Xuechen Li}, \bibinfo{person}{Rohan Taori}, \bibinfo{person}{Tianyi Zhang}, \bibinfo{person}{Ishaan Gulrajani}, \bibinfo{person}{Jimmy Ba}, \bibinfo{person}{Carlos Guestrin}, \bibinfo{person}{Percy Liang}, {and} \bibinfo{person}{Tatsunori~B. Hashimoto}.} \bibinfo{year}{2023}\natexlab{}.
\newblock \showarticletitle{AlpacaFarm: {A} Simulation Framework for Methods that Learn from Human Feedback}. In \bibinfo{booktitle}{\emph{Proc. NeurIPS}}.
\newblock
\urldef\tempurl%
\url{http://papers.nips.cc/paper\_files/paper/2023/hash/5fc47800ee5b30b8777fdd30abcaaf3b-Abstract-Conference.html}
\showURL{%
\tempurl}


\bibitem[\protect\citeauthoryear{Freytag, Maier, and Vossen}{Freytag et~al\mbox{.}}{1994}]%
        {DBLP:books/mk/FreytagMV91}
\bibfield{editor}{\bibinfo{person}{Johann~Christoph Freytag}, \bibinfo{person}{David Maier}, {and} \bibinfo{person}{Gottfried Vossen}} (Eds.). \bibinfo{year}{1994}\natexlab{}.
\newblock \bibinfo{booktitle}{\emph{Query Processing for Advanced Database Systems}}.
\newblock \bibinfo{publisher}{Morgan Kaufmann}.
\newblock
\showISBNx{1-55860-271-2}


\bibitem[\protect\citeauthoryear{Giannakouris and Trummer}{Giannakouris and Trummer}{2024}]%
        {DBLP:conf/sigmod/GiannakourisT24}
\bibfield{author}{\bibinfo{person}{Victor Giannakouris} {and} \bibinfo{person}{Immanuel Trummer}.} \bibinfo{year}{2024}\natexlab{}.
\newblock \showarticletitle{Demonstrating {\(\lambda\)}-Tune: Exploiting Large Language Models for Workload-Adaptive Database System Tuning}. In \bibinfo{booktitle}{\emph{Companion of {SIGMOD/PODS}}}. \bibinfo{publisher}{{ACM}}, \bibinfo{pages}{508--511}.
\newblock
\urldef\tempurl%
\url{https://doi.org/10.1145/3626246.3654751}
\showDOI{\tempurl}


\bibitem[\protect\citeauthoryear{Graefe}{Graefe}{1993}]%
        {DBLP:journals/csur/Graefe93}
\bibfield{author}{\bibinfo{person}{Goetz Graefe}.} \bibinfo{year}{1993}\natexlab{}.
\newblock \showarticletitle{Query Evaluation Techniques for Large Databases}.
\newblock \bibinfo{journal}{\emph{{ACM} Comput. Surv.}} \bibinfo{volume}{25}, \bibinfo{number}{2} (\bibinfo{year}{1993}), \bibinfo{pages}{73--170}.
\newblock
\urldef\tempurl%
\url{https://doi.org/10.1145/152610.152611}
\showDOI{\tempurl}


\bibitem[\protect\citeauthoryear{Graefe, Linville, and Shapiro}{Graefe et~al\mbox{.}}{1994}]%
        {graefe1994sort}
\bibfield{author}{\bibinfo{person}{Goetz Graefe}, \bibinfo{person}{Ann Linville}, {and} \bibinfo{person}{Leonard~D. Shapiro}.} \bibinfo{year}{1994}\natexlab{}.
\newblock \showarticletitle{Sort vs. hash revisited}.
\newblock \bibinfo{journal}{\emph{IEEE Transactions on Knowledge and Data Engineering}} \bibinfo{volume}{6}, \bibinfo{number}{6} (\bibinfo{year}{1994}), \bibinfo{pages}{934--944}.
\newblock


\bibitem[\protect\citeauthoryear{Holtzman, Buys, Du, Forbes, and Choi}{Holtzman et~al\mbox{.}}{2020}]%
        {DBLP:conf/iclr/HoltzmanBDFC20}
\bibfield{author}{\bibinfo{person}{Ari Holtzman}, \bibinfo{person}{Jan Buys}, \bibinfo{person}{Li Du}, \bibinfo{person}{Maxwell Forbes}, {and} \bibinfo{person}{Yejin Choi}.} \bibinfo{year}{2020}\natexlab{}.
\newblock \showarticletitle{The Curious Case of Neural Text Degeneration}. In \bibinfo{booktitle}{\emph{8th International Conference on Learning Representations, {ICLR} 2020, Addis Ababa, Ethiopia, April 26-30, 2020}}. \bibinfo{publisher}{OpenReview.net}.
\newblock
\urldef\tempurl%
\url{https://openreview.net/forum?id=rygGQyrFvH}
\showURL{%
\tempurl}


\bibitem[\protect\citeauthoryear{Hong, Yuan, Zhang, Chen, Dong, Huang, and Huang}{Hong et~al\mbox{.}}{2024}]%
        {DBLP:journals/corr/abs-2406-08426}
\bibfield{author}{\bibinfo{person}{Zijin Hong}, \bibinfo{person}{Zheng Yuan}, \bibinfo{person}{Qinggang Zhang}, \bibinfo{person}{Hao Chen}, \bibinfo{person}{Junnan Dong}, \bibinfo{person}{Feiran Huang}, {and} \bibinfo{person}{Xiao Huang}.} \bibinfo{year}{2024}\natexlab{}.
\newblock \showarticletitle{Next-Generation Database Interfaces: {A} Survey of LLM-based Text-to-SQL}.
\newblock \bibinfo{journal}{\emph{CoRR}}  \bibinfo{volume}{abs/2406.08426} (\bibinfo{year}{2024}).
\newblock
\urldef\tempurl%
\url{https://doi.org/10.48550/ARXIV.2406.08426}
\showDOI{\tempurl}
\showeprint[arXiv]{2406.08426}


\bibitem[\protect\citeauthoryear{Hu, Shen, Wallis, Allen{-}Zhu, Li, Wang, Wang, and Chen}{Hu et~al\mbox{.}}{2022}]%
        {DBLP:conf/iclr/HuSWALWWC22}
\bibfield{author}{\bibinfo{person}{Edward~J. Hu}, \bibinfo{person}{Yelong Shen}, \bibinfo{person}{Phillip Wallis}, \bibinfo{person}{Zeyuan Allen{-}Zhu}, \bibinfo{person}{Yuanzhi Li}, \bibinfo{person}{Shean Wang}, \bibinfo{person}{Lu Wang}, {and} \bibinfo{person}{Weizhu Chen}.} \bibinfo{year}{2022}\natexlab{}.
\newblock \showarticletitle{LoRA: Low-Rank Adaptation of Large Language Models}. In \bibinfo{booktitle}{\emph{Proc. {ICLR}}}. \bibinfo{publisher}{OpenReview.net}.
\newblock
\urldef\tempurl%
\url{https://openreview.net/forum?id=nZeVKeeFYf9}
\showURL{%
\tempurl}


\bibitem[\protect\citeauthoryear{Hussein, Gaber, Elyan, and Jayne}{Hussein et~al\mbox{.}}{2017}]%
        {DBLP:journals/csur/HusseinGEJ17}
\bibfield{author}{\bibinfo{person}{Ahmed Hussein}, \bibinfo{person}{Mohamed~Medhat Gaber}, \bibinfo{person}{Eyad Elyan}, {and} \bibinfo{person}{Chrisina Jayne}.} \bibinfo{year}{2017}\natexlab{}.
\newblock \showarticletitle{Imitation Learning: {A} Survey of Learning Methods}.
\newblock \bibinfo{journal}{\emph{{ACM} Comput. Surv.}} \bibinfo{volume}{50}, \bibinfo{number}{2} (\bibinfo{year}{2017}), \bibinfo{pages}{21:1--21:35}.
\newblock
\urldef\tempurl%
\url{https://doi.org/10.1145/3054912}
\showDOI{\tempurl}


\bibitem[\protect\citeauthoryear{Jarke and Koch}{Jarke and Koch}{1984}]%
        {DBLP:journals/csur/JarkeK84}
\bibfield{author}{\bibinfo{person}{Matthias Jarke} {and} \bibinfo{person}{J{\"{u}}rgen Koch}.} \bibinfo{year}{1984}\natexlab{}.
\newblock \showarticletitle{Query Optimization in Database Systems}.
\newblock \bibinfo{journal}{\emph{{ACM} Comput. Surv.}} \bibinfo{volume}{16}, \bibinfo{number}{2} (\bibinfo{year}{1984}), \bibinfo{pages}{111--152}.
\newblock
\urldef\tempurl%
\url{https://doi.org/10.1145/356924.356928}
\showDOI{\tempurl}


\bibitem[\protect\citeauthoryear{Jiang, Sablayrolles, Mensch, Bamford, Chaplot, Casas, Bressand, Lengyel, Lample, Saulnier, et~al\mbox{.}}{Jiang et~al\mbox{.}}{2023a}]%
        {jiang2023mistral}
\bibfield{author}{\bibinfo{person}{Albert~Q Jiang}, \bibinfo{person}{Alexandre Sablayrolles}, \bibinfo{person}{Arthur Mensch}, \bibinfo{person}{Chris Bamford}, \bibinfo{person}{Devendra~Singh Chaplot}, \bibinfo{person}{Diego de~las Casas}, \bibinfo{person}{Florian Bressand}, \bibinfo{person}{Gianna Lengyel}, \bibinfo{person}{Guillaume Lample}, \bibinfo{person}{Lucile Saulnier}, {et~al\mbox{.}}} \bibinfo{year}{2023}\natexlab{a}.
\newblock \showarticletitle{Mistral 7B}.
\newblock \bibinfo{journal}{\emph{arXiv preprint arXiv:2310.06825}} (\bibinfo{year}{2023}).
\newblock


\bibitem[\protect\citeauthoryear{Jiang, Sablayrolles, Mensch, Bamford, Chaplot, de~Las~Casas, Bressand, Lengyel, Lample, Saulnier, Lavaud, Lachaux, Stock, Scao, Lavril, Wang, Lacroix, and Sayed}{Jiang et~al\mbox{.}}{2023b}]%
        {DBLP:journals/corr/abs-2310-06825}
\bibfield{author}{\bibinfo{person}{Albert~Q. Jiang}, \bibinfo{person}{Alexandre Sablayrolles}, \bibinfo{person}{Arthur Mensch}, \bibinfo{person}{Chris Bamford}, \bibinfo{person}{Devendra~Singh Chaplot}, \bibinfo{person}{Diego de Las~Casas}, \bibinfo{person}{Florian Bressand}, \bibinfo{person}{Gianna Lengyel}, \bibinfo{person}{Guillaume Lample}, \bibinfo{person}{Lucile Saulnier}, \bibinfo{person}{L{\'{e}}lio~Renard Lavaud}, \bibinfo{person}{Marie{-}Anne Lachaux}, \bibinfo{person}{Pierre Stock}, \bibinfo{person}{Teven~Le Scao}, \bibinfo{person}{Thibaut Lavril}, \bibinfo{person}{Thomas Wang}, \bibinfo{person}{Timoth{\'{e}}e Lacroix}, {and} \bibinfo{person}{William~El Sayed}.} \bibinfo{year}{2023}\natexlab{b}.
\newblock \showarticletitle{Mistral 7B}.
\newblock \bibinfo{journal}{\emph{CoRR}}  \bibinfo{volume}{abs/2310.06825} (\bibinfo{year}{2023}).
\newblock
\urldef\tempurl%
\url{https://doi.org/10.48550/ARXIV.2310.06825}
\showDOI{\tempurl}
\showeprint[arXiv]{2310.06825}


\bibitem[\protect\citeauthoryear{Jiang, Wang, Shen, Kim, and Kim}{Jiang et~al\mbox{.}}{2024}]%
        {jiang2024surveylargelanguagemodels}
\bibfield{author}{\bibinfo{person}{Juyong Jiang}, \bibinfo{person}{Fan Wang}, \bibinfo{person}{Jiasi Shen}, \bibinfo{person}{Sungju Kim}, {and} \bibinfo{person}{Sunghun Kim}.} \bibinfo{year}{2024}\natexlab{}.
\newblock \bibinfo{title}{A Survey on Large Language Models for Code Generation}.
\newblock
\newblock
\showeprint[arxiv]{2406.00515}~[cs.CL]
\urldef\tempurl%
\url{https://arxiv.org/abs/2406.00515}
\showURL{%
\tempurl}


\bibitem[\protect\citeauthoryear{Kim, Reiner, and Batory}{Kim et~al\mbox{.}}{1985}]%
        {DBLP:books/sp/KimRB85}
\bibfield{editor}{\bibinfo{person}{Won Kim}, \bibinfo{person}{David~S. Reiner}, {and} \bibinfo{person}{Don~S. Batory}} (Eds.). \bibinfo{year}{1985}\natexlab{}.
\newblock \bibinfo{booktitle}{\emph{Query Processing in Database Systems}}.
\newblock \bibinfo{publisher}{Springer}.
\newblock
\showISBNx{3-540-13831-5}


\bibitem[\protect\citeauthoryear{Kojima, Gu, Reid, Matsuo, and Iwasawa}{Kojima et~al\mbox{.}}{2022}]%
        {DBLP:conf/nips/KojimaGRMI22}
\bibfield{author}{\bibinfo{person}{Takeshi Kojima}, \bibinfo{person}{Shixiang~Shane Gu}, \bibinfo{person}{Machel Reid}, \bibinfo{person}{Yutaka Matsuo}, {and} \bibinfo{person}{Yusuke Iwasawa}.} \bibinfo{year}{2022}\natexlab{}.
\newblock \showarticletitle{Large Language Models are Zero-Shot Reasoners}. In \bibinfo{booktitle}{\emph{Proc. NeurIPS}}, \bibfield{editor}{\bibinfo{person}{Sanmi Koyejo}, \bibinfo{person}{S.~Mohamed}, \bibinfo{person}{A.~Agarwal}, \bibinfo{person}{Danielle Belgrave}, \bibinfo{person}{K.~Cho}, {and} \bibinfo{person}{A.~Oh}} (Eds.).
\newblock
\urldef\tempurl%
\url{http://papers.nips.cc/paper\_files/paper/2022/hash/8bb0d291acd4acf06ef112099c16f326-Abstract-Conference.html}
\showURL{%
\tempurl}


\bibitem[\protect\citeauthoryear{Krishnan, Yang, Goldberg, Hellerstein, and Stoica}{Krishnan et~al\mbox{.}}{2018}]%
        {DBLP:journals/corr/abs-1808-03196}
\bibfield{author}{\bibinfo{person}{Sanjay Krishnan}, \bibinfo{person}{Zongheng Yang}, \bibinfo{person}{Ken Goldberg}, \bibinfo{person}{Joseph~M. Hellerstein}, {and} \bibinfo{person}{Ion Stoica}.} \bibinfo{year}{2018}\natexlab{}.
\newblock \showarticletitle{Learning to Optimize Join Queries With Deep Reinforcement Learning}.
\newblock \bibinfo{journal}{\emph{CoRR}}  \bibinfo{volume}{abs/1808.03196} (\bibinfo{year}{2018}).
\newblock
\showeprint[arXiv]{1808.03196}
\urldef\tempurl%
\url{http://arxiv.org/abs/1808.03196}
\showURL{%
\tempurl}


\bibitem[\protect\citeauthoryear{Lao, Wang, Li, Wang, Zhang, Cheng, Chen, Tang, and Wang}{Lao et~al\mbox{.}}{2024}]%
        {DBLP:journals/pvldb/LaoWLWZCCTW24}
\bibfield{author}{\bibinfo{person}{Jiale Lao}, \bibinfo{person}{Yibo Wang}, \bibinfo{person}{Yufei Li}, \bibinfo{person}{Jianping Wang}, \bibinfo{person}{Yunjia Zhang}, \bibinfo{person}{Zhiyuan Cheng}, \bibinfo{person}{Wanghu Chen}, \bibinfo{person}{Mingjie Tang}, {and} \bibinfo{person}{Jianguo Wang}.} \bibinfo{year}{2024}\natexlab{}.
\newblock \showarticletitle{GPTuner: {A} Manual-Reading Database Tuning System via GPT-Guided Bayesian Optimization}.
\newblock \bibinfo{journal}{\emph{Proc. {VLDB} Endow.}} \bibinfo{volume}{17}, \bibinfo{number}{8} (\bibinfo{year}{2024}), \bibinfo{pages}{1939--1952}.
\newblock
\urldef\tempurl%
\url{https://www.vldb.org/pvldb/vol17/p1939-tang.pdf}
\showURL{%
\tempurl}


\bibitem[\protect\citeauthoryear{Leis, Gubichev, Mirchev, Boncz, Kemper, and Neumann}{Leis et~al\mbox{.}}{2015}]%
        {DBLP:journals/pvldb/LeisGMBK015}
\bibfield{author}{\bibinfo{person}{Viktor Leis}, \bibinfo{person}{Andrey Gubichev}, \bibinfo{person}{Atanas Mirchev}, \bibinfo{person}{Peter~A. Boncz}, \bibinfo{person}{Alfons Kemper}, {and} \bibinfo{person}{Thomas Neumann}.} \bibinfo{year}{2015}\natexlab{}.
\newblock \showarticletitle{How Good Are Query Optimizers, Really?}
\newblock \bibinfo{journal}{\emph{Proc. {VLDB}}} \bibinfo{volume}{9}, \bibinfo{number}{3} (\bibinfo{year}{2015}), \bibinfo{pages}{204--215}.
\newblock
\urldef\tempurl%
\url{https://doi.org/10.14778/2850583.2850594}
\showDOI{\tempurl}


\bibitem[\protect\citeauthoryear{Leis, Radke, Gubichev, Mirchev, Boncz, Kemper, and Neumann}{Leis et~al\mbox{.}}{2018}]%
        {DBLP:conf/job/Leis18}
\bibfield{author}{\bibinfo{person}{Viktor Leis}, \bibinfo{person}{Bernhard Radke}, \bibinfo{person}{Andrey Gubichev}, \bibinfo{person}{Atanas Mirchev}, \bibinfo{person}{Peter Boncz}, \bibinfo{person}{Alfons Kemper}, {and} \bibinfo{person}{Thomas Neumann}.} \bibinfo{year}{2018}\natexlab{}.
\newblock \showarticletitle{Query optimization through the looking glass, and what we found running the join order benchmark}.
\newblock \bibinfo{journal}{\emph{The VLDB Journal}}  \bibinfo{volume}{27} (\bibinfo{year}{2018}), \bibinfo{pages}{643--668}.
\newblock


\bibitem[\protect\citeauthoryear{Leviathan, Kalman, and Matias}{Leviathan et~al\mbox{.}}{2023}]%
        {DBLP:conf/icml/LeviathanKM23}
\bibfield{author}{\bibinfo{person}{Yaniv Leviathan}, \bibinfo{person}{Matan Kalman}, {and} \bibinfo{person}{Yossi Matias}.} \bibinfo{year}{2023}\natexlab{}.
\newblock \showarticletitle{Fast Inference from Transformers via Speculative Decoding}. In \bibinfo{booktitle}{\emph{Proc. {ICML}}} \emph{(\bibinfo{series}{Proceedings of Machine Learning Research})}, Vol.~\bibinfo{volume}{202}. \bibinfo{publisher}{{PMLR}}, \bibinfo{pages}{19274--19286}.
\newblock
\urldef\tempurl%
\url{https://proceedings.mlr.press/v202/leviathan23a.html}
\showURL{%
\tempurl}


\bibitem[\protect\citeauthoryear{Li, Hui, Qu, Yang, Li, Li, Wang, Qin, Geng, Huo, Zhou, Ma, Li, Chang, Huang, Cheng, and Li}{Li et~al\mbox{.}}{2023}]%
        {DBLP:conf/nips/LiHQYLLWQGHZ0LC23}
\bibfield{author}{\bibinfo{person}{Jinyang Li}, \bibinfo{person}{Binyuan Hui}, \bibinfo{person}{Ge Qu}, \bibinfo{person}{Jiaxi Yang}, \bibinfo{person}{Binhua Li}, \bibinfo{person}{Bowen Li}, \bibinfo{person}{Bailin Wang}, \bibinfo{person}{Bowen Qin}, \bibinfo{person}{Ruiying Geng}, \bibinfo{person}{Nan Huo}, \bibinfo{person}{Xuanhe Zhou}, \bibinfo{person}{Chenhao Ma}, \bibinfo{person}{Guoliang Li}, \bibinfo{person}{Kevin~Chen{-}Chuan Chang}, \bibinfo{person}{Fei Huang}, \bibinfo{person}{Reynold Cheng}, {and} \bibinfo{person}{Yongbin Li}.} \bibinfo{year}{2023}\natexlab{}.
\newblock \showarticletitle{Can {LLM} Already Serve as {A} Database Interface? {A} BIg Bench for Large-Scale Database Grounded Text-to-SQLs}. In \bibinfo{booktitle}{\emph{Proc. NeurIPS}}.
\newblock
\urldef\tempurl%
\url{http://papers.nips.cc/paper\_files/paper/2023/hash/83fc8fab1710363050bbd1d4b8cc0021-Abstract-Datasets\_and\_Benchmarks.html}
\showURL{%
\tempurl}


\bibitem[\protect\citeauthoryear{Li, He, Yashar, Cui, Ge, Zhang, Fainman, Zhang, and Chaudhuri}{Li et~al\mbox{.}}{2024a}]%
        {DBLP:journals/pacmmod/LiHYCGZF0C24}
\bibfield{author}{\bibinfo{person}{Peng Li}, \bibinfo{person}{Yeye He}, \bibinfo{person}{Dror Yashar}, \bibinfo{person}{Weiwei Cui}, \bibinfo{person}{Song Ge}, \bibinfo{person}{Haidong Zhang}, \bibinfo{person}{Danielle~Rifinski Fainman}, \bibinfo{person}{Dongmei Zhang}, {and} \bibinfo{person}{Surajit Chaudhuri}.} \bibinfo{year}{2024}\natexlab{a}.
\newblock \showarticletitle{Table-GPT: Table Fine-tuned {GPT} for Diverse Table Tasks}.
\newblock \bibinfo{journal}{\emph{Proc. {ACM} Manag. Data}} \bibinfo{volume}{2}, \bibinfo{number}{3} (\bibinfo{year}{2024}), \bibinfo{pages}{176}.
\newblock
\urldef\tempurl%
\url{https://doi.org/10.1145/3654979}
\showDOI{\tempurl}


\bibitem[\protect\citeauthoryear{Li, Yuan, Wang, Cong, and Bing}{Li et~al\mbox{.}}{2024b}]%
        {DBLP:journals/corr/abs-2404-12872}
\bibfield{author}{\bibinfo{person}{Zhaodonghui Li}, \bibinfo{person}{Haitao Yuan}, \bibinfo{person}{Huiming Wang}, \bibinfo{person}{Gao Cong}, {and} \bibinfo{person}{Lidong Bing}.} \bibinfo{year}{2024}\natexlab{b}.
\newblock \showarticletitle{{LLM-R2:} {A} Large Language Model Enhanced Rule-based Rewrite System for Boosting Query Efficiency}.
\newblock \bibinfo{journal}{\emph{CoRR}}  \bibinfo{volume}{abs/2404.12872} (\bibinfo{year}{2024}).
\newblock
\urldef\tempurl%
\url{https://doi.org/10.48550/ARXIV.2404.12872}
\showDOI{\tempurl}
\showeprint[arXiv]{2404.12872}


\bibitem[\protect\citeauthoryear{Loshchilov and Hutter}{Loshchilov and Hutter}{2019}]%
        {DBLP:conf/iclr/LoshchilovH19}
\bibfield{author}{\bibinfo{person}{Ilya Loshchilov} {and} \bibinfo{person}{Frank Hutter}.} \bibinfo{year}{2019}\natexlab{}.
\newblock \showarticletitle{Decoupled Weight Decay Regularization}. In \bibinfo{booktitle}{\emph{Proc. {ICLR}}}. \bibinfo{publisher}{OpenReview.net}.
\newblock
\urldef\tempurl%
\url{https://openreview.net/forum?id=Bkg6RiCqY7}
\showURL{%
\tempurl}


\bibitem[\protect\citeauthoryear{Marcus, Negi, Mao, Tatbul, Alizadeh, and Kraska}{Marcus et~al\mbox{.}}{2021}]%
        {DBLP:conf/sigmod/MarcusNMTAK21}
\bibfield{author}{\bibinfo{person}{Ryan Marcus}, \bibinfo{person}{Parimarjan Negi}, \bibinfo{person}{Hongzi Mao}, \bibinfo{person}{Nesime Tatbul}, \bibinfo{person}{Mohammad Alizadeh}, {and} \bibinfo{person}{Tim Kraska}.} \bibinfo{year}{2021}\natexlab{}.
\newblock \showarticletitle{Bao: Making Learned Query Optimization Practical}. In \bibinfo{booktitle}{\emph{Proc. {SIGMOD}}}. \bibinfo{publisher}{{ACM}}, \bibinfo{pages}{1275--1288}.
\newblock
\urldef\tempurl%
\url{https://doi.org/10.1145/3448016.3452838}
\showDOI{\tempurl}


\bibitem[\protect\citeauthoryear{Marcus, Negi, Mao, Zhang, Alizadeh, Kraska, Papaemmanouil, and Tatbul}{Marcus et~al\mbox{.}}{2019}]%
        {DBLP:journals/pvldb/MarcusNMZAKPT19}
\bibfield{author}{\bibinfo{person}{Ryan Marcus}, \bibinfo{person}{Parimarjan Negi}, \bibinfo{person}{Hongzi Mao}, \bibinfo{person}{Chi Zhang}, \bibinfo{person}{Mohammad Alizadeh}, \bibinfo{person}{Tim Kraska}, \bibinfo{person}{Olga Papaemmanouil}, {and} \bibinfo{person}{Nesime Tatbul}.} \bibinfo{year}{2019}\natexlab{}.
\newblock \showarticletitle{Neo: {A} Learned Query Optimizer}.
\newblock \bibinfo{journal}{\emph{Proc. {VLDB} Endow.}} \bibinfo{volume}{12}, \bibinfo{number}{11} (\bibinfo{year}{2019}), \bibinfo{pages}{1705--1718}.
\newblock
\urldef\tempurl%
\url{https://doi.org/10.14778/3342263.3342644}
\showDOI{\tempurl}


\bibitem[\protect\citeauthoryear{Min, Lyu, Holtzman, Artetxe, Lewis, Hajishirzi, and Zettlemoyer}{Min et~al\mbox{.}}{2022}]%
        {DBLP:conf/emnlp/MinLHALHZ22}
\bibfield{author}{\bibinfo{person}{Sewon Min}, \bibinfo{person}{Xinxi Lyu}, \bibinfo{person}{Ari Holtzman}, \bibinfo{person}{Mikel Artetxe}, \bibinfo{person}{Mike Lewis}, \bibinfo{person}{Hannaneh Hajishirzi}, {and} \bibinfo{person}{Luke Zettlemoyer}.} \bibinfo{year}{2022}\natexlab{}.
\newblock \showarticletitle{Rethinking the Role of Demonstrations: What Makes In-Context Learning Work?}. In \bibinfo{booktitle}{\emph{Proc. {EMNLP}}}. \bibinfo{publisher}{Association for Computational Linguistics}, \bibinfo{pages}{11048--11064}.
\newblock
\urldef\tempurl%
\url{https://doi.org/10.18653/V1/2022.EMNLP-MAIN.759}
\showDOI{\tempurl}


\bibitem[\protect\citeauthoryear{Mou, Li, Zhang, Wang, and Jin}{Mou et~al\mbox{.}}{2016}]%
        {DBLP:conf/aaai/MouLZWJ16}
\bibfield{author}{\bibinfo{person}{Lili Mou}, \bibinfo{person}{Ge Li}, \bibinfo{person}{Lu Zhang}, \bibinfo{person}{Tao Wang}, {and} \bibinfo{person}{Zhi Jin}.} \bibinfo{year}{2016}\natexlab{}.
\newblock \showarticletitle{Convolutional Neural Networks over Tree Structures for Programming Language Processing}. In \bibinfo{booktitle}{\emph{Proc. {AAAI}}}. \bibinfo{publisher}{{AAAI} Press}, \bibinfo{pages}{1287--1293}.
\newblock
\urldef\tempurl%
\url{https://doi.org/10.1609/AAAI.V30I1.10139}
\showDOI{\tempurl}


\bibitem[\protect\citeauthoryear{OpenAI, :, Jaech, Kalai, Lerer, Richardson, El-Kishky, Low, Helyar, Madry, Beutel, Carney, Iftimie, Karpenko, Passos, Neitz, Prokofiev, Wei, Tam, Bennett, Kumar, Saraiva, Vallone, Duberstein, Kondrich, Mishchenko, Applebaum, Jiang, Nair, Zoph, Ghorbani, Rossen, Sokolowsky, Barak, McGrew, Minaiev, Hao, Baker, Houghton, McKinzie, Eastman, Lugaresi, Bassin, Hudson, Li, de~Bourcy, Voss, Shen, Zhang, Koch, Orsinger, Hesse, Fischer, Chan, Roberts, Kappler, Levy, Selsam, Dohan, Farhi, Mely, Robinson, Tsipras, Li, Oprica, Freeman, Zhang, Wong, Proehl, Cheung, Mitchell, Wallace, Ritter, Mays, Wang, Such, Raso, Leoni, Tsimpourlas, Song, von Lohmann, Sulit, Salmon, Parascandolo, Chabot, Zhao, Brockman, Leclerc, Salman, Bao, Sheng, Andrin, Bagherinezhad, Ren, Lightman, Chung, Kivlichan, O'Connell, Osband, Gilaberte, Akkaya, Kostrikov, Sutskever, Kofman, Pachocki, Lennon, Wei, Harb, Twore, Feng, Yu, Weng, Tang, Yu, Candela, Palermo, Parish, Heidecke, Hallman, Rizzo, Gordon, Uesato, Uesato,
  Ward, Huizinga, Wang, Chen, Xiao, Singhal, Nguyen, Cobbe, Shi, Wood, Rimbach, Gu-Lemberg, GuLemberg, Liu, Lu, Stone, Yu, Ahmad, Yang, Liu, Maksin, Ho, Fedus, Weng, Li, McCallum, Held, Kuhn, Kondraciuk, Kaiser, Metz, Boyd, Trebacz, Joglekar, Chen, Tintor, Meyer, Jones, Kaufer, Schwarzer, Shah, Yatbaz, Guan, Xu, Yan, Glaese, Chen, Chen, Lampe, Malek, Wang, Fradin, McClay, Pavlov, Wang, Wang, Murati, Bavarian, Rohaninejad, McAleese, Chowdhury, Chowdhury, Ryder, Tezak, Brown, Nachum, Boiko, Murk, Watkins, Chao, Ashbourne, Izmailov, Zhokhov, Dias, Arora, Lin, Lopes, Gaon, Miyara, Leike, Hwang, Garg, Brown, James, Shu, Cheu, Greene, Jain, Altman, Toizer, Toyer, Miserendino, Agarwal, Hernandez, Baker, McKinney, Yan, Zhao, Hu, Santurkar, Chaudhuri, Zhang, Fu, Papay, Lin, Balaji, Sanjeev, Sidor, Broda, Clark, Wang, Gordon, Sanders, Patwardhan, Sottiaux, Degry, Dimson, Zheng, Garipov, Stasi, Bansal, Creech, Peterson, Eloundou, Qi, Kosaraju, Monaco, Pong, Fomenko, Zheng, Zhou, McCabe, Zaremba, Dubois, Lu, Chen, Cha,
  Bai, He, Zhang, Wang, Shao, and Li}{OpenAI et~al\mbox{.}}{2024}]%
        {openai2024openaio1card}
\bibfield{author}{\bibinfo{person}{OpenAI}, \bibinfo{person}{:}, \bibinfo{person}{Aaron Jaech}, \bibinfo{person}{Adam Kalai}, \bibinfo{person}{Adam Lerer}, \bibinfo{person}{Adam Richardson}, \bibinfo{person}{Ahmed El-Kishky}, \bibinfo{person}{Aiden Low}, \bibinfo{person}{Alec Helyar}, \bibinfo{person}{Aleksander Madry}, \bibinfo{person}{Alex Beutel}, \bibinfo{person}{Alex Carney}, \bibinfo{person}{Alex Iftimie}, \bibinfo{person}{Alex Karpenko}, \bibinfo{person}{Alex~Tachard Passos}, \bibinfo{person}{Alexander Neitz}, \bibinfo{person}{Alexander Prokofiev}, \bibinfo{person}{Alexander Wei}, \bibinfo{person}{Allison Tam}, \bibinfo{person}{Ally Bennett}, \bibinfo{person}{Ananya Kumar}, \bibinfo{person}{Andre Saraiva}, \bibinfo{person}{Andrea Vallone}, \bibinfo{person}{Andrew Duberstein}, \bibinfo{person}{Andrew Kondrich}, \bibinfo{person}{Andrey Mishchenko}, \bibinfo{person}{Andy Applebaum}, \bibinfo{person}{Angela Jiang}, \bibinfo{person}{Ashvin Nair}, \bibinfo{person}{Barret Zoph}, \bibinfo{person}{Behrooz
  Ghorbani}, \bibinfo{person}{Ben Rossen}, \bibinfo{person}{Benjamin Sokolowsky}, \bibinfo{person}{Boaz Barak}, \bibinfo{person}{Bob McGrew}, \bibinfo{person}{Borys Minaiev}, \bibinfo{person}{Botao Hao}, \bibinfo{person}{Bowen Baker}, \bibinfo{person}{Brandon Houghton}, \bibinfo{person}{Brandon McKinzie}, \bibinfo{person}{Brydon Eastman}, \bibinfo{person}{Camillo Lugaresi}, \bibinfo{person}{Cary Bassin}, \bibinfo{person}{Cary Hudson}, \bibinfo{person}{Chak~Ming Li}, \bibinfo{person}{Charles de Bourcy}, \bibinfo{person}{Chelsea Voss}, \bibinfo{person}{Chen Shen}, \bibinfo{person}{Chong Zhang}, \bibinfo{person}{Chris Koch}, \bibinfo{person}{Chris Orsinger}, \bibinfo{person}{Christopher Hesse}, \bibinfo{person}{Claudia Fischer}, \bibinfo{person}{Clive Chan}, \bibinfo{person}{Dan Roberts}, \bibinfo{person}{Daniel Kappler}, \bibinfo{person}{Daniel Levy}, \bibinfo{person}{Daniel Selsam}, \bibinfo{person}{David Dohan}, \bibinfo{person}{David Farhi}, \bibinfo{person}{David Mely}, \bibinfo{person}{David Robinson},
  \bibinfo{person}{Dimitris Tsipras}, \bibinfo{person}{Doug Li}, \bibinfo{person}{Dragos Oprica}, \bibinfo{person}{Eben Freeman}, \bibinfo{person}{Eddie Zhang}, \bibinfo{person}{Edmund Wong}, \bibinfo{person}{Elizabeth Proehl}, \bibinfo{person}{Enoch Cheung}, \bibinfo{person}{Eric Mitchell}, \bibinfo{person}{Eric Wallace}, \bibinfo{person}{Erik Ritter}, \bibinfo{person}{Evan Mays}, \bibinfo{person}{Fan Wang}, \bibinfo{person}{Felipe~Petroski Such}, \bibinfo{person}{Filippo Raso}, \bibinfo{person}{Florencia Leoni}, \bibinfo{person}{Foivos Tsimpourlas}, \bibinfo{person}{Francis Song}, \bibinfo{person}{Fred von Lohmann}, \bibinfo{person}{Freddie Sulit}, \bibinfo{person}{Geoff Salmon}, \bibinfo{person}{Giambattista Parascandolo}, \bibinfo{person}{Gildas Chabot}, \bibinfo{person}{Grace Zhao}, \bibinfo{person}{Greg Brockman}, \bibinfo{person}{Guillaume Leclerc}, \bibinfo{person}{Hadi Salman}, \bibinfo{person}{Haiming Bao}, \bibinfo{person}{Hao Sheng}, \bibinfo{person}{Hart Andrin}, \bibinfo{person}{Hessam
  Bagherinezhad}, \bibinfo{person}{Hongyu Ren}, \bibinfo{person}{Hunter Lightman}, \bibinfo{person}{Hyung~Won Chung}, \bibinfo{person}{Ian Kivlichan}, \bibinfo{person}{Ian O'Connell}, \bibinfo{person}{Ian Osband}, \bibinfo{person}{Ignasi~Clavera Gilaberte}, \bibinfo{person}{Ilge Akkaya}, \bibinfo{person}{Ilya Kostrikov}, \bibinfo{person}{Ilya Sutskever}, \bibinfo{person}{Irina Kofman}, \bibinfo{person}{Jakub Pachocki}, \bibinfo{person}{James Lennon}, \bibinfo{person}{Jason Wei}, \bibinfo{person}{Jean Harb}, \bibinfo{person}{Jerry Twore}, \bibinfo{person}{Jiacheng Feng}, \bibinfo{person}{Jiahui Yu}, \bibinfo{person}{Jiayi Weng}, \bibinfo{person}{Jie Tang}, \bibinfo{person}{Jieqi Yu}, \bibinfo{person}{Joaquin~Quiñonero Candela}, \bibinfo{person}{Joe Palermo}, \bibinfo{person}{Joel Parish}, \bibinfo{person}{Johannes Heidecke}, \bibinfo{person}{John Hallman}, \bibinfo{person}{John Rizzo}, \bibinfo{person}{Jonathan Gordon}, \bibinfo{person}{Jonathan Uesato}, \bibinfo{person}{Jonathan Uesato},
  \bibinfo{person}{Jonathan Ward}, \bibinfo{person}{Joost Huizinga}, \bibinfo{person}{Julie Wang}, \bibinfo{person}{Kai Chen}, \bibinfo{person}{Kai Xiao}, \bibinfo{person}{Karan Singhal}, \bibinfo{person}{Karina Nguyen}, \bibinfo{person}{Karl Cobbe}, \bibinfo{person}{Katy Shi}, \bibinfo{person}{Kayla Wood}, \bibinfo{person}{Kendra Rimbach}, \bibinfo{person}{Keren Gu-Lemberg}, \bibinfo{person}{Keren GuLemberg}, \bibinfo{person}{Kevin Liu}, \bibinfo{person}{Kevin Lu}, \bibinfo{person}{Kevin Stone}, \bibinfo{person}{Kevin Yu}, \bibinfo{person}{Lama Ahmad}, \bibinfo{person}{Lauren Yang}, \bibinfo{person}{Leo Liu}, \bibinfo{person}{Leon Maksin}, \bibinfo{person}{Leyton Ho}, \bibinfo{person}{Liam Fedus}, \bibinfo{person}{Lilian Weng}, \bibinfo{person}{Linden Li}, \bibinfo{person}{Lindsay McCallum}, \bibinfo{person}{Lindsey Held}, \bibinfo{person}{Lorenz Kuhn}, \bibinfo{person}{Lukas Kondraciuk}, \bibinfo{person}{Lukasz Kaiser}, \bibinfo{person}{Luke Metz}, \bibinfo{person}{Madelaine Boyd}, \bibinfo{person}{Maja
  Trebacz}, \bibinfo{person}{Manas Joglekar}, \bibinfo{person}{Mark Chen}, \bibinfo{person}{Marko Tintor}, \bibinfo{person}{Mason Meyer}, \bibinfo{person}{Matt Jones}, \bibinfo{person}{Matt Kaufer}, \bibinfo{person}{Max Schwarzer}, \bibinfo{person}{Meghan Shah}, \bibinfo{person}{Mehmet Yatbaz}, \bibinfo{person}{Melody Guan}, \bibinfo{person}{Mengyuan Xu}, \bibinfo{person}{Mengyuan Yan}, \bibinfo{person}{Mia Glaese}, \bibinfo{person}{Mianna Chen}, \bibinfo{person}{Mianna Chen}, \bibinfo{person}{Michael Lampe}, \bibinfo{person}{Michael Malek}, \bibinfo{person}{Michele Wang}, \bibinfo{person}{Michelle Fradin}, \bibinfo{person}{Mike McClay}, \bibinfo{person}{Mikhail Pavlov}, \bibinfo{person}{Miles Wang}, \bibinfo{person}{Mingxuan Wang}, \bibinfo{person}{Mira Murati}, \bibinfo{person}{Mo Bavarian}, \bibinfo{person}{Mostafa Rohaninejad}, \bibinfo{person}{Nat McAleese}, \bibinfo{person}{Neil Chowdhury}, \bibinfo{person}{Neil Chowdhury}, \bibinfo{person}{Nick Ryder}, \bibinfo{person}{Nikolas Tezak},
  \bibinfo{person}{Noam Brown}, \bibinfo{person}{Ofir Nachum}, \bibinfo{person}{Oleg Boiko}, \bibinfo{person}{Oleg Murk}, \bibinfo{person}{Olivia Watkins}, \bibinfo{person}{Patrick Chao}, \bibinfo{person}{Paul Ashbourne}, \bibinfo{person}{Pavel Izmailov}, \bibinfo{person}{Peter Zhokhov}, \bibinfo{person}{Rachel Dias}, \bibinfo{person}{Rahul Arora}, \bibinfo{person}{Randall Lin}, \bibinfo{person}{Rapha~Gontijo Lopes}, \bibinfo{person}{Raz Gaon}, \bibinfo{person}{Reah Miyara}, \bibinfo{person}{Reimar Leike}, \bibinfo{person}{Renny Hwang}, \bibinfo{person}{Rhythm Garg}, \bibinfo{person}{Robin Brown}, \bibinfo{person}{Roshan James}, \bibinfo{person}{Rui Shu}, \bibinfo{person}{Ryan Cheu}, \bibinfo{person}{Ryan Greene}, \bibinfo{person}{Saachi Jain}, \bibinfo{person}{Sam Altman}, \bibinfo{person}{Sam Toizer}, \bibinfo{person}{Sam Toyer}, \bibinfo{person}{Samuel Miserendino}, \bibinfo{person}{Sandhini Agarwal}, \bibinfo{person}{Santiago Hernandez}, \bibinfo{person}{Sasha Baker}, \bibinfo{person}{Scott McKinney},
  \bibinfo{person}{Scottie Yan}, \bibinfo{person}{Shengjia Zhao}, \bibinfo{person}{Shengli Hu}, \bibinfo{person}{Shibani Santurkar}, \bibinfo{person}{Shraman~Ray Chaudhuri}, \bibinfo{person}{Shuyuan Zhang}, \bibinfo{person}{Siyuan Fu}, \bibinfo{person}{Spencer Papay}, \bibinfo{person}{Steph Lin}, \bibinfo{person}{Suchir Balaji}, \bibinfo{person}{Suvansh Sanjeev}, \bibinfo{person}{Szymon Sidor}, \bibinfo{person}{Tal Broda}, \bibinfo{person}{Aidan Clark}, \bibinfo{person}{Tao Wang}, \bibinfo{person}{Taylor Gordon}, \bibinfo{person}{Ted Sanders}, \bibinfo{person}{Tejal Patwardhan}, \bibinfo{person}{Thibault Sottiaux}, \bibinfo{person}{Thomas Degry}, \bibinfo{person}{Thomas Dimson}, \bibinfo{person}{Tianhao Zheng}, \bibinfo{person}{Timur Garipov}, \bibinfo{person}{Tom Stasi}, \bibinfo{person}{Trapit Bansal}, \bibinfo{person}{Trevor Creech}, \bibinfo{person}{Troy Peterson}, \bibinfo{person}{Tyna Eloundou}, \bibinfo{person}{Valerie Qi}, \bibinfo{person}{Vineet Kosaraju}, \bibinfo{person}{Vinnie Monaco},
  \bibinfo{person}{Vitchyr Pong}, \bibinfo{person}{Vlad Fomenko}, \bibinfo{person}{Weiyi Zheng}, \bibinfo{person}{Wenda Zhou}, \bibinfo{person}{Wes McCabe}, \bibinfo{person}{Wojciech Zaremba}, \bibinfo{person}{Yann Dubois}, \bibinfo{person}{Yinghai Lu}, \bibinfo{person}{Yining Chen}, \bibinfo{person}{Young Cha}, \bibinfo{person}{Yu Bai}, \bibinfo{person}{Yuchen He}, \bibinfo{person}{Yuchen Zhang}, \bibinfo{person}{Yunyun Wang}, \bibinfo{person}{Zheng Shao}, {and} \bibinfo{person}{Zhuohan Li}.} \bibinfo{year}{2024}\natexlab{}.
\newblock \bibinfo{title}{OpenAI o1 System Card}.
\newblock
\newblock
\showeprint[arxiv]{2412.16720}~[cs.AI]
\urldef\tempurl%
\url{https://arxiv.org/abs/2412.16720}
\showURL{%
\tempurl}


\bibitem[\protect\citeauthoryear{OpenAI}{OpenAI}{2023}]%
        {DBLP:journals/corr/abs-2303-08774}
\bibfield{author}{\bibinfo{person}{OpenAI}.} \bibinfo{year}{2023}\natexlab{}.
\newblock \showarticletitle{{GPT-4} Technical Report}.
\newblock \bibinfo{journal}{\emph{CoRR}}  \bibinfo{volume}{abs/2303.08774} (\bibinfo{year}{2023}).
\newblock
\urldef\tempurl%
\url{https://doi.org/10.48550/ARXIV.2303.08774}
\showDOI{\tempurl}
\showeprint[arXiv]{2303.08774}


\bibitem[\protect\citeauthoryear{Ouyang, Wu, Jiang, Almeida, Wainwright, Mishkin, Zhang, Agarwal, Slama, Ray, Schulman, Hilton, Kelton, Miller, Simens, Askell, Welinder, Christiano, Leike, and Lowe}{Ouyang et~al\mbox{.}}{2022}]%
        {DBLP:conf/nips/Ouyang0JAWMZASR22}
\bibfield{author}{\bibinfo{person}{Long Ouyang}, \bibinfo{person}{Jeffrey Wu}, \bibinfo{person}{Xu Jiang}, \bibinfo{person}{Diogo Almeida}, \bibinfo{person}{Carroll~L. Wainwright}, \bibinfo{person}{Pamela Mishkin}, \bibinfo{person}{Chong Zhang}, \bibinfo{person}{Sandhini Agarwal}, \bibinfo{person}{Katarina Slama}, \bibinfo{person}{Alex Ray}, \bibinfo{person}{John Schulman}, \bibinfo{person}{Jacob Hilton}, \bibinfo{person}{Fraser Kelton}, \bibinfo{person}{Luke Miller}, \bibinfo{person}{Maddie Simens}, \bibinfo{person}{Amanda Askell}, \bibinfo{person}{Peter Welinder}, \bibinfo{person}{Paul~F. Christiano}, \bibinfo{person}{Jan Leike}, {and} \bibinfo{person}{Ryan Lowe}.} \bibinfo{year}{2022}\natexlab{}.
\newblock \showarticletitle{Training language models to follow instructions with human feedback}. In \bibinfo{booktitle}{\emph{Proc. NeurIPS}}.
\newblock
\urldef\tempurl%
\url{http://papers.nips.cc/paper\_files/paper/2022/hash/b1efde53be364a73914f58805a001731-Abstract-Conference.html}
\showURL{%
\tempurl}


\bibitem[\protect\citeauthoryear{Penedo, Malartic, Hesslow, Cojocaru, Alobeidli, Cappelli, Pannier, Almazrouei, and Launay}{Penedo et~al\mbox{.}}{2023}]%
        {DBLP:conf/nips/PenedoMHCACPAL23}
\bibfield{author}{\bibinfo{person}{Guilherme Penedo}, \bibinfo{person}{Quentin Malartic}, \bibinfo{person}{Daniel Hesslow}, \bibinfo{person}{Ruxandra Cojocaru}, \bibinfo{person}{Hamza Alobeidli}, \bibinfo{person}{Alessandro Cappelli}, \bibinfo{person}{Baptiste Pannier}, \bibinfo{person}{Ebtesam Almazrouei}, {and} \bibinfo{person}{Julien Launay}.} \bibinfo{year}{2023}\natexlab{}.
\newblock \showarticletitle{The RefinedWeb Dataset for Falcon {LLM:} Outperforming Curated Corpora with Web Data Only}. In \bibinfo{booktitle}{\emph{Proc. NeurIPS}}.
\newblock
\urldef\tempurl%
\url{http://papers.nips.cc/paper\_files/paper/2023/hash/fa3ed726cc5073b9c31e3e49a807789c-Abstract-Datasets\_and\_Benchmarks.html}
\showURL{%
\tempurl}


\bibitem[\protect\citeauthoryear{Poess, Smith, Kollar, and Larson}{Poess et~al\mbox{.}}{2002}]%
        {DBLP:conf/tpcds/Poess02}
\bibfield{author}{\bibinfo{person}{Meikel Poess}, \bibinfo{person}{Bryan Smith}, \bibinfo{person}{Lubor Kollar}, {and} \bibinfo{person}{Paul Larson}.} \bibinfo{year}{2002}\natexlab{}.
\newblock \showarticletitle{Tpc-ds, taking decision support benchmarking to the next level}. In \bibinfo{booktitle}{\emph{Proceedings of the 2002 ACM SIGMOD international conference on Management of data}}. \bibinfo{pages}{582--587}.
\newblock


\bibitem[\protect\citeauthoryear{Rafailov, Sharma, Mitchell, Manning, Ermon, and Finn}{Rafailov et~al\mbox{.}}{2023}]%
        {DBLP:conf/nips/RafailovSMMEF23}
\bibfield{author}{\bibinfo{person}{Rafael Rafailov}, \bibinfo{person}{Archit Sharma}, \bibinfo{person}{Eric Mitchell}, \bibinfo{person}{Christopher~D. Manning}, \bibinfo{person}{Stefano Ermon}, {and} \bibinfo{person}{Chelsea Finn}.} \bibinfo{year}{2023}\natexlab{}.
\newblock \showarticletitle{Direct Preference Optimization: Your Language Model is Secretly a Reward Model}. In \bibinfo{booktitle}{\emph{Proc. NeurIPS}}.
\newblock
\urldef\tempurl%
\url{http://papers.nips.cc/paper\_files/paper/2023/hash/a85b405ed65c6477a4fe8302b5e06ce7-Abstract-Conference.html}
\showURL{%
\tempurl}


\bibitem[\protect\citeauthoryear{Roziere, Gehring, Gloeckle, Sootla, Gat, Tan, Adi, Liu, Remez, Rapin, et~al\mbox{.}}{Roziere et~al\mbox{.}}{2023}]%
        {roziere2023code}
\bibfield{author}{\bibinfo{person}{Baptiste Roziere}, \bibinfo{person}{Jonas Gehring}, \bibinfo{person}{Fabian Gloeckle}, \bibinfo{person}{Sten Sootla}, \bibinfo{person}{Itai Gat}, \bibinfo{person}{Xiaoqing~Ellen Tan}, \bibinfo{person}{Yossi Adi}, \bibinfo{person}{Jingyu Liu}, \bibinfo{person}{Tal Remez}, \bibinfo{person}{J{\'e}r{\'e}my Rapin}, {et~al\mbox{.}}} \bibinfo{year}{2023}\natexlab{}.
\newblock \showarticletitle{Code llama: Open foundation models for code}.
\newblock \bibinfo{journal}{\emph{arXiv preprint arXiv:2308.12950}} (\bibinfo{year}{2023}).
\newblock


\bibitem[\protect\citeauthoryear{Selinger, Astrahan, Chamberlin, Lorie, and Price}{Selinger et~al\mbox{.}}{1979}]%
        {DBLP:conf/sigmod/SelingerACLP79}
\bibfield{author}{\bibinfo{person}{Patricia~G. Selinger}, \bibinfo{person}{Morton~M. Astrahan}, \bibinfo{person}{Donald~D. Chamberlin}, \bibinfo{person}{Raymond~A. Lorie}, {and} \bibinfo{person}{Thomas~G. Price}.} \bibinfo{year}{1979}\natexlab{}.
\newblock \showarticletitle{Access Path Selection in a Relational Database Management System}. In \bibinfo{booktitle}{\emph{Proc. {SIGMOD}}}, \bibfield{editor}{\bibinfo{person}{Philip~A. Bernstein}} (Ed.). \bibinfo{publisher}{{ACM}}, \bibinfo{pages}{23--34}.
\newblock
\urldef\tempurl%
\url{https://doi.org/10.1145/582095.582099}
\showDOI{\tempurl}


\bibitem[\protect\citeauthoryear{Silver, Huang, Maddison, Guez, Sifre, Van Den~Driessche, Schrittwieser, Antonoglou, Panneershelvam, Lanctot, et~al\mbox{.}}{Silver et~al\mbox{.}}{2016}]%
        {silver2016mastering}
\bibfield{author}{\bibinfo{person}{David Silver}, \bibinfo{person}{Aja Huang}, \bibinfo{person}{Chris~J Maddison}, \bibinfo{person}{Arthur Guez}, \bibinfo{person}{Laurent Sifre}, \bibinfo{person}{George Van Den~Driessche}, \bibinfo{person}{Julian Schrittwieser}, \bibinfo{person}{Ioannis Antonoglou}, \bibinfo{person}{Veda Panneershelvam}, \bibinfo{person}{Marc Lanctot}, {et~al\mbox{.}}} \bibinfo{year}{2016}\natexlab{}.
\newblock \showarticletitle{Mastering the game of Go with deep neural networks and tree search}.
\newblock \bibinfo{journal}{\emph{nature}} \bibinfo{volume}{529}, \bibinfo{number}{7587} (\bibinfo{year}{2016}), \bibinfo{pages}{484--489}.
\newblock


\bibitem[\protect\citeauthoryear{Stiennon, Ouyang, Wu, Ziegler, Lowe, Voss, Radford, Amodei, and Christiano}{Stiennon et~al\mbox{.}}{2020}]%
        {DBLP:conf/nips/StiennonO0ZLVRA20}
\bibfield{author}{\bibinfo{person}{Nisan Stiennon}, \bibinfo{person}{Long Ouyang}, \bibinfo{person}{Jeffrey Wu}, \bibinfo{person}{Daniel~M. Ziegler}, \bibinfo{person}{Ryan Lowe}, \bibinfo{person}{Chelsea Voss}, \bibinfo{person}{Alec Radford}, \bibinfo{person}{Dario Amodei}, {and} \bibinfo{person}{Paul~F. Christiano}.} \bibinfo{year}{2020}\natexlab{}.
\newblock \showarticletitle{Learning to summarize with human feedback}. In \bibinfo{booktitle}{\emph{Proc. NeurIPS}}.
\newblock
\urldef\tempurl%
\url{https://proceedings.neurips.cc/paper/2020/hash/1f89885d556929e98d3ef9b86448f951-Abstract.html}
\showURL{%
\tempurl}


\bibitem[\protect\citeauthoryear{Sui, Zhou, Zhou, Han, and Zhang}{Sui et~al\mbox{.}}{2024}]%
        {DBLP:conf/wsdm/SuiZZH024}
\bibfield{author}{\bibinfo{person}{Yuan Sui}, \bibinfo{person}{Mengyu Zhou}, \bibinfo{person}{Mingjie Zhou}, \bibinfo{person}{Shi Han}, {and} \bibinfo{person}{Dongmei Zhang}.} \bibinfo{year}{2024}\natexlab{}.
\newblock \showarticletitle{Table Meets {LLM:} Can Large Language Models Understand Structured Table Data? {A} Benchmark and Empirical Study}. In \bibinfo{booktitle}{\emph{Proc. {WSDM}}}. \bibinfo{publisher}{{ACM}}, \bibinfo{pages}{645--654}.
\newblock
\urldef\tempurl%
\url{https://doi.org/10.1145/3616855.3635752}
\showDOI{\tempurl}


\bibitem[\protect\citeauthoryear{Sun, Arik, Nakhost, Dai, Sinha, Yin, and Pfister}{Sun et~al\mbox{.}}{2023}]%
        {DBLP:journals/corr/abs-2306-00739}
\bibfield{author}{\bibinfo{person}{Ruoxi Sun}, \bibinfo{person}{Sercan~{\"{O}}. Arik}, \bibinfo{person}{Hootan Nakhost}, \bibinfo{person}{Hanjun Dai}, \bibinfo{person}{Rajarishi Sinha}, \bibinfo{person}{Pengcheng Yin}, {and} \bibinfo{person}{Tomas Pfister}.} \bibinfo{year}{2023}\natexlab{}.
\newblock \showarticletitle{SQL-PaLM: Improved Large Language Model Adaptation for Text-to-SQL}.
\newblock \bibinfo{journal}{\emph{CoRR}}  \bibinfo{volume}{abs/2306.00739} (\bibinfo{year}{2023}).
\newblock
\urldef\tempurl%
\url{https://doi.org/10.48550/ARXIV.2306.00739}
\showDOI{\tempurl}
\showeprint[arXiv]{2306.00739}


\bibitem[\protect\citeauthoryear{Tai, Socher, and Manning}{Tai et~al\mbox{.}}{2015}]%
        {DBLP:conf/acl/TaiSM15}
\bibfield{author}{\bibinfo{person}{Kai~Sheng Tai}, \bibinfo{person}{Richard Socher}, {and} \bibinfo{person}{Christopher~D. Manning}.} \bibinfo{year}{2015}\natexlab{}.
\newblock \showarticletitle{Improved Semantic Representations From Tree-Structured Long Short-Term Memory Networks}. In \bibinfo{booktitle}{\emph{Proceedings of the 53rd Annual Meeting of the Association for Computational Linguistics and the 7th International Joint Conference on Natural Language Processing of the Asian Federation of Natural Language Processing, {ACL} 2015, July 26-31, 2015, Beijing, China, Volume 1: Long Papers}}. \bibinfo{publisher}{The Association for Computer Linguistics}, \bibinfo{pages}{1556--1566}.
\newblock
\urldef\tempurl%
\url{https://doi.org/10.3115/V1/P15-1150}
\showDOI{\tempurl}


\bibitem[\protect\citeauthoryear{Touvron, Lavril, Izacard, Martinet, Lachaux, Lacroix, Rozi{\`{e}}re, Goyal, Hambro, Azhar, Rodriguez, Joulin, Grave, and Lample}{Touvron et~al\mbox{.}}{2023a}]%
        {DBLP:journals/corr/abs-2302-13971}
\bibfield{author}{\bibinfo{person}{Hugo Touvron}, \bibinfo{person}{Thibaut Lavril}, \bibinfo{person}{Gautier Izacard}, \bibinfo{person}{Xavier Martinet}, \bibinfo{person}{Marie{-}Anne Lachaux}, \bibinfo{person}{Timoth{\'{e}}e Lacroix}, \bibinfo{person}{Baptiste Rozi{\`{e}}re}, \bibinfo{person}{Naman Goyal}, \bibinfo{person}{Eric Hambro}, \bibinfo{person}{Faisal Azhar}, \bibinfo{person}{Aur{\'{e}}lien Rodriguez}, \bibinfo{person}{Armand Joulin}, \bibinfo{person}{Edouard Grave}, {and} \bibinfo{person}{Guillaume Lample}.} \bibinfo{year}{2023}\natexlab{a}.
\newblock \showarticletitle{LLaMA: Open and Efficient Foundation Language Models}.
\newblock \bibinfo{journal}{\emph{CoRR}}  \bibinfo{volume}{abs/2302.13971} (\bibinfo{year}{2023}).
\newblock
\urldef\tempurl%
\url{https://doi.org/10.48550/ARXIV.2302.13971}
\showDOI{\tempurl}
\showeprint[arXiv]{2302.13971}


\bibitem[\protect\citeauthoryear{Touvron, Martin, Stone, Albert, Almahairi, Babaei, Bashlykov, Batra, Bhargava, Bhosale, et~al\mbox{.}}{Touvron et~al\mbox{.}}{2023b}]%
        {touvron2023llama}
\bibfield{author}{\bibinfo{person}{Hugo Touvron}, \bibinfo{person}{Louis Martin}, \bibinfo{person}{Kevin Stone}, \bibinfo{person}{Peter Albert}, \bibinfo{person}{Amjad Almahairi}, \bibinfo{person}{Yasmine Babaei}, \bibinfo{person}{Nikolay Bashlykov}, \bibinfo{person}{Soumya Batra}, \bibinfo{person}{Prajjwal Bhargava}, \bibinfo{person}{Shruti Bhosale}, {et~al\mbox{.}}} \bibinfo{year}{2023}\natexlab{b}.
\newblock \showarticletitle{Llama 2: Open foundation and fine-tuned chat models, 2023}.
\newblock \bibinfo{journal}{\emph{URL https://arxiv. org/abs/2307.09288}} (\bibinfo{year}{2023}).
\newblock


\bibitem[\protect\citeauthoryear{Trummer}{Trummer}{2023}]%
        {DBLP:journals/pvldb/Trummer23}
\bibfield{author}{\bibinfo{person}{Immanuel Trummer}.} \bibinfo{year}{2023}\natexlab{}.
\newblock \showarticletitle{Demonstrating {GPT-DB:} Generating Query-Specific and Customizable Code for {SQL} Processing with {GPT-4}}.
\newblock \bibinfo{journal}{\emph{Proc. {VLDB} Endow.}} \bibinfo{volume}{16}, \bibinfo{number}{12} (\bibinfo{year}{2023}), \bibinfo{pages}{4098--4101}.
\newblock
\urldef\tempurl%
\url{https://doi.org/10.14778/3611540.3611630}
\showDOI{\tempurl}


\bibitem[\protect\citeauthoryear{Urban and Binnig}{Urban and Binnig}{2024}]%
        {DBLP:conf/cidr/UrbanB24}
\bibfield{author}{\bibinfo{person}{Matthias Urban} {and} \bibinfo{person}{Carsten Binnig}.} \bibinfo{year}{2024}\natexlab{}.
\newblock \showarticletitle{{CAESURA:} Language Models as Multi-Modal Query Planners}. In \bibinfo{booktitle}{\emph{Proc. {CIDR}}}. \bibinfo{publisher}{www.cidrdb.org}.
\newblock
\urldef\tempurl%
\url{https://www.cidrdb.org/cidr2024/papers/p14-urban.pdf}
\showURL{%
\tempurl}


\bibitem[\protect\citeauthoryear{Vaswani, Shazeer, Parmar, Uszkoreit, Jones, Gomez, Kaiser, and Polosukhin}{Vaswani et~al\mbox{.}}{2017}]%
        {DBLP:conf/nips/VaswaniSPUJGKP17}
\bibfield{author}{\bibinfo{person}{Ashish Vaswani}, \bibinfo{person}{Noam Shazeer}, \bibinfo{person}{Niki Parmar}, \bibinfo{person}{Jakob Uszkoreit}, \bibinfo{person}{Llion Jones}, \bibinfo{person}{Aidan~N. Gomez}, \bibinfo{person}{Lukasz Kaiser}, {and} \bibinfo{person}{Illia Polosukhin}.} \bibinfo{year}{2017}\natexlab{}.
\newblock \showarticletitle{Attention is All you Need}. In \bibinfo{booktitle}{\emph{Proc. {NIPS}}}. \bibinfo{pages}{5998--6008}.
\newblock
\urldef\tempurl%
\url{https://proceedings.neurips.cc/paper/2017/hash/3f5ee243547dee91fbd053c1c4a845aa-Abstract.html}
\showURL{%
\tempurl}


\bibitem[\protect\citeauthoryear{Wang, Ivison, Dasigi, Hessel, Khot, Chandu, Wadden, MacMillan, Smith, Beltagy, and Hajishirzi}{Wang et~al\mbox{.}}{2023}]%
        {DBLP:conf/nips/WangIDHKCWMSBH23}
\bibfield{author}{\bibinfo{person}{Yizhong Wang}, \bibinfo{person}{Hamish Ivison}, \bibinfo{person}{Pradeep Dasigi}, \bibinfo{person}{Jack Hessel}, \bibinfo{person}{Tushar Khot}, \bibinfo{person}{Khyathi~Raghavi Chandu}, \bibinfo{person}{David Wadden}, \bibinfo{person}{Kelsey MacMillan}, \bibinfo{person}{Noah~A. Smith}, \bibinfo{person}{Iz Beltagy}, {and} \bibinfo{person}{Hannaneh Hajishirzi}.} \bibinfo{year}{2023}\natexlab{}.
\newblock \showarticletitle{How Far Can Camels Go? Exploring the State of Instruction Tuning on Open Resources}. In \bibinfo{booktitle}{\emph{Proc. NeurIPS}}.
\newblock
\urldef\tempurl%
\url{http://papers.nips.cc/paper\_files/paper/2023/hash/ec6413875e4ab08d7bc4d8e225263398-Abstract-Datasets\_and\_Benchmarks.html}
\showURL{%
\tempurl}


\bibitem[\protect\citeauthoryear{Wei, Bosma, Zhao, Guu, Yu, Lester, Du, Dai, and Le}{Wei et~al\mbox{.}}{2022a}]%
        {DBLP:conf/iclr/WeiBZGYLDDL22}
\bibfield{author}{\bibinfo{person}{Jason Wei}, \bibinfo{person}{Maarten Bosma}, \bibinfo{person}{Vincent~Y. Zhao}, \bibinfo{person}{Kelvin Guu}, \bibinfo{person}{Adams~Wei Yu}, \bibinfo{person}{Brian Lester}, \bibinfo{person}{Nan Du}, \bibinfo{person}{Andrew~M. Dai}, {and} \bibinfo{person}{Quoc~V. Le}.} \bibinfo{year}{2022}\natexlab{a}.
\newblock \showarticletitle{Finetuned Language Models are Zero-Shot Learners}. In \bibinfo{booktitle}{\emph{Proc. {ICLR}}}. \bibinfo{publisher}{OpenReview.net}.
\newblock
\urldef\tempurl%
\url{https://openreview.net/forum?id=gEZrGCozdqR}
\showURL{%
\tempurl}


\bibitem[\protect\citeauthoryear{Wei, Tay, Bommasani, Raffel, Zoph, Borgeaud, Yogatama, Bosma, Zhou, Metzler, Chi, Hashimoto, Vinyals, Liang, Dean, and Fedus}{Wei et~al\mbox{.}}{2022b}]%
        {wei2022emergent}
\bibfield{author}{\bibinfo{person}{Jason Wei}, \bibinfo{person}{Yi Tay}, \bibinfo{person}{Rishi Bommasani}, \bibinfo{person}{Colin Raffel}, \bibinfo{person}{Barret Zoph}, \bibinfo{person}{Sebastian Borgeaud}, \bibinfo{person}{Dani Yogatama}, \bibinfo{person}{Maarten Bosma}, \bibinfo{person}{Denny Zhou}, \bibinfo{person}{Donald Metzler}, \bibinfo{person}{Ed~H. Chi}, \bibinfo{person}{Tatsunori Hashimoto}, \bibinfo{person}{Oriol Vinyals}, \bibinfo{person}{Percy Liang}, \bibinfo{person}{Jeff Dean}, {and} \bibinfo{person}{William Fedus}.} \bibinfo{year}{2022}\natexlab{b}.
\newblock \showarticletitle{Emergent Abilities of Large Language Models}.
\newblock \bibinfo{journal}{\emph{Transactions on Machine Learning Research}} (\bibinfo{year}{2022}).
\newblock
\showISSN{2835-8856}
\urldef\tempurl%
\url{https://openreview.net/forum?id=yzkSU5zdwD}
\showURL{%
\tempurl}
\newblock
\shownote{Survey Certification.}


\bibitem[\protect\citeauthoryear{Wu, Wan, Zhang, Sui, Wei, Zhao, Xu, and Jin}{Wu et~al\mbox{.}}{2024}]%
        {DBLP:journals/pacmmod/Wu00SWZ0024}
\bibfield{author}{\bibinfo{person}{Yang Wu}, \bibinfo{person}{Yao Wan}, \bibinfo{person}{Hongyu Zhang}, \bibinfo{person}{Yulei Sui}, \bibinfo{person}{Wucai Wei}, \bibinfo{person}{Wei Zhao}, \bibinfo{person}{Guandong Xu}, {and} \bibinfo{person}{Hai Jin}.} \bibinfo{year}{2024}\natexlab{}.
\newblock \showarticletitle{Automated Data Visualization from Natural Language via Large Language Models: An Exploratory Study}.
\newblock \bibinfo{journal}{\emph{Proc. {ACM} Manag. Data}} \bibinfo{volume}{2}, \bibinfo{number}{3} (\bibinfo{year}{2024}), \bibinfo{pages}{115}.
\newblock
\urldef\tempurl%
\url{https://doi.org/10.1145/3654992}
\showDOI{\tempurl}


\bibitem[\protect\citeauthoryear{Xue, Jiang, Shi, Cheng, Chen, Yang, Zhang, He, Zhang, Wei, Zhao, Zhou, Qi, Yi, Liu, and Chen}{Xue et~al\mbox{.}}{2023}]%
        {DBLP:journals/corr/abs-2312-17449}
\bibfield{author}{\bibinfo{person}{Siqiao Xue}, \bibinfo{person}{Caigao Jiang}, \bibinfo{person}{Wenhui Shi}, \bibinfo{person}{Fangyin Cheng}, \bibinfo{person}{Keting Chen}, \bibinfo{person}{Hongjun Yang}, \bibinfo{person}{Zhiping Zhang}, \bibinfo{person}{Jianshan He}, \bibinfo{person}{Hongyang Zhang}, \bibinfo{person}{Ganglin Wei}, \bibinfo{person}{Wang Zhao}, \bibinfo{person}{Fan Zhou}, \bibinfo{person}{Danrui Qi}, \bibinfo{person}{Hong Yi}, \bibinfo{person}{Shaodong Liu}, {and} \bibinfo{person}{Faqiang Chen}.} \bibinfo{year}{2023}\natexlab{}.
\newblock \showarticletitle{{DB-GPT:} Empowering Database Interactions with Private Large Language Models}.
\newblock \bibinfo{journal}{\emph{CoRR}}  \bibinfo{volume}{abs/2312.17449} (\bibinfo{year}{2023}).
\newblock
\urldef\tempurl%
\url{https://doi.org/10.48550/ARXIV.2312.17449}
\showDOI{\tempurl}
\showeprint[arXiv]{2312.17449}


\bibitem[\protect\citeauthoryear{Yang, Chiang, Luan, Mittal, Luo, and Stoica}{Yang et~al\mbox{.}}{2022}]%
        {DBLP:conf/sigmod/YangC0MLS22}
\bibfield{author}{\bibinfo{person}{Zongheng Yang}, \bibinfo{person}{Wei{-}Lin Chiang}, \bibinfo{person}{Sifei Luan}, \bibinfo{person}{Gautam Mittal}, \bibinfo{person}{Michael Luo}, {and} \bibinfo{person}{Ion Stoica}.} \bibinfo{year}{2022}\natexlab{}.
\newblock \showarticletitle{Balsa: Learning a Query Optimizer Without Expert Demonstrations}. In \bibinfo{booktitle}{\emph{Proc. {SIGMOD}}}. \bibinfo{publisher}{{ACM}}, \bibinfo{pages}{931--944}.
\newblock
\urldef\tempurl%
\url{https://doi.org/10.1145/3514221.3517885}
\showDOI{\tempurl}


\bibitem[\protect\citeauthoryear{Yu, Chai, Li, and Liu}{Yu et~al\mbox{.}}{2022}]%
        {DBLP:journals/pvldb/YuC0L22}
\bibfield{author}{\bibinfo{person}{Xiang Yu}, \bibinfo{person}{Chengliang Chai}, \bibinfo{person}{Guoliang Li}, {and} \bibinfo{person}{Jiabin Liu}.} \bibinfo{year}{2022}\natexlab{}.
\newblock \showarticletitle{Cost-based or Learning-based? {A} Hybrid Query Optimizer for Query Plan Selection}.
\newblock \bibinfo{journal}{\emph{Proc. {VLDB} Endow.}} \bibinfo{volume}{15}, \bibinfo{number}{13} (\bibinfo{year}{2022}), \bibinfo{pages}{3924--3936}.
\newblock
\urldef\tempurl%
\url{https://doi.org/10.14778/3565838.3565846}
\showDOI{\tempurl}


\bibitem[\protect\citeauthoryear{Yu, Li, Chai, and Tang}{Yu et~al\mbox{.}}{2020}]%
        {DBLP:conf/icde/Yu0C020}
\bibfield{author}{\bibinfo{person}{Xiang Yu}, \bibinfo{person}{Guoliang Li}, \bibinfo{person}{Chengliang Chai}, {and} \bibinfo{person}{Nan Tang}.} \bibinfo{year}{2020}\natexlab{}.
\newblock \showarticletitle{Reinforcement Learning with Tree-LSTM for Join Order Selection}. In \bibinfo{booktitle}{\emph{Proc. {ICDE}}}. \bibinfo{publisher}{{IEEE}}, \bibinfo{pages}{1297--1308}.
\newblock
\urldef\tempurl%
\url{https://doi.org/10.1109/ICDE48307.2020.00116}
\showDOI{\tempurl}


\bibitem[\protect\citeauthoryear{Zhang, Roller, Goyal, Artetxe, Chen, Chen, Dewan, Diab, Li, Lin, Mihaylov, Ott, Shleifer, Shuster, Simig, Koura, Sridhar, Wang, and Zettlemoyer}{Zhang et~al\mbox{.}}{2022}]%
        {DBLP:journals/corr/abs-2205-01068}
\bibfield{author}{\bibinfo{person}{Susan Zhang}, \bibinfo{person}{Stephen Roller}, \bibinfo{person}{Naman Goyal}, \bibinfo{person}{Mikel Artetxe}, \bibinfo{person}{Moya Chen}, \bibinfo{person}{Shuohui Chen}, \bibinfo{person}{Christopher Dewan}, \bibinfo{person}{Mona~T. Diab}, \bibinfo{person}{Xian Li}, \bibinfo{person}{Xi~Victoria Lin}, \bibinfo{person}{Todor Mihaylov}, \bibinfo{person}{Myle Ott}, \bibinfo{person}{Sam Shleifer}, \bibinfo{person}{Kurt Shuster}, \bibinfo{person}{Daniel Simig}, \bibinfo{person}{Punit~Singh Koura}, \bibinfo{person}{Anjali Sridhar}, \bibinfo{person}{Tianlu Wang}, {and} \bibinfo{person}{Luke Zettlemoyer}.} \bibinfo{year}{2022}\natexlab{}.
\newblock \showarticletitle{{OPT:} Open Pre-trained Transformer Language Models}.
\newblock \bibinfo{journal}{\emph{CoRR}}  \bibinfo{volume}{abs/2205.01068} (\bibinfo{year}{2022}).
\newblock
\urldef\tempurl%
\url{https://doi.org/10.48550/ARXIV.2205.01068}
\showDOI{\tempurl}
\showeprint[arXiv]{2205.01068}


\bibitem[\protect\citeauthoryear{Zhao, Zhou, Li, Tang, Wang, Hou, Min, Zhang, Zhang, Dong, Du, Yang, Chen, Chen, Jiang, Ren, Li, Tang, Liu, Liu, Nie, and Wen}{Zhao et~al\mbox{.}}{2024}]%
        {zhao2024surveylargelanguagemodels}
\bibfield{author}{\bibinfo{person}{Wayne~Xin Zhao}, \bibinfo{person}{Kun Zhou}, \bibinfo{person}{Junyi Li}, \bibinfo{person}{Tianyi Tang}, \bibinfo{person}{Xiaolei Wang}, \bibinfo{person}{Yupeng Hou}, \bibinfo{person}{Yingqian Min}, \bibinfo{person}{Beichen Zhang}, \bibinfo{person}{Junjie Zhang}, \bibinfo{person}{Zican Dong}, \bibinfo{person}{Yifan Du}, \bibinfo{person}{Chen Yang}, \bibinfo{person}{Yushuo Chen}, \bibinfo{person}{Zhipeng Chen}, \bibinfo{person}{Jinhao Jiang}, \bibinfo{person}{Ruiyang Ren}, \bibinfo{person}{Yifan Li}, \bibinfo{person}{Xinyu Tang}, \bibinfo{person}{Zikang Liu}, \bibinfo{person}{Peiyu Liu}, \bibinfo{person}{Jian-Yun Nie}, {and} \bibinfo{person}{Ji-Rong Wen}.} \bibinfo{year}{2024}\natexlab{}.
\newblock \bibinfo{title}{A Survey of Large Language Models}.
\newblock
\newblock
\showeprint[arxiv]{2303.18223}~[cs.CL]
\urldef\tempurl%
\url{https://arxiv.org/abs/2303.18223}
\showURL{%
\tempurl}


\bibitem[\protect\citeauthoryear{Zhu, Chen, Ding, Chen, Pfadler, Wu, and Zhou}{Zhu et~al\mbox{.}}{2023}]%
        {DBLP:journals/pvldb/ZhuCDCPWZ23}
\bibfield{author}{\bibinfo{person}{Rong Zhu}, \bibinfo{person}{Wei Chen}, \bibinfo{person}{Bolin Ding}, \bibinfo{person}{Xingguang Chen}, \bibinfo{person}{Andreas Pfadler}, \bibinfo{person}{Ziniu Wu}, {and} \bibinfo{person}{Jingren Zhou}.} \bibinfo{year}{2023}\natexlab{}.
\newblock \showarticletitle{Lero: {A} Learning-to-Rank Query Optimizer}.
\newblock \bibinfo{journal}{\emph{Proc. {VLDB} Endow.}} \bibinfo{volume}{16}, \bibinfo{number}{6} (\bibinfo{year}{2023}), \bibinfo{pages}{1466--1479}.
\newblock
\urldef\tempurl%
\url{https://doi.org/10.14778/3583140.3583160}
\showDOI{\tempurl}


\bibitem[\protect\citeauthoryear{Ziegler, Stiennon, Wu, Brown, Radford, Amodei, Christiano, and Irving}{Ziegler et~al\mbox{.}}{2019}]%
        {DBLP:journals/corr/abs-1909-08593}
\bibfield{author}{\bibinfo{person}{Daniel~M. Ziegler}, \bibinfo{person}{Nisan Stiennon}, \bibinfo{person}{Jeffrey Wu}, \bibinfo{person}{Tom~B. Brown}, \bibinfo{person}{Alec Radford}, \bibinfo{person}{Dario Amodei}, \bibinfo{person}{Paul~F. Christiano}, {and} \bibinfo{person}{Geoffrey Irving}.} \bibinfo{year}{2019}\natexlab{}.
\newblock \showarticletitle{Fine-Tuning Language Models from Human Preferences}.
\newblock \bibinfo{journal}{\emph{CoRR}}  \bibinfo{volume}{abs/1909.08593} (\bibinfo{year}{2019}).
\newblock
\showeprint[arXiv]{1909.08593}
\urldef\tempurl%
\url{http://arxiv.org/abs/1909.08593}
\showURL{%
\tempurl}


\end{thebibliography}

\clearpage

\end{document}